\documentclass[twocolumn,rmp,aps,showpacs,amssymb,amsmath,amsmath,superscriptaddress]{revtex4-1}

\usepackage{graphicx}
\usepackage[normalem]{ulem}
\usepackage{epstopdf}
\usepackage{bm}
\usepackage{hyperref}
\usepackage{comment}
\usepackage{enumerate}
\usepackage{color}
\usepackage{ulem}
\usepackage[abs]{overpic}

\hypersetup{%
   pdfpagemode=None, 
   pdfstartpage=1,
   pdfmenubar=true,
   pdftoolbar=true,
   colorlinks = true,
   linkcolor=blue,
   citecolor=blue,
   urlcolor=blue,
   bookmarksopen=false
 }

\newcommand{\be}{\begin{equation}}
\newcommand{\ee}{\end{equation}}
\newcommand{\as}{a_{\mathrm{3D}}}

\newcommand{\mbf}{\mathbf}
\newcommand{\mrm}{\mathrm}

\newcommand{\br}[0]{\mathbf r}
\newcommand{\bR}[0]{\mathbf R}

\begin{document}
\title{Cold hybrid ion-atom systems}

\author{Micha\l~Tomza}
\affiliation{Faculty of Physics$\mbox{,}$  University of Warsaw$\mbox{,}$ Pasteura 5$\mbox{,}$ 02-093 Warsaw$\mbox{,}$ Poland}
\author{Krzysztof Jachymski}
\affiliation{Institute for Theoretical Physics III \& Center for Integrated Quantum Science and Technology$\mbox{,}$ University of Stuttgart$\mbox{,}$ Pfaffenwaldring 57$\mbox{,}$ 70550 Stuttgart$\mbox{,}$ Germany}
\altaffiliation{current address: Institute of Quantum Control (PGI-8), Forschungszentrum J\"ulich, D-52425 J\"{u}lich, Germany}
\author{Rene Gerritsma}
\affiliation{Van der Waals-Zeeman Institute$\mbox{,}$ Institute of Physics$\mbox{,}$ University of Amsterdam$\mbox{,}$ Science Park 904$\mbox{,}$ 1098 XH Amsterdam$\mbox{,}$ The Netherlands}
\author{Antonio Negretti}
\affiliation{Zentrum f\"ur Optische Quantentechnologien and The Hamburg Centre for Ultrafast Imaging$\mbox{,}$ Universit\"at Hamburg$\mbox{,}$  Luruper Chaussee 149$\mbox{,}$ 22761 Hamburg$\mbox{,}$ Germany}
\author{Tommaso Calarco}
\affiliation{Institute for Complex Quantum Systems \& Center for Integrated Quantum Science and Technology$\mbox{,}$ Universit\"at Ulm$\mbox{,}$ Albert-Einstein-Allee 11$\mbox{,}$ 89075 Ulm$\mbox{,}$ Germany}
\altaffiliation{current address: Institute of Quantum Control (PGI-8), Forschungszentrum J\"ulich, D-52425 J\"{u}lich, Germany}
\author{Zbigniew~Idziaszek}
\affiliation{Faculty of Physics$\mbox{,}$  University of Warsaw$\mbox{,}$ Pasteura 5$\mbox{,}$ 02-093 Warsaw$\mbox{,}$ Poland}
\author{Paul S. Julienne}
\affiliation{Joint Quantum Institute$\mbox{,}$ University of Maryland and National Institute of Standards and Technology$\mbox{,}$ College Park$\mbox{,}$ Maryland 20742$\mbox{,}$ USA}

\date{\today}

\begin{abstract}

Hybrid systems of laser-cooled trapped ions and ultracold atoms combined in a single experimental setup have recently emerged as a new platform for fundamental research in quantum physics. This paper reviews the theoretical and experimental progress in research on cold hybrid ion-atom systems which aim to combine the best features of the two well-established fields. We provide a broad overview of the theoretical description of ion-atom mixtures and their applications, and report on advances in experiments with ions trapped in Paul or dipole traps overlapped with a cloud of cold atoms, and with ions directly produced in a Bose-Einstein condensate. We start with microscopic models describing the electronic structure, interactions, and collisional physics of ion-atom systems at low and ultralow temperatures, including radiative and non-radiative charge transfer processes and their control with magnetically tunable Feshbach resonances. Then we describe the relevant experimental techniques and the intrinsic properties of hybrid systems. In particular, we  discuss the impact of the micromotion of ions in Paul traps on ion-atom hybrid systems. Next, we review recent proposals for using ions immersed in ultracold gases for studying cold collisions, chemistry, many-body physics, quantum simulation, and quantum computation and their experimental realizations. In the last part we focus on the formation of molecular ions via spontaneous radiative association, photoassociation, magnetoassociation, and sympathetic cooling. We discuss applications and prospects of cold molecular ions for cold controlled chemistry and precision spectroscopy.

\end{abstract}


\maketitle

\tableofcontents

\section{Introduction}
\label{sec:introduction}

Cold and ultracold controllable atomic systems attract great interest because the quantum nature of the world is visibly manifested at ultralow temperatures and research on such systems gives new insights into the quantum theory of matter and matter-light interactions. Such an understanding is crucial for the progress of many areas of physics as well as future quantum technologies. 

Ions trapping techniques were developed in the 1950s by Hans Dehmelt and Wolfgang Paul, who shared the 1989 Nobel Prize in physics for their work. The two most commonly employed types of ion traps are radio-frequency quadrupole traps, or Paul traps~\cite{PaulRMP90}, and Penning
traps~\cite{PenningP36,DehmeltAAMP67}~ (invented by Dehmelt, but named after Frans Penning), which rely on static electric and magnetic fields. Penning traps are generally not used in combination with cold buffer gas cooling~\cite{ItanoPS95}, since the orbital motion of the trapped ions would not be stable under collisions. Instead, buffer gas cooling has been used extensively in Paul traps~\cite{ItanoPS95}, although quantum gases have been employed only recently. 

The advent of laser cooling has made it possible to manipulate the internal and external degrees of freedom of trapped ions with unprecedented precision~\cite{LeibfriedRMP03}. Resolved sideband cooling and optical pumping allow the preparation of pure quantum states of motion and electron configuration. Manipulation with electromagnetic waves makes it possible to prepare arbitrary quantum states, while the Coulomb interaction between the ions also allows for the generation of entanglement. The internal state of the
ions can be read out projectively by fluorescence detection, while laser coupling between the internal states and the motion, followed by internal state detection, allows for full characterization of the ionic motional state as well. The development of these ground-breaking experimental methods, which enable measuring and manipulation of individual quantum systems,  was awarded the Nobel Prize in Physics in 2012 to David Wineland~\cite{WinelandRMP13}~ (shared with Serge Haroche).

Trapped ions provide highly controllable quantum systems with long-range interactions. Coulomb crystals of up to tens of ions with full control over their motional and internal degrees of freedom are now available with the potential to produce larger systems in the future. The exquisite control and precision offered by ion traps make them of key importance to many fields of physics. Prime examples include metrology and atomic clocks~\cite{LudlowRMP15}, quantum information processing~\cite{HaeffnerPR08,WinelandPSC09,SingerRMP10,MonroeSCI13}, the study of few- and many-body quantum physics~\cite{LeibfriedRMP03,BlattNP12,SchneiderRPP12,ZhangNat17}, cold chemistry~\cite{WillitschIRPC12} and precision spectroscopy and tests of fundamental physics~\cite{SafronovaARX18}. 

The development of methods to cool and trap neutral atoms with laser light has resulted in recent decades in the birth and successful advances of the field of cold and ultracold matter~\cite{WeinerRMP99} and was awarded the Nobel prize in Physics in 1997 to Steven Chu, Claude Cohen-Tannoudji and William D. Phillips~\cite{CohenRMP98,ChuRMP98,PhillipsRMP98}. These discoveries were followed by the first observation of Bose-Einstein condensation in dilute atomic gases in 1995 to Eric A. Cornell, Wolfgang Ketterle and Carl E. Wieman~\cite{KetterleRMP02,CornellRMP02}. 

Nowadays, ultracold Bose and Fermi gases find applications in many areas of physics. In analogy to trapped ions, ultracold atoms are also a highly controllable and scalable quantum system, but with typically short-range van der Waals interactions~\cite{BlochRMP08}. Notwithstanding, the interactions among ultracold atoms, described in terms of scattering lengths, can be controlled via magnetically tunable Feshbach resonances~\cite{ChinRMP10}, field-induced dipole-dipole interactions, e.g.~by means of Rydberg excitations~\cite{SaffmanRMP10}, or by shaping the external trap, which can lead to confinement-induced resonances~\cite{OlshaniiPRL98,PetrovPRL2000}.  
Furthermore, optical lattices, consisting of thousands of optical microtraps created by interfering laser beams~\cite{LewensteinAP07}, provide the opportunity to create a perfect periodic potential for the atoms with variable dimensionality and geometry. Experiments on trapped ultracold atoms connect quantum optics and atomic physics with condensed-matter and solid-state physics~\cite{LewensteinAP07}.  Prominent examples of applications of ultracold atoms are the study of many-body quantum physics~\cite{BlochRMP08}, quantum simulation~\cite{BlochNP12,BernienNat17}, quantum information processing~\cite{MonroeNature02}, and  metrology~\cite{UdemNature02,LudlowRMP15}. 
Further research opportunities in the quantum domain bridging the atomic and condensed matter physics have been opened with the realization of more complex systems such as ultracold molecules~\cite{CarrNJP09,QuemenerCR12,BohnScience17} as well as setups containing multiple species.

\begin{figure}[tb]
\begin{center}
\includegraphics[width=0.85\columnwidth]{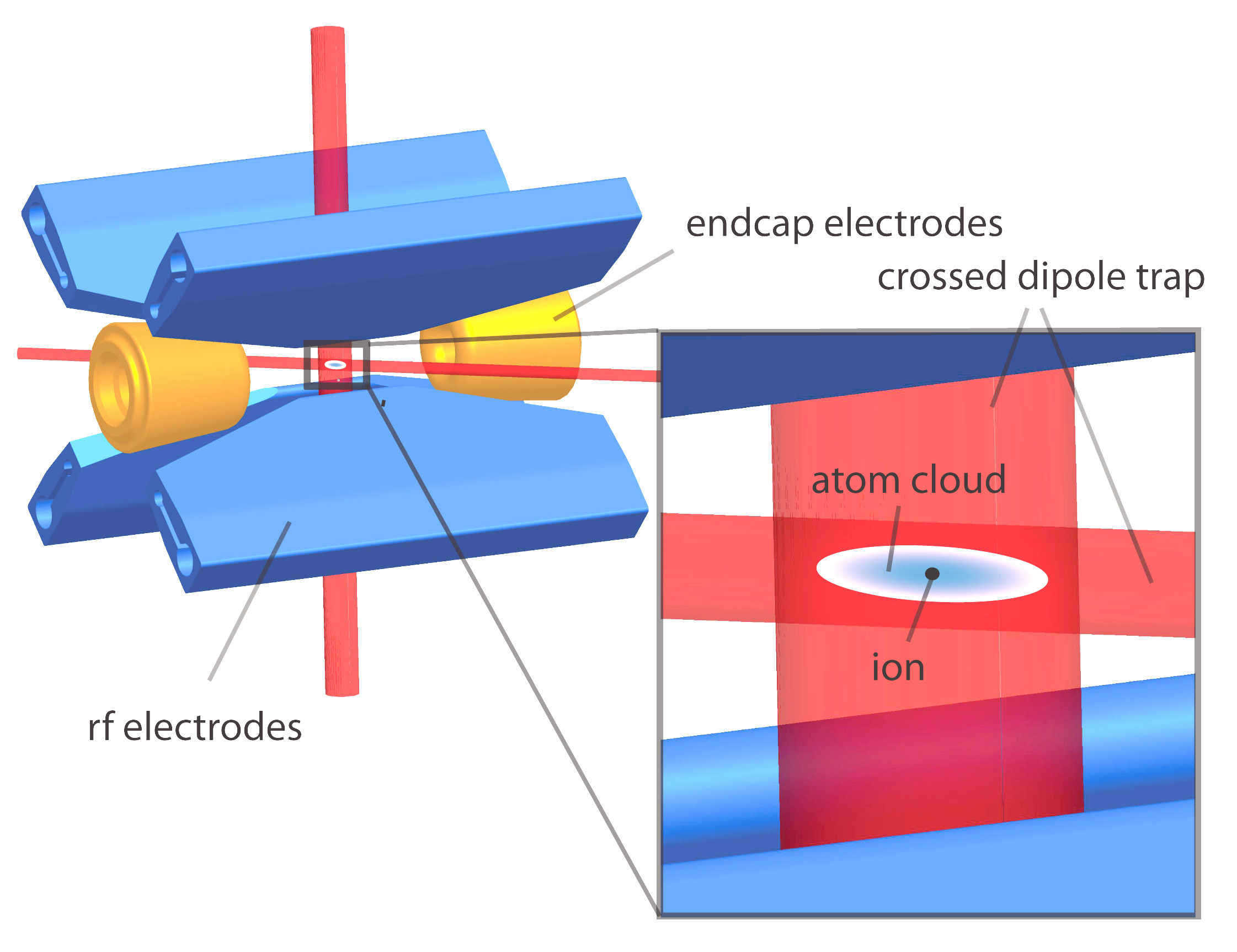}
\end{center}
\caption{Schematic representation of an example hybrid ion-atom experiment: a linear radiofrequency Paul trap is used to store ions, whereas the ultracold atoms are confined in a crossed optical dipole trap. By precisely overlapping the positions of these two traps, a single ion can be immersed into the center of the ultracold neutral atom cloud.
From~\cite{SchmidPRL10}.}
\label{fig1_SchmidPRL10}
\end{figure}

Cold hybrid ion-atom systems have emerged on the intersection of the well-established fields described above: trapped ions and ultracold atoms, combining the two in a single experimental setup (see Fig.~\ref{fig1_SchmidPRL10}). These experiments aim to inherit the most important advantages of both subsystems, while complementing each other, and to show new emerging features providing a platform for both fundamental research in quantum physics and upcoming quantum technologies. The expected features of the hybrid system are: the high controllability with spatial localization and addressability of ionic systems, the long coherence time and extraordinary scalability of the atomic systems, along with intermediate-range and tunable ion-atom interactions as the link between the two. In addition to these, the complementary features of the two systems, like the long-ranged phonon-mediated interactions of trapped ions that are absent in neutral ultracold matter, render the combined system unique and appealing. To fully benefit from such advantages, however, it is essential to understand the fundamental collisional properties of ions and atoms first, since they determine the prospects for coherence and control, but also for decoherence and losses.  

\subsection{Scope and content of the review}

The purpose of this review is twofold: Firstly, it aims to provide a broad overview of the relatively new field of cold ion-atom systems, thus providing a complete picture of the latest advances and underlying techniques both from an experimental and theoretical viewpoint. Secondly, it aims to convey, especially to the newcomers in the field, the most relevant challenges to be addressed in the near future in order to reach the quantum regime as well as new research directions and applications that can arise if these are overcome in the laboratory. 

To this end, the review is divided into five main sections with a final summary of the major issues addressed in the review together with an outlook on future research developments. We begin with a theoretical overview of ultracold ion-atom physics, starting from the basics of two-body interactions and collisions, describing the scattering processes, and then discussing the many-body problem of an ion impurity immersed in a quantum bath (Sec. \ref{sec:theor}). Subsequently, we present the opportunities for tuning the collisional properties of ion-atom systems via external magnetic fields as well as using the trapping potential (Sec. \ref{sec:control}). We then move to the experimental implementations of hybrid systems, especially those based on the combination of radio-frequency and optical traps which are the most common in the laboratories, their specific features and implications for the system dynamics.  We also discuss other approaches such as optical trapping of ions, Rydberg excitations and photoionization (Sec. \ref{sec:Exper}). The next section describes the experimental studies of the collisional properties and theoretical proposals for applications of ion-atom systems for quantum simulation, computation, and detection of many-body correlations (Sec. \ref{sec:app}). The last section focuses on cold molecular ions and their applications in cold chemistry and spectroscopy (Sec. \ref{sec:formation}). 

Let us note that the history of research on cold hybrid ion-atom systems is less than two decades long. Nonetheless, research on ion-neutral interactions and collisions at higher temperatures~\cite{GarciaRMP73} and in astronomical conditions~\cite{SmithCR92} has much longer tradition. Similarly, related research on collisions of neutral atom and molecules with electrons~\cite{InokutiRMP71} started many decades ago. Such systems, however, albeit very interesting, are out of the scope of the present review and will be not treated. 
Finally, we note that some reviews on ion-atom systems are already available~\cite{HarterCP14,WillitschFermi15,Willitsch2017,CoteAAMOP16}.

\section{Theoretical background}
\label{sec:theor}

In this section, we present the theory of ion-atom interactions and collisions at cold and ultracold temperatures. We discuss the electronic structure, quantum chemistry approach to interaction energy calculations, and coupled-channel scattering calculations. We characterize ultracold ion-atom scattering, which includes reactive charge-transfer processes. We also present the quantum defect theory approach to ion-atom collisions. In the last part we consider the many-body problem of an ion immersed in a quantum gas.

\subsection{Ion-neutral interactions}

The interaction between ionic and neutral particles is dominated by the induction component, which can be understood in terms of the interaction of the charge of an ion with the electronic cloud of a neutral partner~\cite{Stone,Israelachvili2011}. Usually, induction-dominated interactions are much stronger than interactions of the van der Waals type. 
The leading long-range part of the ion-neutral interaction potential for structureless particles scales with the interparticle distance $R$ as $-1/R^4$, to be compared with $-1/R^6$ and $-1/R^3$ for van der Waals and dipole-dipole  interactions, respectively~\cite{HapkaCC17}. 

The interaction energy within the Born-Oppenheimer approximation is formally defined by
\begin{equation}\label{eq:V_AB+}
V_{A^++B}=E_{A^++B}-E_{A^+}-E_B\,,
\end{equation}
where $E_{A^++B}$ is the total energy of an interacting complex, and $E_{A^+}$ and $E_B$ are the total energies of separated monomers. The interaction energy can be calculated using Eq.~\eqref{eq:V_AB+} with total energies obtained by solving the many-body Schr\"odinger equation for specific electronic Hamiltonians with methods of quantum chemistry~\cite{Helgaker,HelgakerCR12}. Another possibility is to employ perturbation theory~\cite{JeziorskiCR94}. 

For two-atom systems, the interaction potential is a one-dimensional curve (or a set of curves if many spin configurations are possible), whereas for larger systems the interaction potential is a multidimensional surface (or a set of surfaces). Two regions of potential energy curves and surfaces can be distinguished: at small internuclear distances (short range) the atomic wavefunctions overlap significantly and the strong interactions are non-universally described by quantum chemistry and refereed as ''chemical forces'', whereas at large internuclear distances (long range)  the atomic wavefunctions do not overlap and the interactions are universally described by a multipole expansion of the electrostatic interaction within perturbation theory. The long-range interaction coefficients of the multipole expansion can be expressed using the electronic properties of the monomers. The long-range part of the interaction potential is especially important for cold and ultracold physics and chemistry because quantum reflection and tunneling take place at relatively large internuclear distances in the quantum regime.     
The knowledge of potential energy curves and surfaces is crucial for scattering studies.

\subsubsection{Atomic ion and atom}

\begin{figure}[tb]
\begin{center}
\includegraphics[width=0.85\columnwidth]{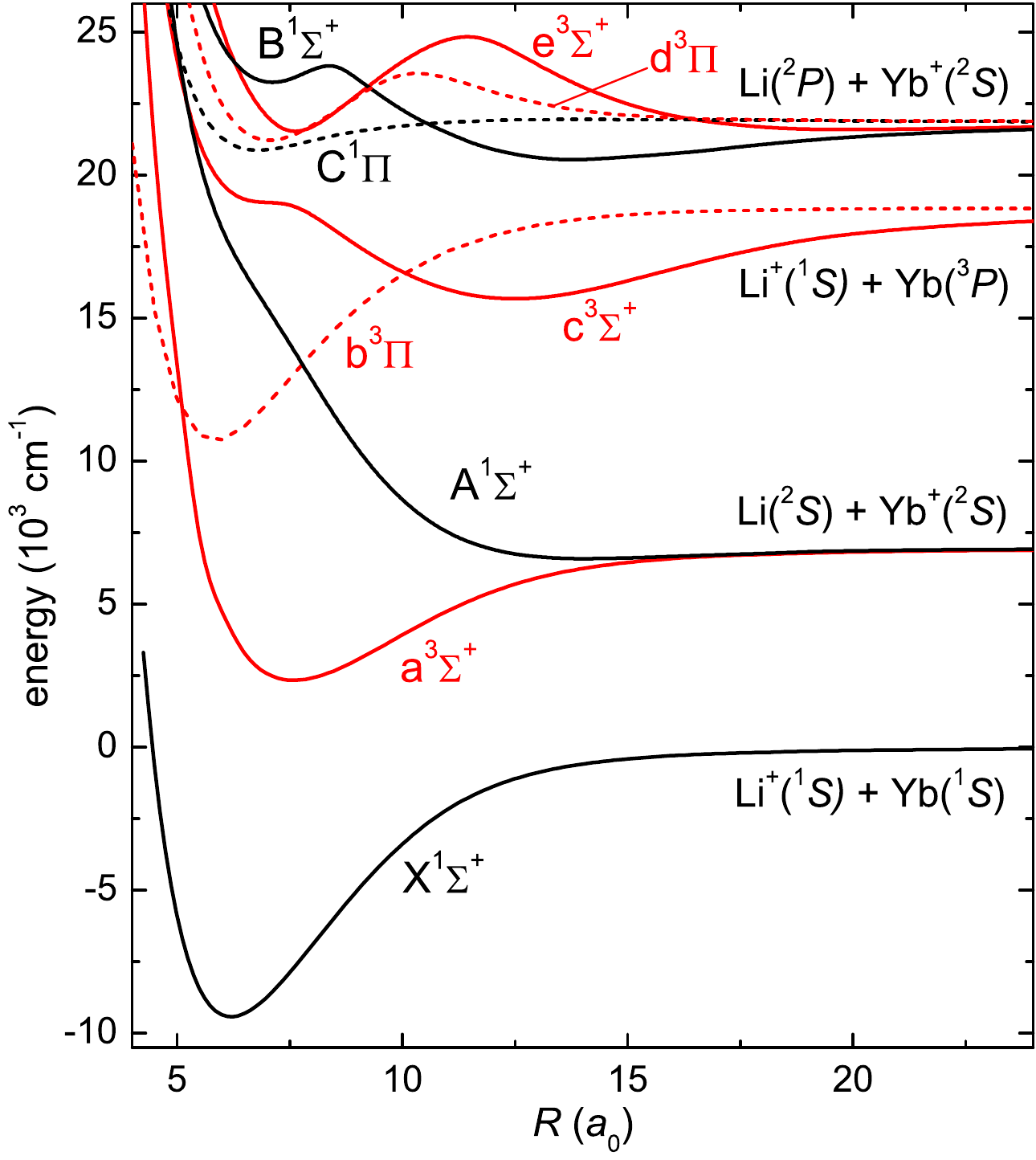}
\end{center}
\caption{Non-relativistic molecular potential energy curves of the (Li+Yb)$^+$ ion-atom system. From~\cite{TomzaPRA15a}.}
\label{fig:LiYb+_curves}
\end{figure}
\begin{figure}[tb]
\begin{center}
\includegraphics[width=0.85\columnwidth]{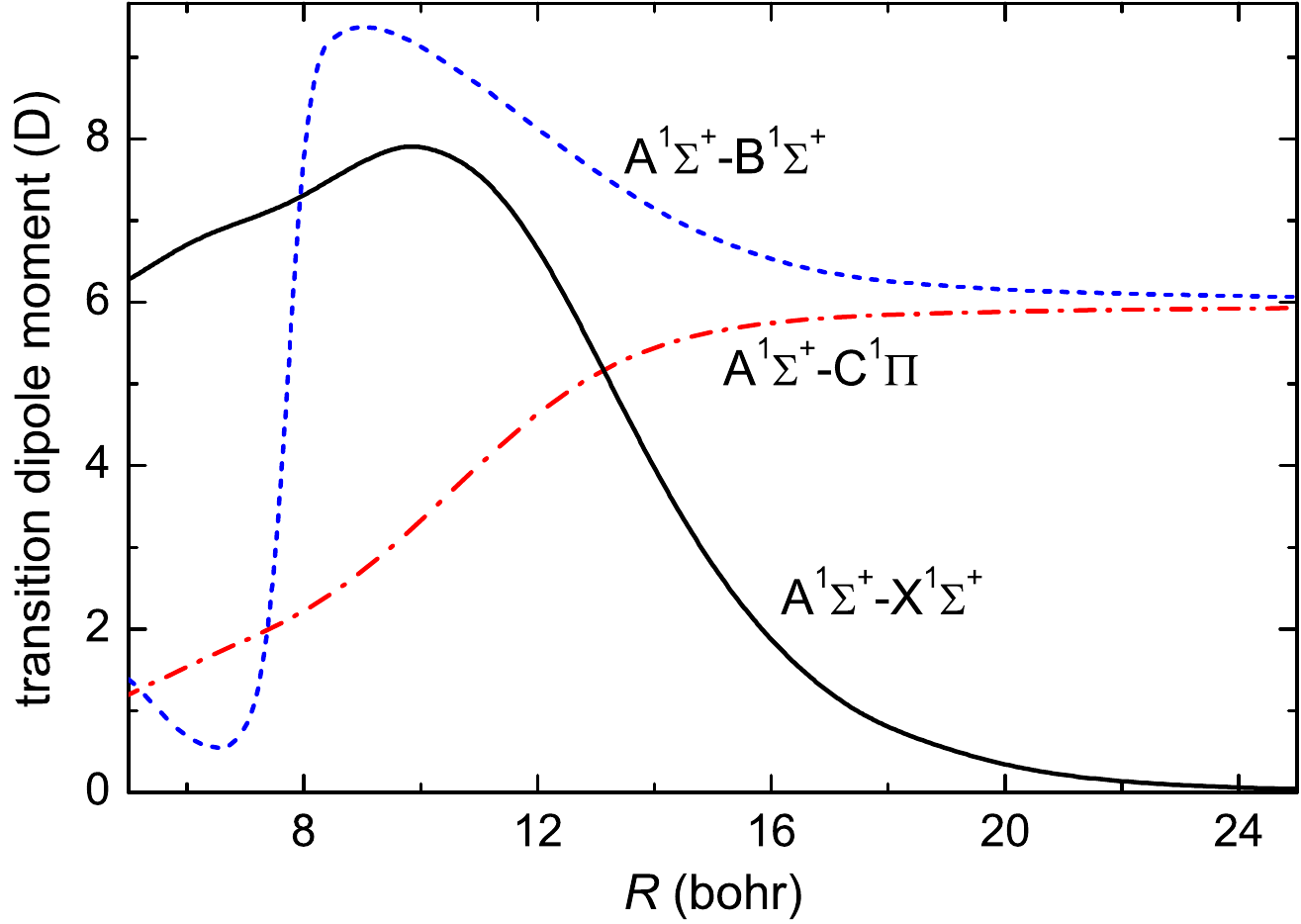}
\end{center}
\caption{Transition electric dipole moments between singlet states of the (Li+Yb)$^+$ ion-atom system. Adapted from~\cite{TomzaPRA15a}.}
\label{fig:LiYb+_dip}
\end{figure}

We start with the simplest charge-neutral system, consisting of an atomic ion and an atom.
Exemplary potential energy curves and transition electric dipole moments for the (Li+Yb)$^+$ ion-atom system are presented in Fig.~\ref{fig:LiYb+_curves} and Fig.~\ref{fig:LiYb+_dip}, respectively~\cite{TomzaPRA15a}.
In order to obtain the presented energy spectrum, quantum chemical techniques such as coupled-cluster and configuration-interaction methods were used to account for correlation energy on top of mean-field Hartree-Fock calculations with Gaussian-type orbital basis sets~\cite{Helgaker}. We will use this example to discuss the general characteristic features of ion-atom interactions.

For all ion-atom systems there exist two families of electronic states associated with two possible arrangements of the charge at the dissociation threshold. The relative position of the lowest electronic states in the two possible arrangements depends on the ionization potentials or electron affinities of the monomers involved. In the case of (Li+Yb)$^+$, at large internuclear distances, the charge can be localized at lithium: Li$^+$+Yb or ytterbium: Li+Yb$^+$. The Li$^+$+Yb arrangement is the absolute ground state of the (Li+Yb)$^+$ system because the ionization potential of the Li atom is smaller than the one of the Yb atom. The charge arrangements and interaction induced charge transfer processes will be discussed in detail in Sec.~\ref{sec:charge_transfer}.   

The long-range part of the interaction potential between an $S$-state ion and an $S$-state atom in the electronic ground state is given by~\cite{HapkaCC17}
\begin{equation}\label{eq:atom_long-range}
V(R)\approx-\frac{C^\textrm{ind}_{4}}{R^4}-\frac{C^\textrm{ind}_{6}}{R^6}-\frac{C^\textrm{disp}_{6}}{R^6}+\dots\,,
\end{equation}  
and it does not depend on the total spin of the resulting molecular electronic state of the $\Sigma$ symmetry. The leading long-range induction coefficient is given by
\begin{equation}\label{eq:Cnind}
C^{\mathrm{ind}}_{4}=\frac{1}{2}q^2\alpha^\textrm{atom}\,,
\end{equation} 
where $q$ is the charge of the ion and $\alpha^\mathrm{atom}$ is the static electric dipole polarizability of the atom. The $-C_4^{\mathrm{ind}}/R^4$ term can be interpreted as the interaction between the charge of the ion and the induced electric dipole moment of the atom.
The next long-range induction coefficient in Eq.~\eqref{eq:atom_long-range} is
\begin{equation}\label{eq:Cnind2}
C^{\mathrm{ind}}_{6}=\frac{1}{2}q^2\beta^\textrm{atom} \,,
\end{equation} 
where $\beta^\textrm{atom}$ is the static electric quadrupole polarizability of the atom. The $-C_6^{\mathrm{ind}}/R^6$ term can be interpreted as the interaction between the charge of the ion and the induced electric quadrupole moment of the atom. The last term in Eq.~\eqref{eq:atom_long-range} describes the dispersion interaction and the long-range dispersion coefficient is given by
\begin{equation}\label{eq:atom_C6}
C^\textrm{disp}_{6}=\frac{3}{\pi}\int_0^\infty  \alpha^{\textrm{ion}}(i\omega)\alpha^\textrm{atom}(i\omega)d\omega\,,
\end{equation}  
where ${\alpha}^{\textrm{atom(ion)}}(i\omega)$ is the dynamic electric dipole polarizability of the atom (ion) at imaginary frequency. The dispersion term results from the interaction between instantaneous dipole-induced dipole moments of the ion and atom arising due to quantum fluctuations.

When an ion or an atom has a non-zero orbital angular momentum, e.g.~in the excited electronic state, the long-range part of the interaction potential takes a more complex form~\cite{KrychPRA11}. For an atom in the $^{2S+1}L$ electronic state with a non-zero orbital angular momentum ($L>0$) interacting with an $S$-state atomic ion, the long-range part of the interaction potential is
\begin{equation}\label{eq:atom_ex_long-range}
V(R)\approx-\frac{C^\textrm{elst}_{3}}{R^3}-\frac{C^\textrm{ind}_{4}}{R^4}-\frac{C^\textrm{elst}_{5}}{R^5}-\frac{C^\textrm{ind}_{6}}{R^6}-\frac{C^\textrm{disp}_{6}}{R^6}+\dots\,,
\end{equation} 
and it does not depend on the total electronic spin of the resulting molecular electronic state, but it depends on the projection of the total orbital angular momentum onto the internuclear axis $|\Lambda|$. The new terms appearing in the expression above, $-C^\textrm{elst}_{3}/R^3$ and $-C^\textrm{elst}_{5}/R^5$, describe the electrostatic interaction between the charge of the ion and the quadrupole and hexadecapole moments of the atom, respectively. The long-range coefficients of Eq.~\eqref{eq:atom_ex_long-range} are given by~\cite{KrychPRA11}
\begin{equation}
\begin{split}
C_3^{\rm elst} =& q(-1)^{1+L-\Lambda}
\begin{pmatrix}
 L & 2 & L \cr -\Lambda & 0& \Lambda \cr
\end{pmatrix}
\langle^{2S+1}L||Q_2||^{2S+1}L\rangle\,, \\
C_4^{\rm ind} =& \frac{1}{2}q^2\left(\alpha_0^\mathrm{atom} + \frac{3\Lambda^2-6}{6}\alpha_2^\mathrm{atom}\right)\,, \\
C_5^{\rm elst} =& q(-1)^{1+L-\Lambda}
\begin{pmatrix} 
L& 4 &L \cr -\Lambda& 0&\Lambda \cr
\end{pmatrix}
\langle^{2S+1}L||Q_4||^{2S+1}L\rangle\,,\\
C_6^{\rm ind} =& \frac{1}{2}q^2\beta_{zz,zz}^\mathrm{atom}\,,\\
C_6^{\rm disp} =& \frac{3}{\pi}\int_0^\infty\alpha_0^{\rm ion}(i\omega)
\Big(\alpha_0^\mathrm{atom}(i\omega)+\\ &\qquad \qquad \qquad \qquad +  \frac{3\Lambda^2-6}{12}\alpha_2^\mathrm{atom}(i\omega)\Big){\rm d}\omega\,,
\end{split}
\end{equation}
where the expressions in round brackets are $3j$ symbols, $\langle^{2S+1}L||Q_{2(4)}||^{2S+1}L\rangle$ is the reduced matrix element of the quadrupole (hexadecapole) moment, and $\alpha_0(i\omega)$ and $\alpha_2(i\omega)$ are the scalar and tensor components of the dynamic electric dipole polarizability at imaginary frequency of the atom in the $^{2S+1}L$ state.

\begin{table*}[tb!]
\caption{Atomic ion-atom systems ($A$+$B$)$^+$ investigated theoretically in the context of cold or ultracold studies. The charge configurations relevant for corresponding experimental works are given.} 
\begin{ruledtabular}
\begin{tabular}{lll}
Ion  & Atom  & References   \\
\hline
Be$^+$ & Li & \cite{RakshitPRA11,GhanmiJPB17}\\ 
Be$^+$ & Na/Ka/Rb & \cite{LadjimiMP18}\\ 
Mg$^+$ & Li & \cite{ElOualhaziJCPA16}\\ 
Ca$^+$ & Li & \cite{SaitoPRA17}\\ 
Ca$^+$ & Na & \cite{MakarovPRA03,GacesaPRA16}\\
Ca$^+$ & Rb & \cite{TacconiPCCP11,BelyaevPRA12}\\
Ca/Sr/Ba/Yb$^+$ & Rb & \cite{daSilvaNJP2015}\\ 
Sr$^+$ & Li & \cite{JellaliMP16}\\   
Sr$^+$ & Li/Na/K/Rb/Cs & \cite{AymarJCP11}\\ 
Sr$^+$ & Na & \cite{BellaouiniEPJ18}\\
Ba$^+$ & Rb & \cite{KrychPRA11,KnechtJPB10}\\ 
Yb$^+$ & Rb & \cite{LambPRA12,SayfutyarovaPRA13,McLaughlinJPB14}\\
Yb$^+$ & Li & \cite{TomzaPRA15a,daSilvaNJP2015,Joger2017}\\ 
Ca/Sr/Ba/Yb$^+$ & Cr & \cite{TomzaPRA15b}\\ 
Li$^+$ & Li & \cite{BouzouitaJMS06,BouchelaghemPCCP14,MusialMP15}\\ 
Na$^+$ & Li & \cite{BuenkerPRA15,MusialaAQC18}\\
Na$^+$ & Na & \cite{BewiczMP17,BerricheIJQC13}\\ 
K$^+$ & Li & \cite{BerricheJMS05,GhanmiCMS07,MusialaAQC18}\\
K$^+$ &  Na &\cite{GhanmiJMS06}\\ 
K$^+$ & K & \cite{SkupinJPCA17}\\ 
Rb$^+$ & Li & \cite{GhanmiIJQC11,RakshitJPB16}\\
Rb$^+$ & Na & \cite{GhanmiJMS07,YanPRA13,YanPRA14}\\ 
Rb$^+$ & Rb & \cite{JraijCP03,JyothiPRL16}\\
Cs$^+$ & Li & \cite{KorekCJP06,GhanmiJMS06b,LiMP13,RakshitJPB16}\\ 
Cs$^+$ & Na & \cite{GhanmiJMS06b,RakshitJPB16}\\ 
Cs$^+$ & Cs & \cite{JraijCP05,JamiesonJPB09}\\ 
Be$^+$ & Be & \cite{BanerjeeCPL10,ZhangPCCP11}\\ 
Mg$^+$ & Mg & \cite{LiMP13,AlharzaliJCPB18}\\ 
Ca$^+$ & Ca & \cite{LiMP13,BanerjeeCPL12}\\ 
Yb$^+$ & Ca & \cite{PetrovJCP17}\\
Yb$^+$ & Yb & \cite{ZhangPRA09}\\
Sr$^+$ & Cl & \cite{PuriJCP14}\\  
Ba$^+$ & Cl & \cite{ChenPRA11}\\ 
Dy$^+$ & Cl & \cite{DunningJCP15}\\ 
C$^+$ & Li/Be & \cite{WellsPCCP11}\\ 
H$^+$ & S & \cite{ShenJPB15}.
\end{tabular}
\label{tab:pairs}
\end{ruledtabular}
\end{table*}

The electronic structure data, including the potential energy curves for the ground and excited electronic states, transition electric dipole moments, and matrix elements of the spin-orbit coupling have been calculated for several ion-atom systems listed in Table~\ref{tab:pairs}. Less accurate calculations for alkali-metal molecular ions from the past century are not included in this list.
 
The long-range interaction coefficients were provided by some authors of \textit{ab initio} calculations together with full potential energy curves, e.g.~for (Rb+Ba)$^+$~\cite{KrychPRA11} or (Li+Yb)$^+$~\cite{TomzaPRA15a}. 
The long-range coefficients for Mg$^+$ and Ca$^+$~\cite{MitroyEPJD08}, Sr$^+$~\cite{MitroyPRA08}, Li$^+$ and Be$^+$~\cite{TangJCP10}, and alkali-metal and alkaline-earth-metal ions~\cite{KaurPRA15,Singh2016} interacting with selected neutral atoms were also reported.

\subsubsection{Molecular ion and atom}

The dimensionality of potential energy surfaces for polyatomic systems depends on the number of atoms and symmetries of the system. 
The actual topology and characteristics of potential energy surfaces depends on the properties of interacting species. The potential energy surfaces obtained in \textit{ab initio} quantum chemical calculations can be expanded in spherical basis functions, which is especially convenient for subsequent coupled-channel scattering calculations that make use of the partial wave expansion. 

For the simplest case of a diatomic molecular ion or a polyatomic linear, axially symmetric molecular ion (in the rigid rotor approximation) interacting with an atom, the potential energy surface in Jacobi coordinates $V(R,\theta)$ can be expanded onto the basis of Legendre polynomials $P_\lambda(\cdot)$
\begin{equation}\label{eq:Vn}
V(R,\theta)=\sum_{\lambda=0}^{\lambda_{max}-1} V_{\lambda}(R) P_{\lambda}(\cos\theta) \,,
\end{equation}
where $R$ is the distance between an atom and the center of mass of a molecular ion and $\theta$ is the angle between the axis of the molecular ion and the axis connecting the atom with the center of mass of the molecular ion. Different $V_\lambda$ terms govern inelastic rotational transitions, allowing for changing the molecular rotation by $\Delta j =\pm\lambda$.   

\begin{figure}[tb!]
\begin{center}
\includegraphics[width=0.9\columnwidth]{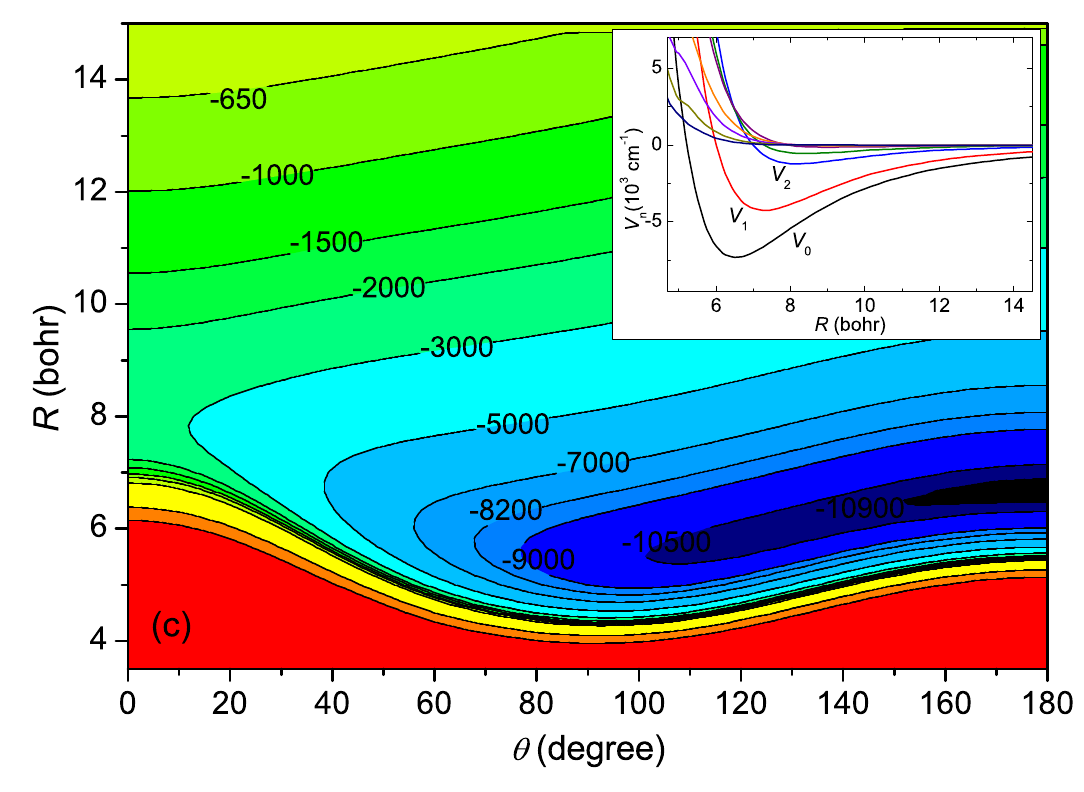}
\end{center}
\caption{The ground-state potential energy surface for the C$_2$H$^-$ molecular anion interacting with the Rb atom. The inset shows the corresponding Legendre components. Adapted from~\cite{TomzaPCCP17}.}
\label{fig:PES}
\end{figure}

An example potential energy surface for a C$_2$H$^-$ molecular anion interacting with a Rb atom is presented in Fig.~\ref{fig:PES}~\cite{TomzaPCCP17}. The corresponding Legendre components, as defined by Eq.~\eqref{eq:Vn}, are shown in the inset. The surface is strongly anisotropic, with the first anisotropic Legendre term $V_1(R)$ almost as large as the isotropic one $V_0(R)$. The large dipole moment (3.22$\,$D) of the molecular ion, related to the localization of the charge on the ending carbon atom, is responsible for the observed anisotropy.

The long-range part of the interaction potential for the simplest case of an axially symmetric molecular ion interacting with an $S$-state atom is~\cite{TomzaPCCP17}
\begin{equation}\label{eq:mol_ion_long-range}
\begin{split}
V(R,\theta)\approx&-\frac{C^\textrm{ind}_{4}}{R^4}-\frac{C^\textrm{ind}_{5,1}}{R^5}\cos\theta-\frac{C^\textrm{ind}_{6,0}}{R^6}-\frac{C^\textrm{disp}_{6,0}}{R^6}\\&-\left(\frac{C^\textrm{ind}_{6,2}}{R^6}+\frac{C^\textrm{disp}_{6,2}}{R^6}\right)P_2(\cos\theta)+\dots\,.
\end{split}
\end{equation}  
The long-range parts of the corresponding Legendre components in Eq.~\eqref{eq:Vn} are
\begin{equation}\label{eq:long}  
\begin{split}
V_0(R)\approx&-\frac{C^\textrm{ind}_{4}}{R^4}-\frac{C^\textrm{ind}_{6,0}}{R^6}-\frac{C^\textrm{disp}_{6,0}}{R^6}+\dots\\
V_1(R)\approx&-\frac{C^\textrm{ind}_{5,1}}{R^5}+\dots\\
V_2(R)\approx&-\frac{C^\textrm{ind}_{6,2}}{R^6}-\frac{ C^\textrm{disp}_{6,2}}{R^6}+\dots\,.
\end{split}
\end{equation} 

The leading long-range induction coefficients are 
\begin{equation}\label{eq:Cnind0}
\begin{split}
C^{\mathrm{ind}}_{4}&=\frac{1}{2}q^2\alpha^\textrm{atom}\,, \\
C^{\mathrm{ind}}_{5,1}&=2d_\textrm{ion}q\alpha^\textrm{atom}\,,  \\
C^{\mathrm{ind}}_{6,0}&=\frac{1}{2}q^2\beta^\textrm{atom}+d^2_\textrm{ion}\alpha^\textrm{atom} \,,\\
C^{\mathrm{ind}}_{6,2}&=2\Theta^\textrm{ion}q\alpha^\textrm{atom}+d^2_\textrm{ion}\alpha^\textrm{atom}\,,\\
\end{split}
\end{equation}  
where $q$ is the charge of the molecular ion, $\alpha^\textrm{atom}$ is the static electric dipole polarizability of the atom, $d_\textrm{ion}$ is the permanent electric dipole moment of the molecular ion, $\Theta^\textrm{ion}$ is the permanent electric quadrupole moment of the molecular ion, and $\beta^\textrm{atom}$ is the static electric quadrupole polarizability of the atom. 
The leading long-range dispersion coefficients are
\begin{equation}\label{eq:Cndisp}
\begin{split}
C^\textrm{disp}_{6,0}&=\frac{3}{\pi}\int_0^\infty  \bar{\alpha}^{\textrm{ion}}(i\omega){\alpha}^\textrm{atom}(i\omega)d\omega\,,\\
C^\textrm{disp}_{6,2}&=\frac{1}{\pi}\int_0^\infty \Delta\alpha^{\textrm{ion}}(i\omega){\alpha}^\textrm{atom}(i\omega)d\omega \,,
\end{split}
\end{equation}  
where ${\alpha}^{\textrm{atom(ion)}}(i\omega)$ is the dynamic polarizability of the atom (ion) at imaginary frequency, and the
average polarizability and polarizability anisotropy are given by $\bar{\alpha}=(\alpha_\parallel+2\alpha_\perp)/3$ and $\Delta\alpha=\alpha_\parallel-\alpha_\perp$ respectively, with $\alpha_\parallel$ and $\alpha_\perp$ being the components of the polarizability tensor parallel and perpendicular to the internuclear axis of the molecular ion.

The new terms appearing in the long-range multipole expansion of the intermolecular interaction energy in Eq.~\eqref{eq:mol_ion_long-range}, as compared to Eq.~\eqref{eq:atom_long-range} and Eq.~\eqref{eq:atom_ex_long-range}, are due to the possible permanent electric dipole and quadrupole moments and the polarizability anisotropy of a molecular ion. The $-C_{5,1}^\mathrm{ind}/R^5\cos\theta$ term describes the interaction between the permanent electric dipole moment of a molecular ion and the induced electric dipole moment of an atom. The first term in $-C_{6,0}^\mathrm{ind}/R^6$ describes the interaction between the charge of a molecular ion and the induced electric quadrupole moment of the atom, whereas the second one describes the interaction between the permanent electric dipole moment of the molecular ion and the higher-order induced electric dipole moment of an atom. The first term in $-C_{6,2}^\mathrm{ind}/R^6$ describes the interaction between the permanent electric quadrupole moment of a molecular ion and the induced electric dipole moment of an atom, and the second one is the same as in $-C_{6,0}^\mathrm{ind}/R^6$. The long-range anisotropy of the dispersion interaction is due to the polarizability anisotropy of a molecular ion. If the atom has a non-zero orbital angular momentum then Eq.~\eqref{eq:mol_ion_long-range} has to be extended to include terms describing the electrostatic interaction between the charge and electric moments of the molecular ion and electric moments of the atom, similarly as in Eq.~\eqref{eq:atom_ex_long-range}.

The potential energy surfaces have been investigated for several molecular ions interacting with atoms in the context of cold or ultracold studies as depicted in Table~\ref{tab:mol}. Considerably large interest in interactions involving helium atom is caused by its potential to act as a versatile sympathetic cooler for molecular ions without undergoing inelastic processes, as described in further sections of this review.

\begin{table*}[tb!]
\caption{Molecular ion-atom systems investigated theoretically in the context of cold or ultracold studies.} 
\begin{ruledtabular}
\begin{tabular}{lll}
Ion & Atom & References   \\
\hline
OH$^-$ & Rb & \cite{GonzalezEPJD08,ByrdPRA13,KasJCP16}\\
MgH$^+$ & Rb & \cite{TacconiEPJD08,TacconiEPJD09} \\
CN$^-$ & Rb/Sr & \cite{MidyaPRA16} \\
BaCl$^+$ & Ca & \cite{StoecklinNP16} \\
OH$^-$/CN$^-$/NCO$^-$/C$_2$H$^-$/C$_4$H$^-$ & Li/Na/K/Rb/Cs & \cite{TomzaPCCP17} \\
OH$^-$/CN$^-$/NCO$^-$/C$_2$H$^-$/C$_4$H$^-$ & Mg/Ca/Sr/Ba & \cite{TomzaPCCP17} \\
OH$^-$ & Li/Na/K/Rb/Cs & \cite{KasJCP17} \\
OH$^-$ & Mg/Ca/Sr/Ba/Be & \cite{KasJCP17} \\
C$_2^-$ & Li/Rb & \cite{Kas2018} \\
OH$^+$/OH$^-$ & He & \cite{MarinettiJTCC06,GonzalezJPB06} \\
CH$^+$ & He & \cite{StoecklinEPJD08,HammamiJMS08} \\
LiH$^-$ & He & \cite{LopezEPJD09} \\
NO$^+$ & He & \cite{StoecklinJCP11} \\
SH$^-$ & He & \cite{BopJCP17} \\
\end{tabular}
\label{tab:mol}
\end{ruledtabular}
\end{table*}

\subsubsection{Atomic ion and molecule}

The short-range part of the interaction potential for an atomic ion interacting with a molecule can, in principle, be similar to the one for a molecular ion interacting with an atom. The essential difference in the interaction potential for these two systems lies in the long-range part.

The long-range part of the interaction potential for the simplest case of an $S$-state atomic ion interacting with an axially symmetric molecule is~\cite{HeijmenMP96,HapkaCC17}
\begin{equation}
\begin{split}\label{eq:at_ion_mol_long-range}
V(R,\theta)\approx&-\frac{C^\textrm{elst}_{2,1}}{R^2}\cos\theta-\frac{C^\textrm{elst}_{3,2}}{R^3}P_2(\cos\theta)-\frac{C^\textrm{elst}_{4,3}}{R^4}P_3(\cos\theta)\\
&-\frac{C^\textrm{ind}_{4,0}}{R^4}-\frac{C^\textrm{ind}_{4,2}}{R^4}P_2(\cos\theta) +\dots\,,
\end{split}
\end{equation} 
where the first three terms describe the electrostatic interaction between the charge of the atomic ion and the permanent electric dipole, quadrupole, and octupole moments
of the molecule, respectively. The last terms represent the induction interaction typical for ion-neutral systems with additional anisotropic part resulting from the anisotropy of the molecular electric dipole polarizability.

The leading long-range electrostatic coefficients are
\begin{equation}
\begin{split}
C^{\mathrm{elst}}_{2,1}&=qd^\textrm{mol}\,, \\
C^{\mathrm{elst}}_{3,2}&=\frac{1}{2}q\Theta^\textrm{mol}\,,  \\
C^{\mathrm{elst}}_{4,3}&=\frac{1}{4}q\Omega^\textrm{mol}\,,\\
\end{split}
\end{equation}  
where $d^\textrm{mol}$, $\Theta^\textrm{mol}$, and $\Omega^\textrm{mol}$ are the permanent electric dipole, quadrupole, and octupole moments of the molecule, respectively. The leading long-range induction coefficients are
\begin{equation}
\begin{split}
C^{\mathrm{ind}}_{4,0}&=\frac{1}{2}q^2\bar{\alpha}^\textrm{mol}\,, \\
C^{\mathrm{ind}}_{4,2}&=\frac{1}{6}q^2\Delta\alpha^\textrm{mol}\,, \\
\end{split}
\end{equation} 
where $\bar{\alpha}^\textrm{mol}$ and $\Delta\alpha^\textrm{mol}$ are the average electric dipole polarizability and polarizability anisotropy of the molecule. 

If an atomic ion with a non-zero orbital angular momentum or a molecular ion interacts with a molecule, then Eq.~\eqref{eq:at_ion_mol_long-range} has to be extended to include terms describing the electrostatic interaction between the charge and electric moments of involved species, resulting in terms similar as in Eqs.~\eqref{eq:atom_ex_long-range} and~\eqref{eq:mol_ion_long-range}.

The potential energy surfaces have been investigated for several atomic ions interacting with molecules in the context of cold or ultracold studies: Li$^+$+D$_2$~\cite{BovinoPRA08}, Ar$^+$+N$_2$\cite{TrippelPRL13}, H$^-$+HCN~\cite{SattaAJ15}.

\subsection{Ion-atom collisions}
Having described the interparticle interactions, we can now turn to the collisional phenomena. 
The Schr\"odinger equation for the relative nuclear motion of two colliding species (e.g.~an ion $A^+$ and an atom $B$) may be written in the form
\begin{equation}
\label{TSchr}
\left(-\frac{\hbar^2}{2\mu}\nabla^2+\mathbf{V}(R,\tau)+\mathbf{H}_\mathrm{mon}(\tau)\right)\mathbf{\Psi}(R,\tau)=E\mathbf{\Psi}(R,\tau)\,,
\end{equation}
where $\hbar^2/(2\mu)\nabla^2$ is the kinetic energy operator, $\mu$ is the reduced mass, $\tau=(\tau_{A^+},\tau_B)$ denotes all coordinates (internal degrees of freedom of the monomers $A^+$ and $B$) except the intermolecular distance $R$, $\mathbf{V}(R,\tau)$ is the interatomic or intermolecular interaction potential, $\mathbf{H}_\mathrm{mon}(\tau)=\mathbf{H}_{A^+}(\tau_{A+})+\mathbf{H}_B(\tau_B)$ is the Hamiltonian describing the internal degrees of freedom of the monomers $A^+$ and $B$, and $\mathbf{\Psi}(R,\tau)$ denotes the total wave function with the energy $E$.

The total wave function can be decomposed into the basis set of $N$ channel functions $\{\Theta_i(\tau)\}_{i=1}^N$
\begin{equation}
\mathbf{\Psi}(R,\tau)=\sum_i\Phi_{i}(R)\Theta_i(\tau)/R\,.
\end{equation}
Channel functions can be or can contain spherical harmonics $Y_{l}^m(\theta,\phi)$ which are eigenstates of the orbital angular momentum operator for end-over-end motion of the two colliding particles around one another. Channels with different values of $l$ are referred to as different partial waves ($l$-waves).

Substituting $\mathbf{\Psi}(R,\tau)$ into the Schr\"odinger equation~\eqref{TSchr} gives a set of coupled equations
\begin{equation}
\label{eq:coupled_Sch_eq}
\left(-\frac{\hbar^2}{2\mu}\frac{\partial^2}{\partial R^2}+\mathbf{W}(R)\right)\mathbf{\Phi}(R)=E\mathbf{\Phi}(R)\,,
\end{equation}
where $\mathbf{\Phi}(R)$ is a matrix of $N$ linearly independent radial solutions and $\mathbf{W}(R)$ is a matrix describing the effective interaction, with elements given by $\mathbf{W}_{ij}(R)=\langle \Theta_i(\tau)|\mathbf{V}(R,\tau)+\mathbf{H}_\mathrm{mon}(\tau) | \Theta_j(\tau)\rangle $.

The set of coupled equations~\eqref{eq:coupled_Sch_eq} can be solved numerically using some propagation method, e.g.~the renormalized Numerov approach~\cite{JohnsonJCP78} or the log-derivative propagator~\cite{JohnsonJCP73}. By imposing the long-range scattering boundary conditions on $\mathbf{\Phi}(R)$ in terms of Bessel functions
\begin{equation}
\label{eq:boundarycond}
\mathbf{\Phi}(R)\stackrel{R\to\infty}{\longrightarrow}\left[\mathbf{J}(R)-\mathbf{N}(R)\mathbf{K}\right]\mathbf{A}\,,
\end{equation}
where $\mathbf{J}(R)$ and $\mathbf{N}(R)$ are diagonal matrices with proper spherical Bessel functions and $\mathbf{A}$ is the normalization constant, one can get the $\mathbf{K}$ reactance matrix, which gives the $\mathbf{S}$ scattering matrix
\begin{equation}
\mathbf{S}=(\mathbf{1}+i\mathbf{K})^{-1}(\mathbf{1}-i\mathbf{K})\,.
\end{equation}
The energy-dependent scattering length $a_n$, partial elastic cross section $\sigma^n_\mathrm{el}$, and partial inelastic cross section $\sigma^n_\mathrm{in}$ for channel $n$ can be directly calculated from the $\mathbf{S}$ matrix
\begin{equation}
\begin{split}
a_n(E)=&\frac{1}{ik_n}\frac{1-S_{nn}}{1+S_{nn}}\,,\\
\sigma_\mathrm{el}^n(E)=& \frac{\pi}{k_n^2}\left|1-S_{nn}\right|^2 \,,\\
\sigma_\mathrm{in}^n(E)=& \frac{\pi}{k_n^2}\left(1-\left|S_{nn}\right|\right)^2 \,,\\
\end{split}
\end{equation} 
where $k_n=\sqrt{2\mu (E-E_n^\infty)/\hbar^2}$ is the channel wave vector and $E_n^\infty$ denotes the threshold energy.
For $k\to 0$, the $s$-wave scattering length becomes energy-independent. 
To get total cross sections, the partial cross sections have to be summed up over partial waves. 
Collision rate coefficients at a given collision energy can be obtained from the cross sections by the simple relation
\begin{equation}
K=\frac{\hbar k}{\mu}\sigma\,.
\end{equation}

For an atomic ion $A^+$ colliding with an atom $B$ we can write the specific form of the Hamiltonian in Eq.~\eqref{TSchr}~\cite{IdziaszekPRA09,IdziaszekNJP11,TomzaPRA15a,TscherbulPRL15} as
\begin{equation}\label{eq:Ham}
\begin{split}
    \hat{H}=&-\frac{\hbar^2}{2\mu}\frac{1}{R}\frac{d^2}{dR^2}R+
    \frac{\mathbf{\hat{l}}^2}{2\mu R^2}+ \hat{V}(R)\\
    &+\hat{V}^\mathrm{ss}(R)+\hat{V}^\mathrm{so}(R)+\hat{H}_{A^+}+\hat{H}_{B}\,,
\end{split}
\end{equation}
where $\mathbf{\hat{l}}$ is the rotational angular momentum operator. $\hat{V}(R)$ is the interaction potential, which depends on the total spin $S$ and its projection $M_S$
\begin{equation}
\hat{V}(R)=\sum_{S,M_S}V_S(R)|S,M_S\rangle\langle S,M_S|\,.
\end{equation}
The relativistic terms $V^\mathrm{ss}(R)$ and $V^\mathrm{so}(R)$ stand for the spin dipole-dipole interaction and the
second-order spin-orbit term, respectively, and are responsible for dipolar relaxation
\begin{equation}
\begin{split}
V^\mathrm{ss}(R)=&-\sqrt{\frac{44\pi}{5}}\frac{\alpha^2}{R^3}\sum_q Y_{2}^q(\hat{R})[\mathbf{\hat{s}}_a \otimes \mathbf{\hat{s}}_b]^{(2)}_q\,,\\
V^\mathrm{so}(R)=&\sqrt{\frac{44\pi}{5}}\lambda_{SO}(R)\sum_q Y_{2}^q(\hat{R})[\mathbf{\hat{s}}_a \otimes \mathbf{\hat{s}}_b]^{(2)}_q\,.
\end{split}
\end{equation}
The atomic Hamiltonian, $\hat{H}_j$ ($j=A^+$, $B$),
including hyperfine and Zeeman interactions, is given by 
\begin{equation}\label{eq:Ham_at}
\hat{H}_j=\zeta_{j}\mathbf{\hat{i}}_{j}\cdot\mathbf{\hat{s}}_{j}
  +g_e\mu_{{B}}\mathbf{\hat{s}}_{j}\cdot\mathbf{B}+g_{j}\mu_{{N}}\mathbf{\hat{i}}_{j}\cdot\mathbf{B}\,,
\end{equation}
with $\mathbf{\hat{s}}_{j}$ and $\mathbf{\hat{i}}_{j}$ the electron and nuclear spin operators, $\zeta_{j}$ denoting the hyperfine coupling constant, $g_{e/j}$  the electron and nuclear $g$ factors, $\mu_{B/N}$ the Bohr and nuclear magnetons, and $\mathbf{B}$ the magnetic field.  

The equations \eqref{TSchr}-\eqref{eq:Ham_at} have the same form as for neutral species and are given here for the sake of completeness. The essential difference between ion-atom and atom-atom scattering results from the different asymptotic form of the polarization potential for ion-atom systems as described in the previous subsection. 

For the the polarization potential
\begin{equation}\label{eq:V_pol}
V(R)=-\frac{C_4}{R^4}\,,
\end{equation}
by equating the potential to the kinetic energy one can define the characteristic interaction length scale $R^\star$
\begin{equation}
R^\star=\sqrt{\frac{2\mu C_4}{\hbar^2}}\,.
\end{equation}
While the scattering length can have any value, $R^\star$ establishes the typical order of magnitude of the scattering length for an ion-atom potential. The typical interaction length scales for ion-atom systems are at least an order of magnitude larger as compared to neutral atom-atom ones. The related energy scale is given by
\begin{equation}
E^\star= \frac{\hbar^2}{2 \mu (R^{\star})^2}\,.
\end{equation}
The characteristic energy scale for ion-atom systems is at least two orders of magnitude smaller as compared to neutral atom-atom systems. This is one of the reasons why reaching the $s$-wave scattering regime for ion-atom systems is more challenging as compared to neutral atom systems. The characteristic lengths and energies for selected ion-atom systems are collected in Table~\ref{tab:characteristic}. The smallest $R^\star$ and the largest $E^\star$ are expected for systems with the smallest reduced masses and atomic polarizabilities, thus such systems are the most favorable for reaching the $s$-wave regime.  

\begin{table}[b!]
\caption{Characteristic length $R^{\star}$, characteristic energy $E^{\star}$, and long-range induction interactions coefficient $C_4$ for selected ion-atom systems.} 
\begin{ruledtabular}
\begin{tabular}{llll}
 & $R^{\star}\,$($a_0$) & $E^{\star}/h\,$(kHz) & $C_4\,$($E_ha_0^4$) \\
\hline
$^{40}$Ca$^{+}$ +  $^6$Li & 1250 &  220.9  &  82 \\
$^{176}$Yb$^{+}$ + $^6$Li & 1319 &  178.2  & 82 \\
$^{40}$Ca$^{+}$ + $^{23}$Na & 2081 & 28.56 & 81   \\
$^{40}$Ca$^{+}$ + $^{87}$Rb & 3989 & 4.143 & 160 \\
$^{88}$Sr$^{+}$ + $^{87}$Rb & 5049 & 1.620 & 160 \\
$^{135}$Ba$^{+}$ + $^{87}$Rb & 5544 & 1.111  & 160 \\
$^{172}$Yb$^{+}$ + $^{87}$Rb & 5793 & 0.9313 & 160 \\
\end{tabular}
\label{tab:characteristic}
\end{ruledtabular}
\end{table}

\begin{figure}[tb!]
\begin{center}
\includegraphics[width=0.9\columnwidth]{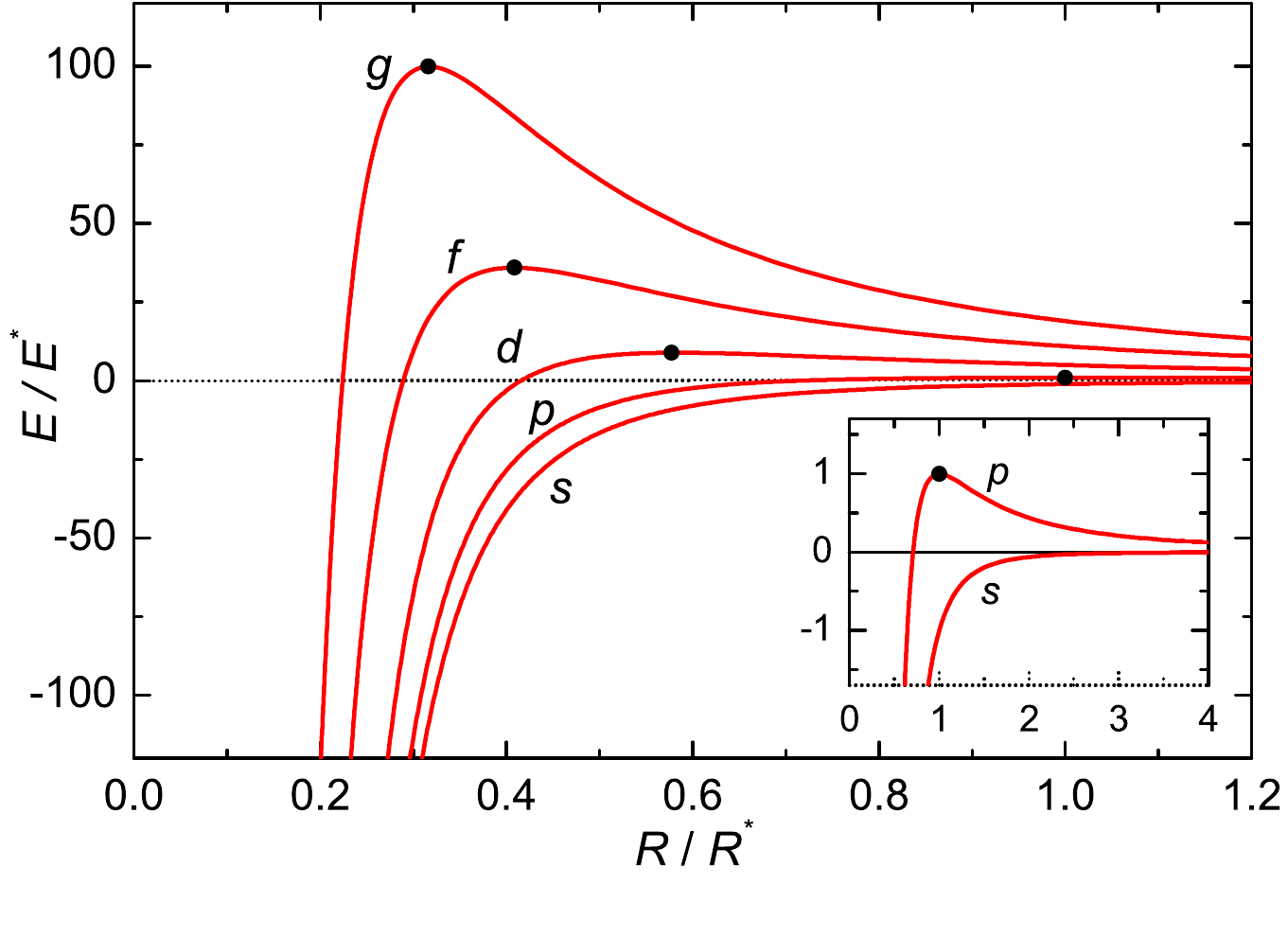}
\end{center}
\caption{Long-range ion-atom potential for the lowest partial waves in units of the characteristic energy $E^\star$ and the characteristic length $R^\star$. Dots show the positions of the maxima of the centrifugal barriers. Inset shows two lowest potentials.}
\label{fig:long-range}
\end{figure}

The energy $E^\star$ defines the height of the centrifugal barrier for the $p$-wave ($l=1$) collision, whereas the length $R^\star$ gives the position of the maximum of this centrifugal barrier. The effective long-range ion-atom potentials for the lowest partial waves in units of the characteristic energy and length are presented in Fig.~\ref{fig:long-range}. The position and height of the centrifugal barrier for higher partial waves are given by
\begin{equation}\label{eq:E_max}
\begin{split}
R^\mathrm{max}_l=&\sqrt{\frac{2}{l(l+1)}}R^\star\,,\\
E^\mathrm{max}_l=&\frac{l^2(l+1)^2}{4}E^\star\,.
\end{split}
\end{equation}

As shown in the previous paragraphs, in addition to the polarization potential one can be dealing with multiple additional interaction terms. Typically the higher order part is dominated by the $1/r^4$ term at low energies, and can become relevant at $E\gtrsim E_6$ with $E_6=\hbar^2/2 \mu R_6^2$ and $R_6=\left(2\mu C_6/\hbar^2\right)^{1/4}$ defined in analogy to $R^\star$. On the other hand, the dipolar term becomes increasingly important as the collision energy is lowered. 

Collisions can be classified as elastic, inelastic, and reactive depending on their possible outcomes:
\begin{itemize}
\item An \textit{elastic} collision is one in which the kinetic energy of the relative motion of the two particles does not change, though their individual kinetic energies and velocities may change. 
\item An \textit{inelastic} collision is one in which energy is transferred between internal
and relative kinetic energy, but the chemical constitution of the collision
partners is unchanged.
\item A \textit{reactive} collision is one in which the products are chemically distinct from
the reactants, usually also involving a change in relative kinetic energy.
\end{itemize}

\begin{figure}[tb!]
\begin{center}
\includegraphics[width=0.9\columnwidth]{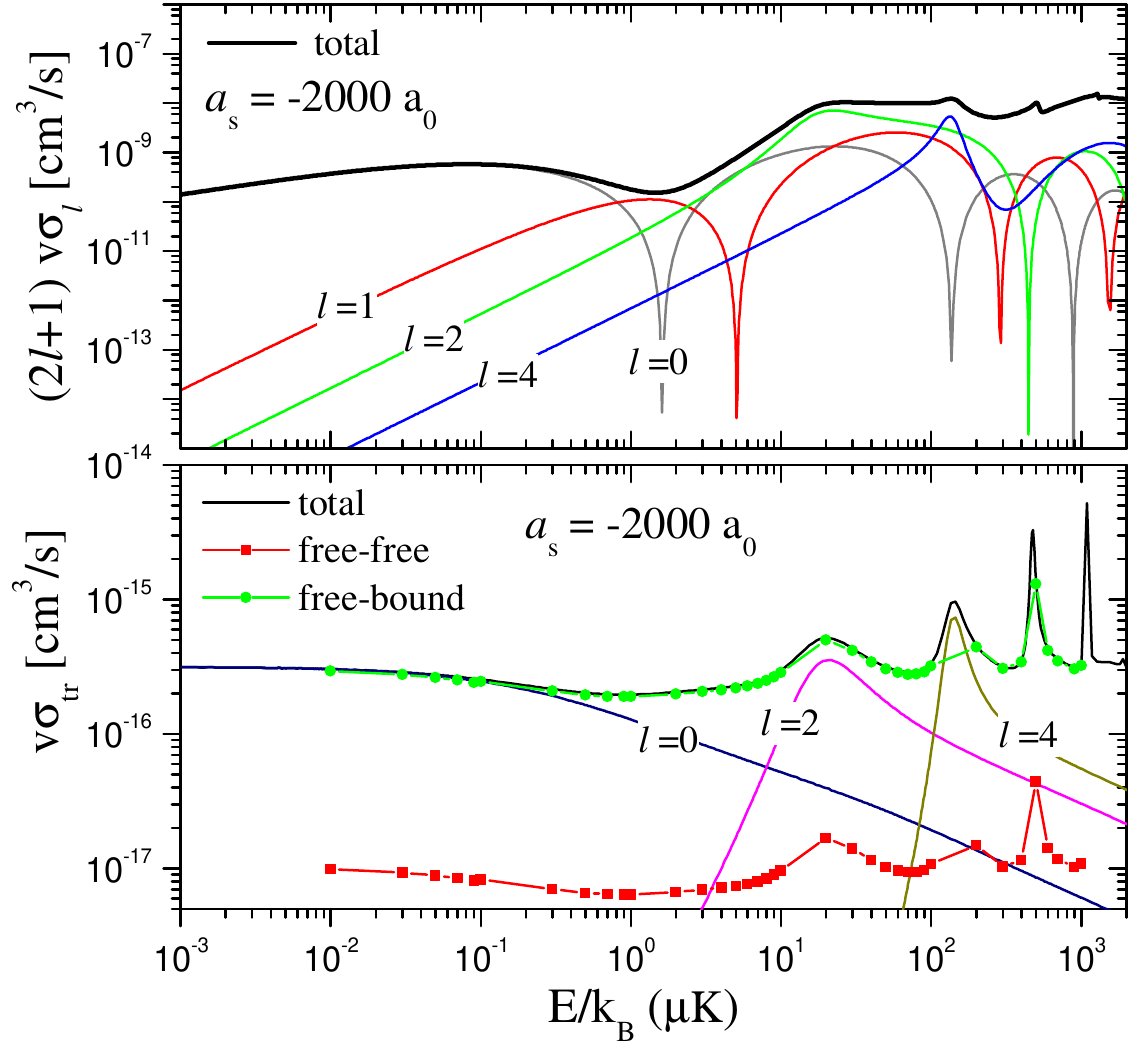}
\end{center}
\caption{Elastic collision rate constant in the singlet channel $A^{1} \Sigma^{+}$ (upper panel) and rate constant of radiative $A^{1} \Sigma^{+} \to X^{1} \Sigma^{+}$ charge transfer (lower panel) versus collision energy in the Ca$^+$+Na system. Both the elastic and the total reactive rate constants are decomposed into selected partial waves. From~\cite{IdziaszekPRA09}.}
\label{fig1_IdziaszekPRA09}
\end{figure}

Elastic collisions are essential for thermalization of the system as well as evaporative and sympathetic cooling. These processes are feasible if the rate constant for elastic collisions are at least two orders of magnitude larger than the rate for inelastic and reactive ones. Inelastic collisions involve processes such as changing spin or rotational state of the colliding partners. The most important reactive collisions for ion-neutral systems are charge transfer processes. The nature of charge transfer processes will be analyzed in the next subsection.

Figure~\ref{fig1_IdziaszekPRA09} shows an example of rate constants for elastic and charge-transfer reactive collisions of Ca$^+$ ions with Na atoms obtained in quantum scattering calculations by~\cite{IdziaszekPRA09}. For energies larger than 1$\,\mu$K already few partial waves contribute to the total rate. For elastic collisions, a single shape resonance from the partial wave $l=4$ is visible, whereas for reactive collisions, sharper resonances, which are characteristic for charge transfer processes, can be observed. The shape resonances are more pronounced for reactive collision rate constants as compared to elastic ones because the charge transfer process is a short-range phenomenon and the enhancement of the tunneling over the centrifugal barrier to the short range significantly increases its probability. At the same time, the rate constants for charge transfer losses are over three orders of magnitude smaller than the rate constants for elastic collisions. This is typical for ion-atom systems with both an ion and an atom in the ground states and with the entrance threshold well separated from other electronic states.

The rate constants for reactive collisions in Fig.~\ref{fig1_IdziaszekPRA09} are additionally decomposed into contributions from free-to-free and free-to-bound transitions, where the former leads to the charge exchange between ions and atoms, whereas the latter leads to the formation of molecular ions. It is characteristic for charge transfer processes in ion-atom systems that the radiative formation of molecular ions is significantly more likely than the radiative charge exchange. The radiative formation of molecular ions will be discussed in detail in Sec.~\ref{sec:charge_transfer} and~\ref{sec:radiative_association}.

In the regime dominated by single $s$-wave collisions and vanishing collision energy, the energy dependence of the cross sections and related rate constants for both elastic and inelastic (reactive) collisions follows the Wigner threshold laws~\cite{WignerPRA48}
\begin{equation}
\begin{split}
\sigma_\mathrm{el}^l & \stackrel{k\to 0}{\sim}  k^{4l}\,,\\
\sigma_\mathrm{in}^l & \stackrel{k\to 0}{\sim} k^{2l-1}\,,
\end{split}
\end{equation}
where $l$ is the angular momentum for the relative motion of the colliding particles. This means that the rate constant for reactive collisions should be energy-independent in the ultracold regime. Such behavior in ion-atom collisions is predicted in theory (cf.~Fig.~\ref{fig1_IdziaszekPRA09}) and expected in experiments (cf.~Sec.~\ref{sec:cold_col}).

\begin{figure}[tb!]
\begin{center}
\includegraphics[width=0.9\columnwidth]{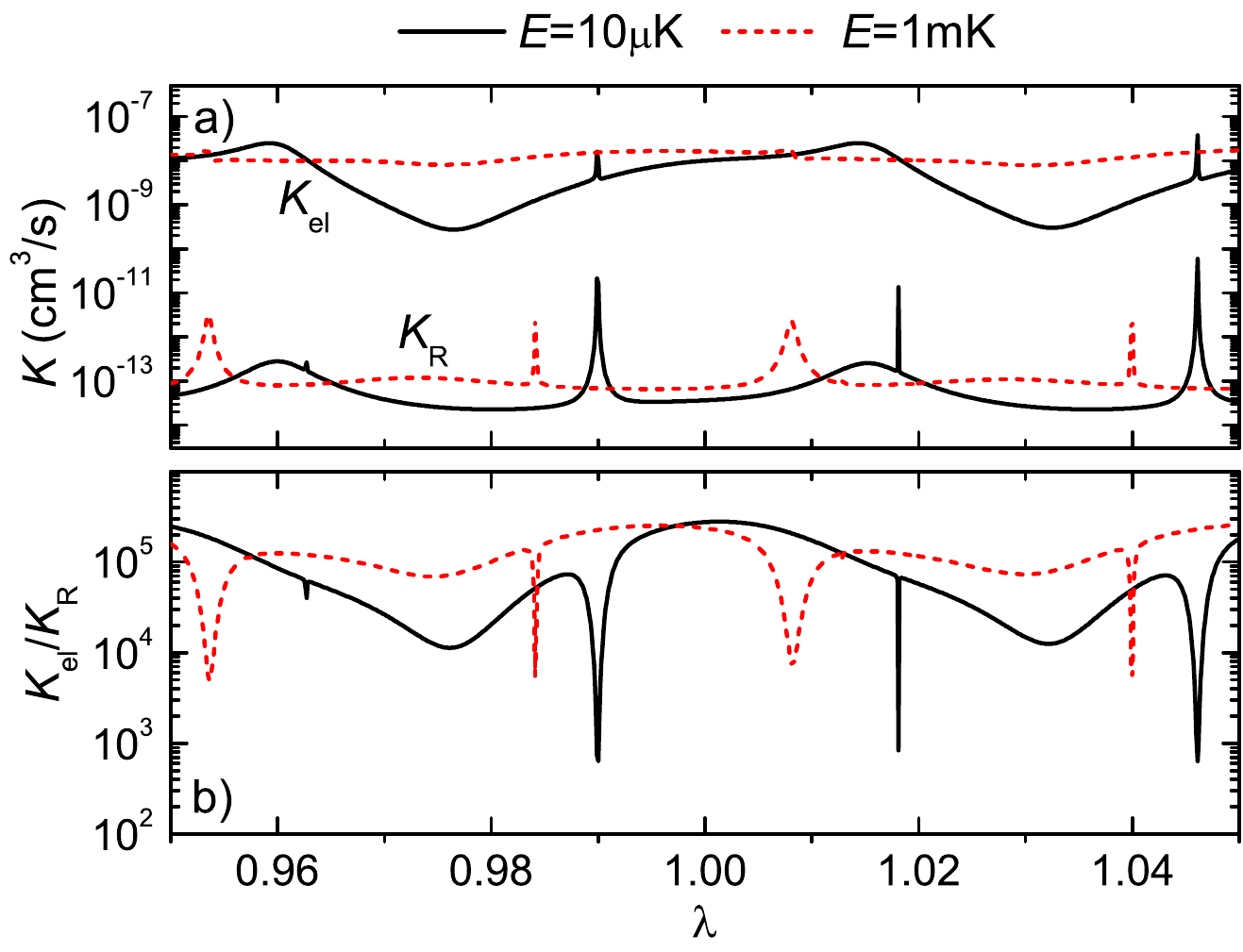}
\end{center}
\caption{Sensitivity of the rate constants for elastic $K_\mathrm{el}$ and reactive $K_\mathrm{R}$ scattering (a) and of the ratio of elastic to reactive rate constants (b) at collision energies of $10\,\mu$K (black solid lines) and $1\,$mK (red dashed lines) in the Yb$^+$+Li system to a scaling factor $\lambda$ applied to the interaction potential, $V(R)\to\lambda\cdot V(R)$. From~\cite{TomzaPRA15a}.}
\label{fig:sensitivity}
\end{figure}

Unfortunately, at the moment even the best state-of-the-art \textit{ab initio} methods do not allow one to predict accurately the scattering lengths, and consequently the exact shape resonance pattern, for collisions between many-electron atoms, ions, or molecules. Therefore, it is important to assess the dependency of the results of scattering calculations of cross sections and rate constants on the accuracy of the interaction potentials and, when possible, to correct them using scattering data from experiments. 
Figure~\ref{fig:sensitivity} shows the sensitivity of the rate constants for Yb$^+$ colliding with Li on scaling the interaction potential by a constant factor, in the range that corresponds to the potential's uncertainty~\cite{TomzaPRA15a}. A weak dependence of the rate constants is observed to only be interrupted by the presence of sharp resonances that occur when bound states of the ion-atom system cross the incoming threshold. Additionally, the ratio of elastic to inelastic rate constants is always significantly larger than 100, which should be sufficient for  successful sympathetic cooling in this system.   

The elastic collisions at intermediate collision energies $E\gtrsim 10E^\star$ can be described by using a semiclassical approximation. \cite{CotePRA00} obtained for the polarization potential the following expression for the elastic cross section
\begin{equation}
\sigma_\mathrm{el}(E)=\pi\left(\frac{4\mu C_4^2}{\hbar^2}\right)^{1/3}\left(1+\frac{\pi^2}{16}\right)\frac{1}{E^{1/3}}\,.
\end{equation}

The inelastic and reactive collisions in the regime of intermediate collision energies and large probability of chemical reaction at short range can be successfully described with the classical \textit{Langevin capture model}~\cite{Levine09}. In this model, rate constants for barrierless and exothermic reactions are governed by the interplay between the long-range part of the interaction potential and the centrifugal barrier. If the collision energy for a given partial wave is larger than the height of the centrifugal barrier, Eq.~\eqref{eq:E_max}, then the trajectory of the ion-atom collision is inward-spiralling and thus reactive. If the collision energy is smaller than the height of the centrifugal barrier, then the trajectory is only weakly modified in a so-called glancing collision. For a pure polarization potential of the form given in Eq.~\eqref{eq:V_pol}, the classical capture cross section is given by
\begin{equation}
\sigma_\mathrm{L}(E)=2\pi\sqrt{\frac{C_4}{E}}\,,
\end{equation}
and the corresponding Langevin rate constant for ion-atom collisions, $K_\mathrm{L}=\frac{\hbar k}{\mu} \sigma_\mathrm{L}$, is independent of the collision energy. The quantum calculations predict the same energy-independence both using the exact coupled-channel and approximate e.g.~random phase approximation approaches. The above classical model can be straightforwardly generalized to the case in which the reaction does not happen at short range with unit probability, but with some finite probability $P_{\rm re}$. In this case the Langevin rate coefficient should be simply rescaled by $P_{\rm re}$. The quantum version of this model, which correctly reproduces low energy scattering properties including shape resonances, will be discussed in Sec.~\ref{subsec:mqdt}.

\subsection{Radiative and non-radiative charge transfer}
\label{sec:charge_transfer}

\begin{figure}[tb!]
\includegraphics[width=0.9\columnwidth]{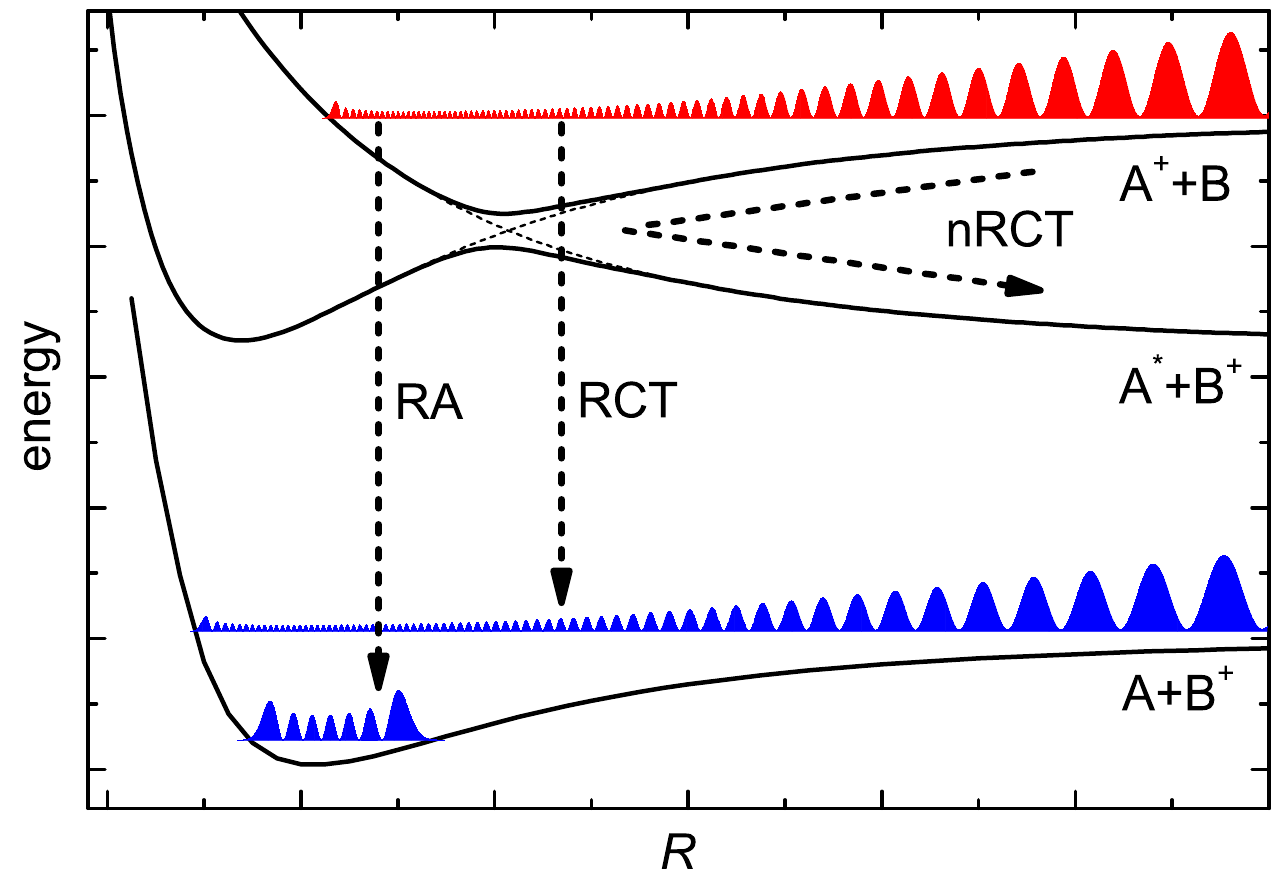}
\caption{Schematic representation of possible charge transfer processes in cold ion-atom collisions:  RA - radiative association, RCT - radiative charge transfer,  nRCT - non-radiative charge transfer. Exemplary electronic states and vibrational wavefunctions are presented.}
\label{fig:CT_scheme}
\end{figure}

For all ion-neutral systems there exist two families of electronic states associated with two possible arrangements of the charge at the dissociation threshold: $A+B^+$ and $A^++B$  ($A+B^-$ and $A^{-}+B$), see Fig.~\ref{fig:CT_scheme}. $A+B^+$ is the absolute electronic ground state if the ionization potential of $B$ is smaller than the ionization potential of~$A$ ($A+B^-$ is the absolute electronic ground state if the electron affinity of $B$ is larger than the electron affinity of $A$). 

If ionic and neutral species are separated by a large distance and interatomic or intermolecular interactions are negligible, then the second arrangement $A^++B$ ($A^-+B$) is also stable. However, if $A^+$ ($A^-$) and $B$ species collide and interact, then the interaction-induced transition electric dipole moment can appear between electronic states associated with the $A^++B$ and $A+B^+$ ($A^{-}+B$ and $A+B^-$) dissociation thresholds (see e.g.~the transition moment between the $X^1\Sigma^+$ and $A^1\Sigma^+$ states of the (Li+Yb)$^+$ ion-atom system in Fig.~\ref{fig:LiYb+_dip}). This can lead to  collision- and interaction-induced spontaneous radiative charge transfer (RCT) processes 
\begin{equation}\label{eq:charge_transfer}
\begin{split}
A^++B\to A+B^+ + h\nu\,,\\
A^-+B\to A+B^- + h\nu\,,
\end{split}
\end{equation}
where the electron is spontaneously transfered from an atom $B$ (a negative ion $A^-$) to a positive ion  $A^+$ (an atom $B$) emitting a photon of energy $h\nu$.

If the charge transfer process of Eq.~\eqref{eq:charge_transfer} is energetically allowed, then also the spontaneous radiative association (RA) driven by the transition between two electronic states can happen  
\begin{equation}\label{eq:assoctaion}
\begin{split}
A^++B\to AB^+ + h\nu\,,\\
A^-+B\to AB^- + h\nu\,,
\end{split}
\end{equation}
where an ion $A^+$ ($A^-$) and an atom $B$ spontaneously form an ionic complex $AB^+$ ($AB^-$) emitting a photon of energy $h\nu$.  

The spontaneous radiative association driven by the transition between rovibrational levels of one electronic state can also happen for all polar complexes $AB^+$ ($AB^-$), however rate constants for such a process are much smaller and usually negligible. 

\begin{figure}[b!]
\includegraphics[width=0.95\columnwidth]{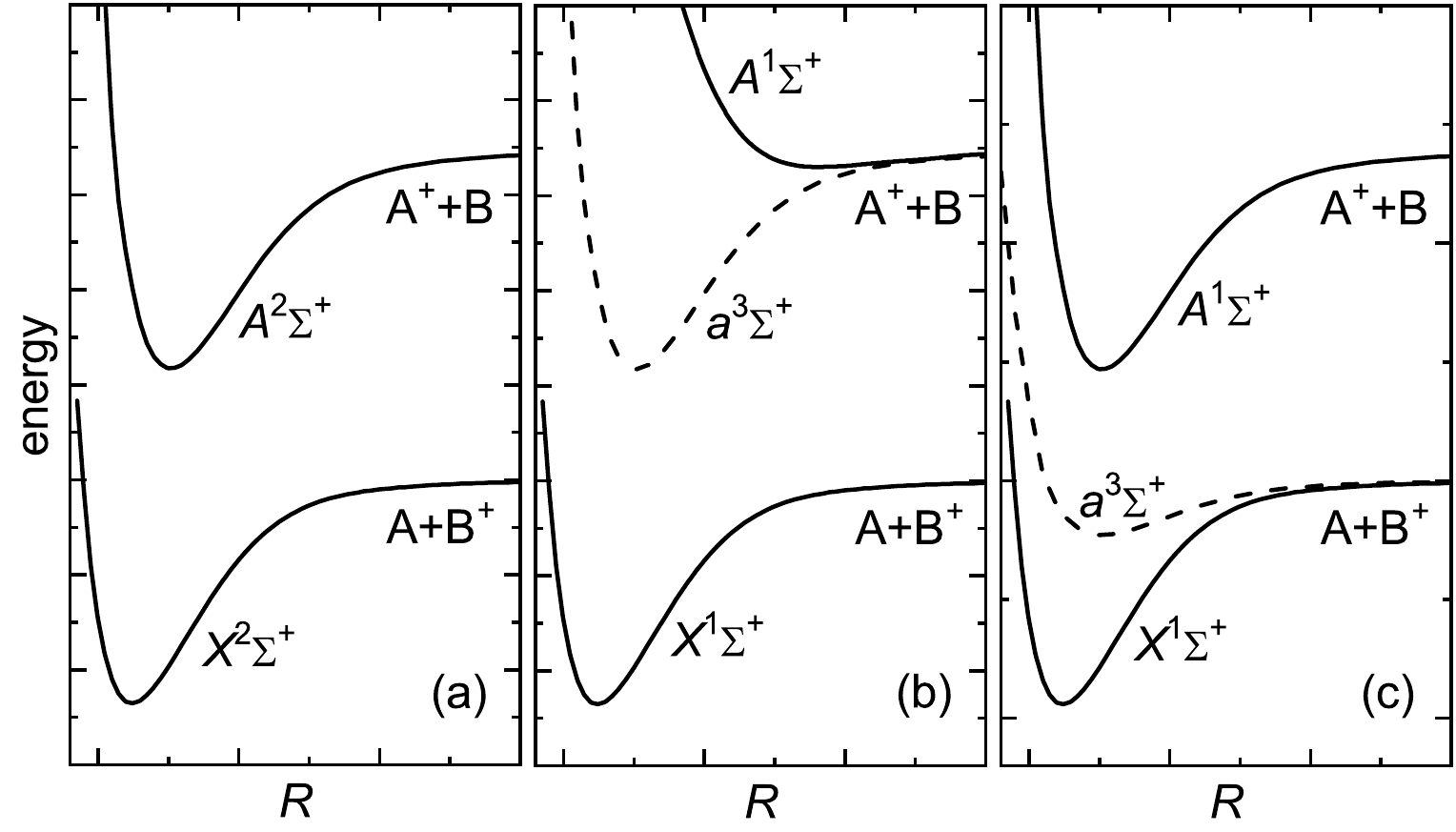}
\caption{Selected possible arrangements of the charge and electronic states in ion-atom systems. Panel (a) is typical for alkali-metal ions interacting with alkali-metal atoms and panels (b) and (c) for alkaline-earth-metal ions interacting with alkali-metal atoms. Magnetically tunable Feshbach resonances are expected for $A^++B$ and $A+B^+$ collisions in panels (b) and (c), respectively.  }
\label{fig:geom}
\end{figure}

Two electronic states associated with the $A^++B$ and $A+B^+$ ($A^{-}+B$ and $A+B^-$) dissociation thresholds can also cross each other at short interatomic or intermolecular distances (see~Fig.~\ref{fig:CT_scheme}). In the case of an atomic ion and an atom, this will lead to an avoided crossing of two states coupled by non-adiabatic coupling, whereas for molecular systems this will be a conical intersection. In both cases, such a crossing can lead to non-radiative charge transfer. 

\begin{table*}[tb!]
\caption{Hybrid heteronuclear ion-atom systems without charge transfer radiative losses among the combinations of the present experimentally accessible atomic ions and ultracold atoms, all in the ground electronic state. From~\cite{TomzaPRA15b}.} 
\begin{ruledtabular}
\begin{tabular}{ll}
Nature of ion/atom &  Systems without charge transfer radiative losses   \\
\hline
open-shell/open-shell 
&  Ba$^+$/Li, Ca$^+$/Cr, Sr$^+$/Cr, Ba$^+$/Cr, Yb$^+$/Cr, Sr$^+$/Dy, Ba$^+$/Dy, Sr$^+$/Er, Ba$^+$/Er\\
open-shell/closed-shell 
& Ca$^+$/Mg, Sr$^+$/Mg, Ba$^+$/Mg, Yb$^+$/Mg, Sr$^+$/Ca, Ba$^+$/Ca, Ba$^+$/Sr, Ca$^+$/Yb, Sr$^+$/Yb, Ba$^+$/Yb\\
closed-shell/open-shell 
& Na$^+$/Li, K$^+$/Li, Rb$^+$/Li, Cs$^+$/Li, K$^+$/Na,  Rb$^+$/Na, Cs$^+$/Na, Rb$^+$/K, Cs$^+$/K, Cs$^+$/Rb  \\
& alkali-metal ion/Cr, alkali-metal ion/Dy, alkali-metal ion/Er \\
closed-shell/closed-shell & alkali-metal ion/alkaline-earth-metal atom (except Li$^+$/Ba)
\end{tabular}
\label{tab:losses}
\end{ruledtabular}
\end{table*}

The charge transfer and association processes can lead to heating and loss of the ion because of the excessive energy release. The list of hybrid heteronuclear ion-atom systems free of charge-transfer radiative losses among the combinations of the present experimentally accessible atomic ions and ultracold atoms was presented by~\cite{TomzaPRA15b} and is collected in Table~\ref{tab:losses}. Related possible arrangements of the charge and electronic states in ion-atom systems are presented in Fig.~\ref{fig:geom}. Standard magnetically tunable Feshbach resonances, which are essential for controlling ion-atom collisions and will be described in Sec.~\ref{secIII:Feshbach}, are expected when both ion and atom are open-shell and more than one spin configuration is possible for related dissociation threshold. 

The radiative and non-radiative charge transfer processes are short-range phenomena because the non-negligible transition electric dipole moments and crossings between electronic states occur at relatively small interatomic or intermolecular distances. Therefore, to predict rate constants for charge transfer processes the full interaction potentials and transition electric dipole moments are needed. For the same reason, charge transfer processes lead mostly to the formation of molecular ions rather than to charge exchange between the ion and atom, as was discussed in the previous section. The formation of molecular ions in such a scenario will be analyzed in detail in Sec.~\ref{sec:radiative_association}.

Spontaneous radiative processes are governed by the Einstein coefficients~(see e.g.~\cite{MakarovPRA03,IdziaszekNJP11,KrychPRA11,TomzaPRA15a} and references therein). For transitions between two bound rovibrational states $vl$ and $v'l'$ (bound-bound), between a scattering
state of energy $E$ and a bound state $v'l'$ (free-bound), and between two scattering states of energies $E$ and $E'$ (free-free), they are given by 
\begin{equation}\label{eq:Einstein}
    \begin{split}
      A_{vl,v'l'} &= \frac{4\alpha^3}{3e^4\hbar^2}
      H_{l} (E_{vl}-E_{v'l'})^3  \Big| \langle\Psi_{vl}|
      d(R)|\Psi_{v'l'}\rangle\Big|^2\,,       \\
      A_{El,v'l'} &= \frac{4\alpha^3}{3e^4\hbar^2}
      H_{l} (E-E_{v'l'})^3  \Big| \langle\Psi_{El}|
      d(R)|\Psi_{v'l'}\rangle\Big|^2\,,       \\
      A_{El,E'l'} &= \frac{4\alpha^3}{3e^4\hbar^2}
      H_{l} (E-E')^3  \Big| \langle\Psi_{El}|
      d(R)|\Psi_{E'l'}\rangle\Big|^2\,,       
    \end{split}    
\end{equation}
respectively. In Eq.~\eqref{eq:Einstein} the primed and unprimed quantities pertain to the ground- and excited-state potentials, respectively, $d(R)$ is the transition electric dipole moment from the ground to the excited electronic state, $\alpha$ the fine structure constant, and $e$ the electron charge. The H\"onl-London factor $H_{l}$ is equal to $(l+1)/(2l+1)$ for the $P$ branch ($l=l'-1$), and to $l/(2l+1)$ for the $R$ branch ($l=l'+1$). In Eq.~\eqref{eq:Einstein}, the scattering states $|\Psi_{El}\rangle$  are energy normalized, whereas the wavefunctions of bound levels are normalized to unity, such that the three types of Einstein coefficients have different dimensions.

Radiative charge transfer at a given collision energy $E$ can be described by the following Fermi golden rule type expression for the rate constant, 
\begin{equation}\label{eq:K_ct}
K_\mathrm{RCT}(E)=\frac{4\pi^2\hbar^2}{\mu k}\sum_{l=0}^\infty(2l+1)\sum_{l'=l\pm 1}
 \int_0^{\varepsilon_\text{max}} A_{El,E'l'} d \varepsilon \,,
\end{equation}
where $\varepsilon=E-E'$. Analogously, 
the rate constant for radiative association is given by
\begin{equation}
  K_\mathrm{RA}(E)=\frac{4\pi^2\hbar^2}{\mu k}\sum_{l=0}^\infty(2l+1)
  \sum_{l'=l\pm 1}\sum_{v'}A_{El,v'l'}\,.
  \label{eq:K_ra}
\end{equation}
The total rate constant for radiative losses at a given collision energy $E$ is the sum of
Eqs.~(\ref{eq:K_ct}) and~(\ref{eq:K_ra}), 
\begin{equation}
  K(E)=K_\mathrm{RCT}(E)+K_\mathrm{RA}(E)\,.
\end{equation}
To get thermal rate constants at a given temperature $T$, the energy-dependent rate constants have to be averaged over a thermal distribution
\begin{equation}
  K(T)=\langle K(E)\rangle_T\,.
\end{equation}
However, for ion-atom systems the energy distribution can be different from the standard Maxwell-Boltzmann one due to the presence of micromotion. This will be discussed in detail in Sec.~\ref{sec:micromotion}.

Radiative and non-radiative charge transfer have been experimentally investigated for several cold ion-atom systems and these results will be presented in Sec.~\ref{sec:cold_col}. It has been also theoretically investigated for several ion-atom systems relevant for ongoing experimental efforts: Ca$^+$+Na~\cite{MakarovPRA03,GacesaPRA16}, Ba$^+$+Rb~\cite{KrychPRA11}, Yb$^+$+Rb~\cite{SayfutyarovaPRA13,McLaughlinJPB14}, Yb$^+$+Li~\cite{TomzaPRA15a,daSilvaNJP2015}, (Rb+Ca/Sr/Ba/Yb)$^+$~\cite{daSilvaNJP2015}, Yb$^+$+Ca~\cite{ZygelmanJPB14,PetrovJCP17}, Li$^+$+Na~\cite{BuenkerPRA15}, Na$^+$+Rb~\cite{YanPRA13,YanPRA14}, Be$^+$+Li~\cite{RakshitPRA11}, H$^+$+Na/K~\cite{WatanabePRA02}, H$^+$+D~\cite{EsryJPB2000,BodoNJP08}. The non-radiative charge transfer driven by non-adiabatic couplings has been theoretically investigated for Ca$^+$+Rb~\cite{TacconiPCCP11,BelyaevPRA12} and Yb$^+$+Rb~\cite{SayfutyarovaPRA13}.

\subsection{Quantum defect theory}
\label{subsec:mqdt}
As demonstrated in the previous sections, the two-body ion-atom problem exhibits a rich variety of multichannel phenomena and represents a challenge for theoretical calculations. On the other hand, ion-atom systems share several universal properties, in particular the form of the long-range interaction, which is typically dominated by a polarization potential which scales as $1/R^{4}$. This becomes particularly important at low temperatures, when the long-range part of the potential, which affects collisions with low momenta, becomes crucial for the dynamics. Multichannel quantum defect theory (MQDT) allows us to make use of the simple form of the long-range potential and to describe the properties of ion-atom systems with minimal numerical effort. While originally developed to describe the electronic structure of atoms, it found widespread applications in atomic and molecular physics~\cite{greene1979general,seaton1983quantum,MiesJCP84,greene1985molecular,gao1998quantum,gao2005multichannel}. The basic idea of MQDT is to make use of the fact that the interaction potential approaches a simple form at large interparticle distances. For atomic and molecular collisions the long-range potential is indeed given by the leading dispersive term and the coupling between different channels typically vanishes. This is especially important at low collision energies $\lesssim E^\star$, since in such a case the collision partners approach each other more slowly and thus the long-range part of the potential has decisive impact on the collision process.
Power-law potentials such as the van der Waals and polarization ones are particularly appealing since one can benefit from analytical solutions of the Schr\"{o}dinger equation at long range.

First treatment of polarization potential using quantum defect methods has been performed by~\cite{WatanabePRA80} in the context of negative ion photodetachment. We will now briefly review the MQDT treatment of low-energy ion-atom collisions following~\cite{MiesJCP84,IdziaszekPRA09,IdziaszekNJP11}. An equivalent treatment based on a slightly different formulation has been developed by Gao and coworkers~\cite{GaoPRL10,LiPRA12,GaoPRA13,LiPRA14}, allowing for additional insights e.g. into the characterization of scattering resonances. Similar approach has been developed also by~\cite{RaabPRA09} to describe the bound states of the polarization potential. 

Consider a general multichannel two-body problem, where the channels account for the hyperfine structure and for different partial waves, described by a close-coupled radial Schr\"odinger equation as in Eq.~\eqref{eq:coupled_Sch_eq}
\begin{equation}
\label{eq:multichanSEQDT}
\frac{\partial^2 \mathbf{\Phi}(R)}{\partial R^2}+\frac{2\mu}{\hbar^2}\left[E\mathbf{I}-\mathbf{W}(R)\right]\mathbf{\Phi}(R)=0\,,
\end{equation}
where $\mathbf{I}$ is the identity matrix and $\mathbf{W}(R)$ is the interaction matrix, which includes couplings between different channels, but reaches the asymptotic form
\begin{equation}
W_{ij}\stackrel{R\to\infty}{\longrightarrow}\left(E_i^\infty+\frac{\hbar^2 l_i (l_i +1)}{2\mu R^2}-\frac{C_4}{R^4}\right)\delta_{ij}\,.
\end{equation}
Here $E_i^\infty$ is the threshold energy of channel $i$. The solution matrix $\mathbf{\Phi}(R)$ can be split into blocks describing energetically open (o) [$E_i^\infty < E$] and closed (c) [$E_i^\infty > E$] channels.
One now replaces the interaction $\mathbf{W}(R)$ with a set of reference potentials $\{V_i(R)\}$ which can be arbitrary as long as they reproduce the behavior of $\mathbf{W}$ at large distances. Two linearly independent solutions of this new system can be denoted as $\hat{f}_i$, $\hat{g}_i$. The solution matrix can then be written as
\begin{equation}
\mathbf{\Phi}(R)\stackrel{R\to\infty}{\longrightarrow}\left[\mathbf{\hat{f}}(R)+\mathbf{\hat{g}}(R)\mathbf{Y}\right]\mathbf{A}\,,
\end{equation} 
where $\mathbf{Y}$ is called the quantum defect matrix and it contains all relevant information about the short-range potential details, and $\mathbf{A}$ gives the amplitudes. Since typically deviations from the asymptotic form of the interaction occur in the region where the potential is much deeper than typical energy scales such as $E^\star$, the $\mathbf{Y}$ matrix is expected to only weakly depend on the collision energy.

The $\hat{f}$, $\hat{g}$ functions can be explicitly connected to long-distance scattering (denoted as $f$, $g$) and bound state ($\phi)$ solutions by means of the MQDT functions $C$, $\lambda$ and $\nu$
\begin{align}
\begin{array}{lll}
f_i(R)    & = & C_i^{-1}(E)\hat{f}_i(R)\,,\\
g_i(R)    & = & C_i(E)[\hat{g}_i(R)+\tan\lambda_i\hat{f}_i(R)]\,,\\
\phi_i(R) & = & \mathcal{N}(E)[\cos\nu_i(E)\hat{f}_i(R)-\sin\nu_i(E)\hat{g}_i(R)]\,,
\end{array}
\end{align}
where $\mathcal{N}(E)$ is the normalization factor.
All the properties of the system are given by the scattering matrix $\mathbf{S}$ and by the structure of the bound states. These can be determined from the quantum defect matrix $\mathbf{Y}$ and the quantum defect functions according to~\cite{MiesJCP84}
\begin{equation}
\label{So}
\mathbf{S}_\mathrm{oo}=e^{i\mathbf{\xi}_\mathrm{oo}}(1+i\mathbf{R}_\mathrm{oo})(1-i\mathbf{R}_\mathrm{oo})^{-1}e^{i\mathbf{\xi}_\mathrm{oo}}\,,
\end{equation}
where $i\boldsymbol{\xi}_\mathrm{oo}$ is a diagonal matrix of phase shifts associated with the reference potentials $\xi_{ij}=\xi_i\delta_{ij}$ and
\begin{equation}
\label{Ro}
\mathbf{R}_\mathrm{oo}=\mathbf{C}^{-1}(E)(\bar{\mathbf{Y}}_\mathrm{oo}^{-1}-\tan\mathbf{\lambda}(E)_\mathrm{oo})^{-1}\mathbf{C}^{-1}(E)\,.
\end{equation}
Here $\bar{\mathbf{Y}}_\mathrm{oo}$ is the open-channel block of the $\mathbf{Y}$ matrix, renormalized due to the presence of closed channels
\begin{equation}
\label{Yren}
\bar{\mathbf{Y}}_\mathrm{oo}=\mathbf{Y}_\mathrm{oo}-\mathbf{Y}_\mathrm{oc}(\tan\mathbf{\nu}(E)_\mathrm{cc}+\mathbf{Y}_\mathrm{cc})^{-1}\mathbf{Y}_\mathrm{co}\,.
\end{equation}
The bound states of the system can be found by making all channels closed and by solving the equation
\begin{equation}
\mathrm{det}\left(\mathbf{Y}+\tan\nu(E)\right)=0\,.
\end{equation}
\cite{GaoPRA13} derived a highly general method allowing to study the resonance spectrum using an expression with very similar form.

The presented formalism is general and it can be applied to any collisional problem. The specific features of ion-atom systems are contained in the quantum defect functions $C$, $\lambda$ and $\nu$ and in the structure of the $\mathbf{Y}$ matrix. 
The quantum defect functions can be obtained by considering the single channel Schr\"{o}dinger equation reduced to the dimensionless form by using $R^\star$, $E^\star$ units (in the remaining part of this section we use these dimensionless units unless otherwise stated)
\begin{equation}
\frac{\partial^2 \Phi(R)}{\partial R^2}+\left(E-\frac{l(l+1)}{R^2}-\frac{1}{R^4}\right)\Phi(R)=0\,.
\end{equation}
For the case of single channel the wave fundtion at short range  can be parametrized using short-range phase $\phi$ which determines the scattering length of the full potential.
The next step is mapping onto the Mathieu equation by substituting $\Phi(R)=\psi(z)z^{1/2}$~\cite{VogtPR1954,OMalleyPR61}
\begin{equation}
\frac{d^2 \psi}{d z^2}-\left(a-2q\cosh (2z)\right)\psi=0\,,
\end{equation}
with $a=(l+1/2)^2$ and $q=\sqrt{E}$. The Mathieu equation can be solved analytically to find the properties of both bound and scattering states. The detailed recipe for finding the quantum defect functions knowing the analytic solutions was given by~\cite{IdziaszekNJP11}.

\begin{figure}[tb!]
\includegraphics*[width=0.9\linewidth]{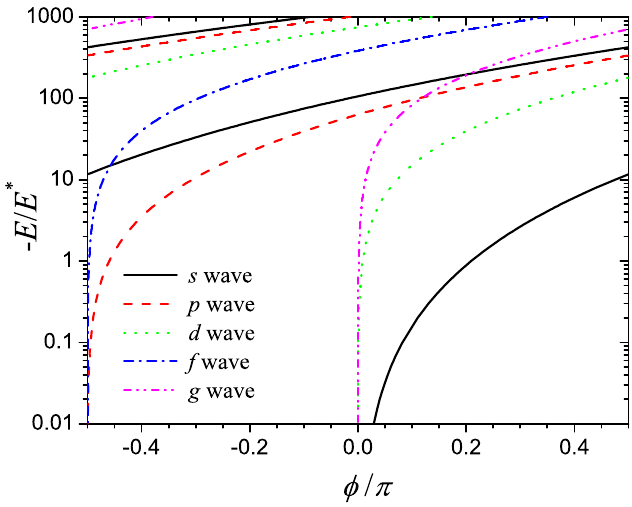}
\caption{Bound state energies of the polarization potential as a function of the short-range phase $\phi$ for the first few lowest partial waves. From \cite{IdziaszekNJP11}.
}
\label{fig6_IdziaszekNJP11}
\end{figure}

The analytic structure of the bound states of the polarization potential has several remarkable properties. Firstly, they have $l=2$ periodicity: for even values of $l$ the bound states appear at threshold when the $s$-wave scattering length diverges ($a=\pm\infty$), while for odd $l$ this happens at $a=0$. Furthermore, the energy of the near-threshold $s$-wave bound states is given by
\begin{equation}
E=-\frac{1}{a^2}+\frac{2\pi}{3}\frac{1}{a^3}+\mathcal{O}\left(1/a^4\right)\,.
\end{equation}
The first term in the above expression describes the universal scaling of the weakly-bound-state energy for contact interactions, while the second is the leading order correction coming from the long-range nature of the potential. An example of the bound-state spectrum is shown in Fig.~\ref{fig6_IdziaszekNJP11}. The $\ell=2$ periodicity is clearly visible at the threshold.

An important part of the MQDT calculation is to determine the quantum defect matrix. This can be done by numerically solving the close-coupled Schr\"{o}dinger equation at short range up to a distance when interchannel couplings become negligible. A simple alternative method is to use a frame transformation technique~\cite{BurkePRL98}. If both atom and ion are in their electronic ground states, the channels states are the hyperfine structure eigenstates characterized by a set of quantum numbers $\left|f_\mathrm{i} m_\mathrm{i} f_\mathrm{a} m_\mathrm{a} l m_l\right>$. At short distances the proper quantum numbers are the molecular ones with total nuclear spin $I=i_\mathrm{i}+i_\mathrm{a}$ and total electron spin $S=s_\mathrm{i}+s_\mathrm{a}$. The quantum defect matrix in the molecular basis is diagonal, and it can be parametrized as
\begin{equation}
Y^{IS}_{\alpha\alpha\prime}=\delta_{\alpha\alpha\prime}(a_\alpha)^{-1}\,,
\end{equation}
where $a_\alpha$ is the $s$-wave scattering length characterizing the channel $\alpha$ for $\phi=0$. Then the $\mathbf{Y}$ matrix in the asymptotic basis can be obtained by the unitary transformation $\mathbf{Y}=\mathbf{UY}^{IS}\mathbf{U}^\dagger$. In this way the problem is parametrized with very few unknown quantities.

MQDT methods can also be extended to take into account reactive processes such as charge exchange and radiative association~\cite{IdziaszekPRA09,LiPRA12,LiPRA14}. It turns out that if the reaction process is barrierless and sufficiently exoergic, i.e., the exit channel threshold is far below the initial state, and if the reaction takes place at length scales $R_0\ll R^\star$, it is possible to benefit from the separation of scales and to derive analytically a number of universal properties of the reaction process that depend only on the form of the long-range potential~\cite{IdziaszekPRL10,GaoPRA11,JachymskiPRL13,JachymskiPRA14}.
In this case a simple and intuitive parametrization of the short-range wave function can be given in a WKB-like form~\cite{IdziaszekPRL10,SakimotoPRA16}
\begin{equation}
\psi(R)\sim \frac{\mathrm{exp}\left[-i\int{k(R)dR}\right]}{\sqrt{k(R)}}-\frac{1-y}{1+y}\frac{\mathrm{exp}\left[i\int{k(R)dR}\right]}{\sqrt{k(R)}}\,,
\end{equation}
where $k(R)$ is the local wave vector. Here the first term describes the flux coming into the reaction region, and the second one is the outgoing flux with amplitude reduced by the factor $\frac{1-y}{1+y}$ where $0\leq y\leq 1$. The short-range reaction probability can be expressed as $P_{\rm re}=4y/(1+y)^2$. In the limit $y\to 1$, there is no outgoing flux and so one can expect that the reaction will not be sensitive to the short-range potential at all. Finite reaction probability results in a series of shape resonances in different partial waves. It is possible to derive their positions and widths based on the analytic structure of the theory~\cite{GaoPRA13,JachymskiPRL13} as well as by a semiclassical treatment~\cite{JachymskiPRA14,SakimotoPRA16}.

\begin{figure}
\includegraphics*[width=0.9\linewidth]{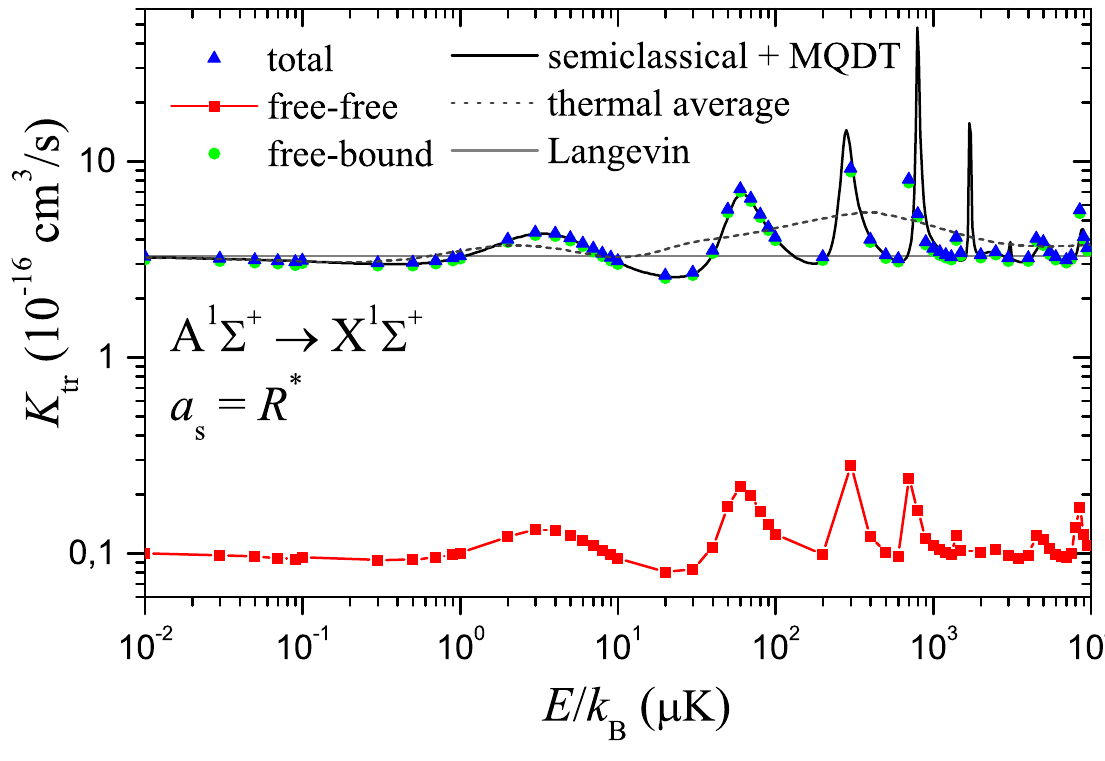}
\caption{Radiative charge transfer rate constants for the Ca$^+$+Na system assuming a scattering length $a=R^\star$. The predictions of the quantum defect model (solid black lines) with fitted reaction probability are compared with the full numerical calculation (blue triangles). The red squares show the contribution of the free-free transitions (charge transfer), while the green circles stand for the free-bound transitions (radiative association). The dashed line gives a thermal average of the rate constants, and thin grey line is the Langevin rate constant. From \cite{IdziaszekNJP11}.
}
\label{fig:mqdtrates}
\end{figure}

At high collision energies, the results derived within this framework agree with the classical Langevin model, which assumes that every classical trajectory falling on the collision center contributes to the reaction with probability $P_{\rm re}$, and it predicts that for an $1/R^{4}$ interaction the rate constant does not depend on the collision energy (obviously, the classical Langevin model does not predict any resonances). At low energies, the rate coefficient approaches the constant $s$-wave limit, in agreement with Wigner threshold laws
\begin{equation}
K_{\rm re}\stackrel{E\to 0}{\longrightarrow}\frac{2h R^\star y}{\mu}\frac{1+s^2}{1+y^2 s^2}\,,
\end{equation}
where $s$ is the scattering length of the reference potential given in units of $R^\star$. 

As expected, in the universal limit $y\to 1$ the rate constant does not depend on $s$. Universal reaction does not lead to any scattering resonances. In practice, ion-atom collisions tend to have a small $y$ parameter $\lesssim 0.01$ when both ion and atom are in the ground electronic state. The predictions of this simple model are in very good agreement with numerical calculations performed for different systems, e.g.~\cite{LaraPRA15}.

Figure~\ref{fig:mqdtrates} shows the rate constants for Ca$^+$+Na charge transfer and radiative association processes calculated using MQDT. A full numerical solution of the close-coupled problem is in excellent agreement with the quantum defect results.

\subsection{Ion impurity in a many-body system}
\label{sec:impurity}


When a charged particle is placed in a polarizeable medium, the induced polarization around it will follow its motion. The mobility and transport properties of the ion can thus be strongly influenced by the structure of the medium as well as the particle-environment interactions. For instance, charge-exchange collisions increase the mobility of the charge in the gas~\cite{DalgarnoPTRS58}. \cite{CotePRL00} found that the total charge mobility of Na$^+$ in its parent gas presents a sharp enhancement as the temperature is reduced, showing a transition from an almost insulating to a conducting state at a few $\mu$K with typical magneto-optical trap densities due to enhanced charge hopping rate. 

Other types of effects can be found when one considers correlations in the medium, especially those originated by interactions. For example, dragging an impurity through the fluid and measuring the dissipation can be used to probe whether a superfluid has been formed. 
In this regard, the first experiments on the dynamics of an ion impurity date back to the late 1950s, when the mobility of an ion in liquid helium was investigated~\cite{WilliamsCJP57,CareriLTPC58,MeyerPR58} and used to probe its superfluid properties~\cite{ReifPR60,RayfieldPR64}. 
An unexpected slow motion of the ion within the liquid helium was observed and explained by a phenomenological model in which electrostriction effects increase the liquid density around the ion, leading to an increased effective mass~\cite{AtkinsPR59,KuperPR61}. \cite{GrossAP62} developed a systematic microscopic quantum theory based on the self-consistent field approximation for an ion impurity coupled to a weakly interacting homogeneous ensemble of bosons and confirmed that the effective ion mass can be much higher than its bare mass.

From the point of view of condensed matter physics, the system of a charged impurity immersed in a many-body environment is the crucial building block for the understanding of solids. Real materials contain multiple interacting electrons and ions, which leads to extremely large Hilbert space, and therefore becoming soon theoretically intractable from an $ab initio$ perspective. However, one can consider a single electron interacting with crystal deformations (phonons) as first done by~\cite{Frohlich1954}, who proposed to describe it with the Hamiltonian
\begin{align}
\label{eq:polaron}
\hat H = E_{\text{env}}^0 + \frac{\hat p_{\rm i}}{2m_{\rm i}}+\sum_{\mathbf k\ne 0}\hbar\omega_{\mathbf{k}}\hat a_{\mathbf{k}}^{\dag} \hat a_{\mathbf{k}} +\sum_{\mathbf{k}}V_{\mathbf{k}} e^{-i{\mathbf{k}}\cdot {\mathbf{r}}}(\hat a_{\mathbf{k}} + \hat a_{-\mathbf{k}}^{\dag}).
\end{align}
Here $\hat p_{\rm i}$ is the momentum of the impurity of mass $m_{\rm i}$, $E_{\text{env}}^0$ is the zero-temperature energy of the environment in which the impurity is immersed (the phonon bath), $\hat a_{\mathbf{k}}$ ($\hat a_{\mathbf{k}}^{\dag}$) denotes the annihilation (creation) operator of the phonon of frequency $\omega_{\mathbf{k}}$, and $V_{\mathbf{k}}$ is the Fourier transform of the impurity-environment interaction, which is essentially two-body but involves environment excitation instead of a real atom. One can now construct a new quasiparticle composed from lattice phonons with momenta $\mathbf{q}$ and the electron with $\mathbf{k}-\mathbf{q}$, which is called the polaron. 
The polaron, which is a quasiparticle dressed by the polarization cloud, is an extremely relevant concept in order to understand transport, optical response, and induced interactions in solid-state materials. Polarons are characterized by their self-energy, effective mass, and response to external (e.g., electric and magnetic) fields~\cite{DevreeseRP2009}.

This kind of description is generic and can be developed both for neutral as well as charged impurities and various types of bath such as thermal gases, phonons in a crystal, or strongly correlated Bose or Fermi gases. In the case of a Bose-Einstein condensate the role of phonons is played by the Bogoliubov excitations, for which an extended Fr\"ohlich model taking into account higher order scattering processes can be derived~\cite{RathPRA13}. The charge of the ion does not lead to any fundamental differences with respect to neutral atoms. The long-range nature of the potential, however, will manifest itself in increased importance of finite energy and finite range effects.
Depending on the ratio $\alpha$ between the impurity-bath interaction strength and the interaction between the bath constituents, the polarons can be divided into two main categories: weak and strong coupling polarons. 
For instance, for an ion immersed in an atomic condensate we can take $\alpha = (R^\star)^4/(a_{\text{3D}}^{\text{aa}}\xi^3)$ with $\xi$ being the healing length of the condensate~\cite{CasteelsJLTP11}. The strong coupling limit (i.e., $\alpha \gg 1$) is typically quite difficult to attain in condensed-matter systems. Neutral impurities in a BEC can reach $\alpha\gg 1$ by means of Feshbach resonances~\cite{TemperePRB09} as recently demonstrated in experiments~\cite{HuPRL16,JorgensenPRL16}, but a charged impurity may lead to the strong-coupling polaronic limit in a more direct way as a consequence of the large characteristic interaction length $R^\star$.

\begin{figure}[tb!]
\includegraphics[width=0.9\linewidth]{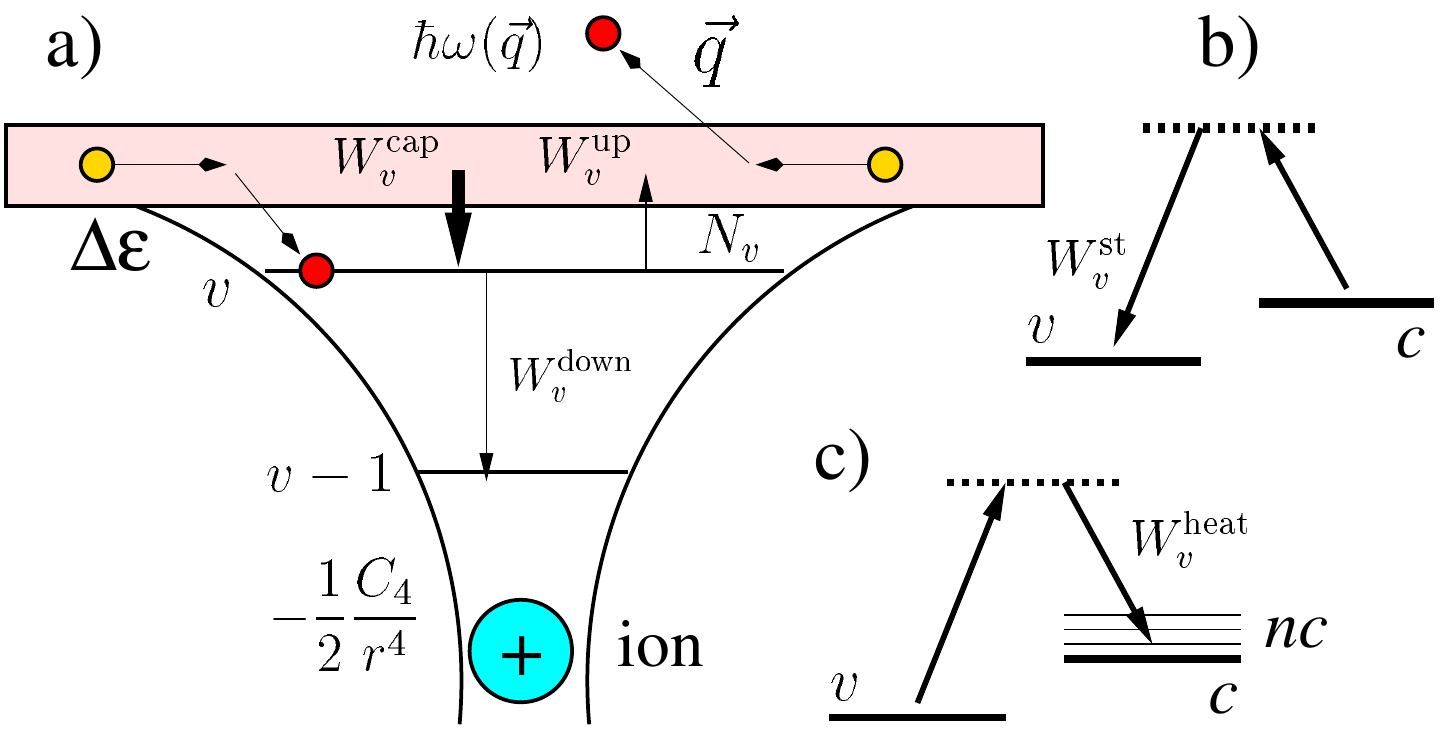}
\caption{Diagrams of an atom capture by an ion. (a) The spontaneous capture in level $v$ is followed by phonon emission (with corresponding rates). (b) The stimulated rate from the condensate $c$ to the bound level $v$ may also produce heating, i.e.~population of noncondensate atoms $nc$ (c). From~\cite{CotePRL02}.}
\label{fig1_CotePRL02}
\end{figure}

Another possible effect which stems from the long-range character of the interactions has been discussed by~\cite{CotePRL02}, who suggested the possible existence of a state in which a large number of ultracold atoms is bound to the ion forming a cluster with complex structure. Such mesoscopic molecular ions can be in principle produced either by collisions, or by spontaneous or laser-stimulated photoassociation (see~Fig.~\ref{fig1_CotePRL02}). For the latter, one could employ a Raman transition with two off-resonant and co-propagating laser beams. In this way, the transition from the condensate to the molecular bound state occurs with negligible momentum transfer to the atoms. It has been theoretically predicted by~\cite{CotePRL02}, by means of a mean-field rate-equation analysis, that hundreds of atoms could be captured in such loosely bound states in about one second. This phenomenon occurs through (superelastic) collisions, in which the excess energy is transferred into a collective excitation, i.e., with the emission of an acoustic Bogoliubov phonon. Formation of such a state would therefore be impossible in a thermal gas. The capture rate of the atoms into the mesoscopic ion $W^{\text{cap}}$ is controlled by the dimensionless parameter $\Xi = 8 \pi n_b a^2 a^{\text{aa}}\mu/m_\mathrm{a}$ with $a^{\text{aa}}$ and $a$ denoting the atom-atom and ion-atom scattering length respectively, and $n_b$ being the Bose gas density. Two main physical regimes were identified:~(i) the phonon-like regime, for which $\Xi\rightarrow \infty$, for which $W^{\text{cap}}\propto \hbar/[m_\mathrm{a} a^3\sqrt{n_b a^{\text{aa}}}]$, meaning that the capturing process is dominated by phonon-assisted transitions; (ii)~binary regime, for which $\Xi\rightarrow 0$ (i.e., $n_b a^{\text{aa}}\ll 1$, dilute condition) and $W^{\text{cap}}\propto \hbar n_b^2 (a^{\text{aa}})^2 a^2/m_\mathrm{a}$ so that the capture process occurs essentially because of three-body recombination events. Other effects such as charge hopping or thermal fluctuations limit the maximum number of captured atoms $N^{\max}$. \cite{CotePRL02} estimated that for sodium atoms at $T\sim 100$ nK and $a \sim 2000 a_0$ one obtains $N_{\upsilon}^{\max} \simeq 600$, and adding more atoms via photoassociation is prohibited by mechanisms similar to the dipole-blockade in Rydberg atoms.

From a different perspective,~\cite{MassignanPRA05} investigated the density modifications that an ion produces in a condensate. They calculated the excess atom number $\Delta N= 4\pi\int_{\mathbb{R}^+}dr\,r^2[n(r) - n_b]$, which is the number of atoms to be (hypothetically) added or subtracted from the atomic ensemble due to the presence of the ion. Here $n_b$ is the atomic density far away from the ion. In this model, the condensate properties (e.g. density, chemical potential) far from the ion remain unaltered by the ion presence. Based on a thermodynamical approach, it was found that in the low density limit $\Delta N = \frac{m}{\mu}\frac{a}{a^{\text{aa}}}$, where a pseudopotential treatment of the ion-atom interaction was assumed. This simple relation is in good agreement with a Gross-Pitaevskii mean-field calculation of the condensate density profile for different bulk densities. It is interesting that $\Delta N$ can be either positive or negative, depending upon the ion-atom phase shift (i.e, short-range behaviour). Typical numbers for $\vert\Delta N\vert$ are of the order of a few hundreds, while the spatial perturbation of the atom density around the ion is found to be on the $\mu$m-range. This implies that the presence of an ion indeed yields appreciable and observable
effects in the density measurements.

\begin{figure}[tb!]
\includegraphics[width=0.9\linewidth]{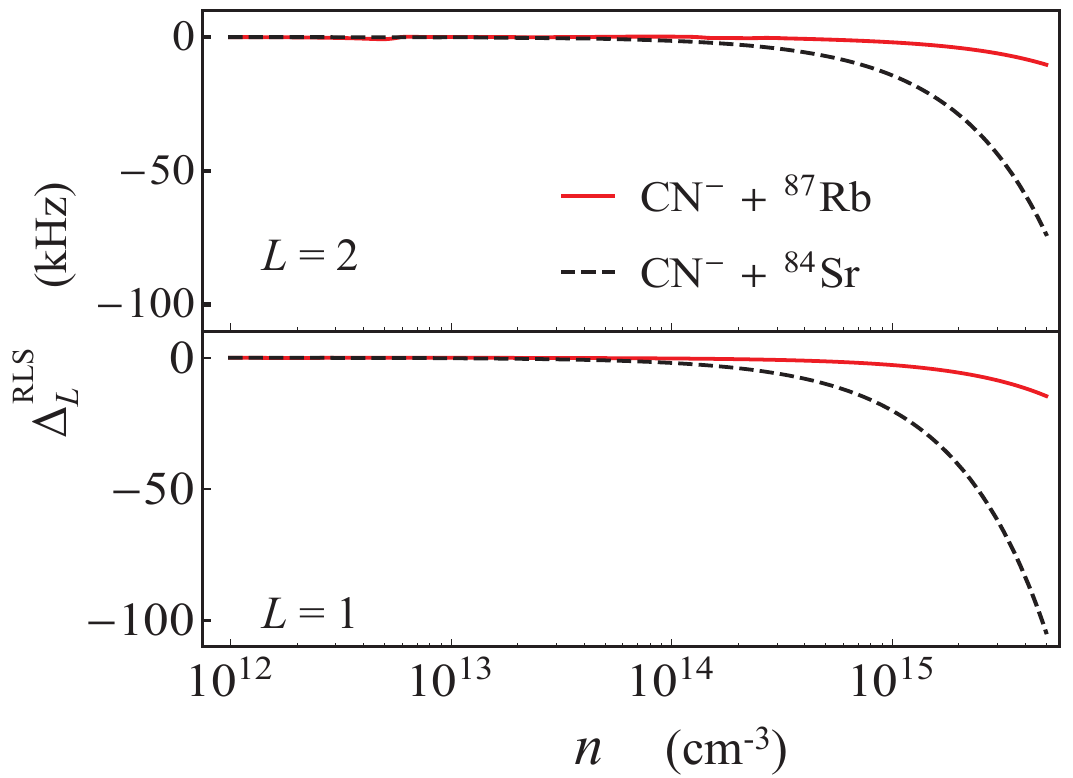}
\caption{Differential Lamb shift (see text for definition) for CN$^-$ in either a $^{87}$Rb (solid lines) or $^{84}$Sr (dashed lines) BEC as a function of the atomic density $n$. The upper panel corresponds to a total impurity-BEC angular momentum $L=2$, whereas the lower panel to $L=1$. From~\cite{MidyaPRA16}.}
\label{fig4_MidyaPRA16}
\end{figure}

Another type of polaron can arise if we consider the case of a molecular ion instead of an atomic ion. In this case its rotational degrees of freedom come into play. Such a problem in its generality has been investigated by~\cite{SchmidtPRL15}, and a new kind of quasi-particle, named the angulon, has been identified. It is the rotational analogue of the polaron, where the rotational motion is dressed with many-body excitations of the environment. \cite{MidyaPRA16} theoretically demonstrated that the signatures of the angulon can be observed in modern experiments with a single CN$^-$ anion in a BEC of $^{87}$Rb or $^{84}$Sr atoms. More precisely, they showed that the rotational Lamb shift and the many-body-induced fine structure resulting from the anisotropic impurity-BEC interaction are measurable. The rotational Lamb shift, which is the rotational analog of the effective mass of the polaron, is shown in Fig.~\ref{fig4_MidyaPRA16}. The differential shift is defined as $\Delta_L^{\text{RLS}}=[E_L(n) - E_0(n)]-[E_L(0) - E_0(0)]$ with $E_L(n)$ being the energy of the state of total angular momentum $L$ in the presence of a phonon bath in the atomic BEC environment at density $n$. For both CN$^-$+Rb and CN$^-$+Sr systems, the differential shift is on the order of several kHz, reaching almost 100 kHz for the latter system at typical condensate densities, all within current experimental resolution. 

\begin{figure}[tb!]
\includegraphics[width=1.0\linewidth]{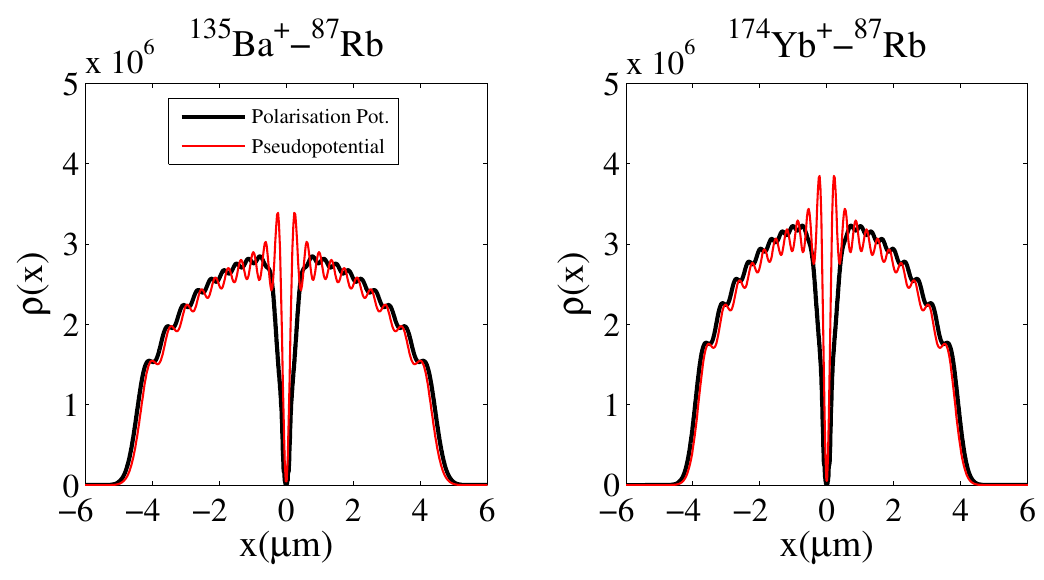}
\caption{The single particle densities of a Tonk-Girardeau gas of 20 particles in the presence of a central ion for $^{135}$Ba$^{+}$+$^{87}$Rb and $^{174}$Yb$^{+}$+$^{87}$Rb systems with a typical trapping frequency, $\omega=70$Hz (thick black line).  The short range phases in each case are chosen to be $\phi_e=\frac{\pi}{4}$ and $\phi_o=-\frac{\pi}{4}$. The thin red line in the plots represents the result of a pseudopotential approximation for the ion, using a large value for the scattering length. From~\cite{GooldPRA10}.}
\label{fig3_GooldPRA10}
\end{figure}

Let us now turn to the case of one-dimensional systems. In this case the existence of a true Bose-Einstein condensate is forbidden and phase fluctuations play more important role in the system properties. In the presence of an ionic impurity, the mean-field description of the static as well as dynamical properties of the atomic system might be questionable and Bogoliubov or perturbation theories are not accurate enough~\cite{MassignanPRA05}. One limiting case of one-dimensional system is the Tonks-Girardeau gas of impenetrable bosons. Behavior of the impurity embedded in the Tonks gas has been studied by~\cite{GooldPRA10}. They tackled the problem by means of quantum defect theory and by employing the Bose-Fermi mapping. The latter eliminates the atom-atom interaction so that the problem reduces to finding the single-particle states in the presence of the static ion-atom potential. Because of the strong repulsion between the atoms, three-body recombination can be neglected, since the density-density correlation function is vanishing on the length scale of the inter-particle distance. As a consequence, the gas density for $N$ particles reduces to $\rho(x) = \sum_{n=0}^{N-1} \vert\psi_n(x)\vert^2$, where $\psi_n(x)$ are the eigenfunctions of the single-particle Hamiltonian. Interestingly, it has been found that the presence of the ion drastically perturbs the atom density by generating a bubble in the trap center, where the ion is positioned, whose size is of the order of one micrometer (Fig.~\ref{fig3_GooldPRA10}), similarly to the prediction by~\cite{MassignanPRA05}. Even though this result might be counterintuitive at first glance, as the ion-atom interaction is attractive at long distances, in this case no relaxation to the molecular bound states is allowed because of the strong atom-atom repulsion, so no build-up of atomic density around the ion is expected.

\begin{figure}[tb!]
\includegraphics[width=0.9\linewidth]{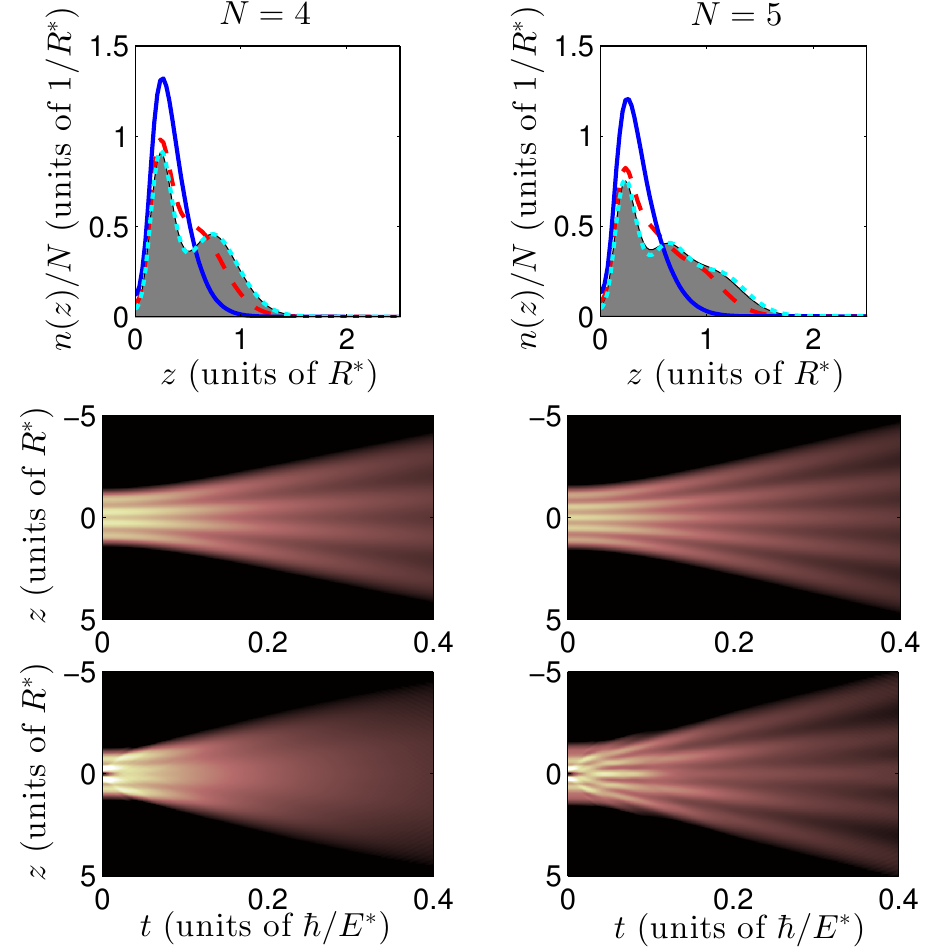}
\caption{Upper panels: Normalised atomic density profiles for different values of the atom-atom interaction strength $g_{\text{1D}}/(E^\star R^\star) = 10, 40, 160$ (solid, dashed, and dotted lines) with the static ion located in the trap center. Besides, the density profile of the Tonk-Girardeau gas as a grey shaded area is shown. Note that we show only the densities along the positive semi-axis, because of the symmetry of the ground state. Middle panels: Free expansion in the waveguide for the TG gas in a harmonic trap (i.e., without ion). Bottom panels: Free expansion in the waveguide for the TG gas after the sudden removal of the ion at time $t=0$.
In all panels the units of the length and energy scales are calculated with respect the atom mass, namely $R^\star = \sqrt{\alpha e^2 m_\mathrm{a}/\hbar^2}$ and $E^\star = \hbar^2/[2 m_\mathrm{a} (R^\star)^2]$, and the trap length is $R^\star/\sqrt{8}$. Adapted from~\cite{SchurerPRA14}.}
\label{fig12-16_SchurerPRA14}
\end{figure}

\begin{figure*}[tb!]
\includegraphics[width=0.95\linewidth]{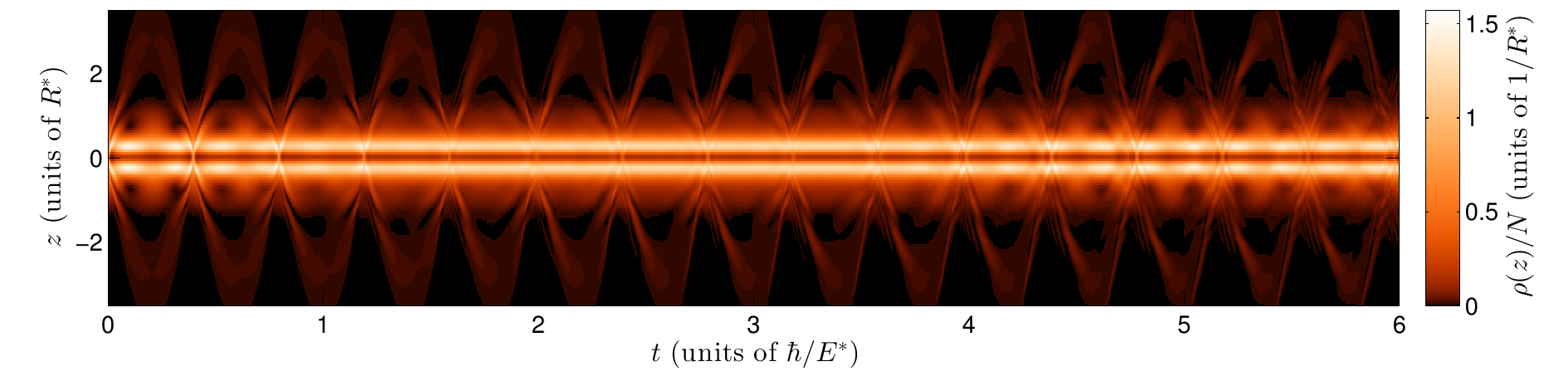}
\caption{Time evolution of the density for $N=10$ atoms after sudden creation of a tightly trapped ion in the center of the atom trap ($\hbar/E^\star = 0.39$ ms for $^{87}$Rb atoms). Adapted from~\cite{SchurerNJP15}.}
\label{fig:secIIIc02}
\end{figure*}

In recent years, a significant effort has been undertaken to simulate in a numerically efficient way the many-body Schr\"odinger equation describing an ion in a one-dimensional atomic cloud.
The ground-state properties of ultracold bosons in interaction with a static (i.e., tightly trapped) ion impurity have been investigated by~\cite{SchurerPRA14} by means of the multiconfigurational time-dependent Hartree method for bosons~\cite{OfirPRA08,CaoJCP13,KroenkeNJP13}. A harmonically trapped ensemble of atoms with a static ion has been considered, with the effective~1D Hamiltonian
\begin{align}
\hat H & =\sum_{j=1}^N\left[
-\frac{\hbar^2}{2 m_\mathrm{a}}\frac{\partial^2}{\partial z_j^2} + V_{\text{trap}}(z_j)+V_{\text{ai}}^{\text{mod}}(z_j)
\right]\nonumber\\
\phantom{=}&
+g_{\text{1D}} \sum_{j<n}\delta(z_j-z_n),
\label{eq:secIV-Hbosons}
\end{align}
and with a fixed atom-atom interaction strength $g_{\text{1D}}$ corresponding to a weakly interacting gas. Here $V_{\text{ai}}^{\text{mod}}(z_j)$ denotes a model potential for the ion-atom interaction, $V_{\text{ai}}^{\text{mod}}(z_j)\to -C_4/z_j^4$ and $ V_{\text{trap}}(z_j)=\frac{1}{2}m\omega z_j^2$ is the harmonic trap with the frequency $\omega$ (see Sec.~\ref{subsec:colltrap} for a discussion on the 1D conditions in ion-atom systems). It has been found that the ion drastically ``slows down" the transition from the weakly interacting regime to the Thomas-Fermi limit as the atom number is increased. Such a behaviour is owed to the fact that the lowest energy states are bound and thus strongly localized near the ion, and the trap eigenstates can only be populated when the interaction energy is sufficiently large, i.e., $N\gg 1$. Furthermore, the presence of the ion strongly affects the momentum distribution of the atoms. These effects become even more pronounced when $g_{\text{1D}}$ is enhanced to attain the Tonks-Girardeau limit. Interestingly, the presence of the ion can influence the atomic gas in different ways depending on the atom number parity. This effect can be understood as follows: since the Hamiltonian~(\ref{eq:secIV-Hbosons}) is symmetric and the atom wavefunction has to vanish at the ion location (i.e., the minimum of the trap center), the atomic density distribution has to be symmetric. Thus, while for an even atom number the density dip characteristic for a gas of impenetrable bosons is observed (see Fig.~\ref{fig3_GooldPRA10} and Fig.~\ref{fig12-16_SchurerPRA14} upper left panel), an additional atom has to be equally distributed between the two sides of the atomic cloud (see Fig.~\ref{fig12-16_SchurerPRA14} upper right panel). Hence, while for an even atom number the ion literally separates the atomic could into two incoherent parts, the additional atom restores the spatial coherence of the system. This effect is observable in the free expansion of the atomic gas within the waveguide after the sudden removal of the ion, as illustrated in Fig.~\ref{fig12-16_SchurerPRA14} (bottom panels). For $N=4$ atoms, the hump-like structure disappears as time evolves, as the two sides of the clouds are incoherent, whereas for $N=5$, the coherence between the two halves is provided by the fifth atom such that the hump-like structure becomes more and more pronounced (see also the case without the ion, middle panel of Fig.~\ref{fig12-16_SchurerPRA14}). Naturally, such reliance on the atom parity can be more easily observed in small atomic samples, i.e. a few tens of atoms, as for larger atom numbers such hump-like structures in the density will likely be smeared over. Strongly interacting systems in reduced dimensions, however, are better attained at low densities and typical Tonks gas experiments can be prepared with $\sim$10-50 atoms in a single wire~\cite{PaganoNP14,MeinertS17}, thus enabling the observation of such phenomena in the laboratory. 

A recent many-body analysis including the ion motion in a harmonic trap has shown that the formation of molecular ions in one dimension crucially depends on the atom number and interatomic interaction $g_{\text{1D}}$~\cite{SchurerArxiv16}. Importantly, there exists a critical atom number, for a fixed $g_{\text{1D}}$, beyond which no atom can be bound to the ion anymore. The effective mass of the ion depends mostly on the number of bound atoms. In addition, an effective trapping potential appears due to the presence of unbound atoms surrounding the molecular ion. Moreover, the actual many-body quantum state exhibits a shell-like structure, in which multiple bound and trapped states are occupied~\cite{GaoPRL10}.

So far, we have discussed mostly static properties and, in a specific many-body study, we have shown that multiple scales involved in the problem induce multi-mode features in the system properties. The latter aspect becomes even more pronounced in time-dependent processes. To be concrete, let us consider an initially harmonically trapped condensate, in which suddenly a single atom is ionized~\cite{SchurerNJP15}. This quenching scenario could be realised, for instance, by preparing an atom of a different species in a tight dipole trap and then shining a laser field interacting only with it (i.e., off-resonant for the rest of the cloud) in order to extract the valence electron. A sufficiently deep optical trap would ensure that the formed ion is still trapped. The time evolution for such a quench process is illustrated in Fig.~\ref{fig:secIIIc02} for $N=10$ atoms and for a trap length of $R^\star/\sqrt{8}$. As it can be seen, for short times the sudden generation of the ion transfers most of the atoms into
the bound states of the ion-atom polarisation potential. The remaining atoms are ejected into the outer regions of the harmonic trap, and then (at $t\approx 0.2\hbar/E^\star$) reflected back to the trap center because of the confinement. In addition to this, there is another faster oscillation in the region of the atoms captured by the ion (holes in Fig.~\ref{fig:secIIIc02} at $z\approx \pm R^\star/2$), which shows a collapse and revival type behaviour. 
A detailed analysis reveals that such a behaviour is connected to the loss and gain of spatial coherence between the inner and outer density fractions. This demonstrates that the interplay between different energy and length scales provides a rather rich correlated dynamics.

The discussed instances of an ion immersed in a structured quantum bath such as the one provided by a degenerate atomic quantum gas illustrate how such compound system can shed light on the general impurity problem as well as improve our understanding of open quantum systems. A number of theoretical and experimental studies can be foreseen in the near future. From the point of view of quantum engineering, the influence of the quantum bath on the internal and motional coherence of the ion should be investigated. Another possible direction is to engineer mediated interactions between ionic impurities in the quantum gas. Finally, one may ask whether the ion can be exploited for sensing the properties of the bath, e.g. its temperature or magnetization (see also Sec.~\ref{subsec:probing}). Given the exquisite control and sensitivity of the ion motion against perturbations, one could envisage that global properties of the bath can be inferred by accurate measurements of the ion trajectory.

\section{Controlling collisions}
\label{sec:control}

In this section we discuss the prospects for controlling the collisional properties of ion-atom systems. We first focus on Feshbach resonances, which can be used to tune the interaction strength once the $s$-wave limit is reached. Then we discuss the role of external confinement, in particular trap- and confinement-induced resonances.

\subsection{Magnetically tunable Feshbach resonances}
\label{secIII:Feshbach}

Magnetic Feshbach resonances are one of the crucial tools available for controlling the interactions of ultracold atoms~\cite{ChinRMP10}. A Feshbach resonance is a generic phenomenon resulting from coupling of the scattering state to a bound state in a closed channel. The position of this bound state relative to the threshold can be controlled if the channel states have different magnetic moments. The coupling mechanism usually comes from the interaction between hyperfine structure states. More exotic resonances resulting from e.g.~weak electronic spin interaction or from interaction-induced variation of the hyperfine coupling have generally a much smaller width $\Delta$ as defined below.

Resonances manifest themselves by the divergence of the $s$-wave scattering length described by the formula~\cite{ChinRMP10}
\begin{equation}
a(B)=a_{\rm bg}\left(1-\frac{\Delta}{B-B_{\rm res}}\right),
\end{equation}
where $\Delta$ is the resonance width, $a_{\rm bg}$ is the background scattering length and $B_{\rm res}$ is the resonance position. Controlling the magnetic field strength allows then for tuning  $a(B)$. For short-range interactions at low collision energies, the scattering length is the most important quantity characterizing the interaction, which makes the resonance extremely useful.

Feshbach resonances can be expected to occur also in ion-atom systems, with no fundamental differences from neutral atoms. However, the relatively long range nature of the interaction results in much smaller characteristic energy scales. While atomic scattering is dominated by the $s$-wave already at mK temperatures, for ion-atom systems the $s$-wave regime is typically reached at sub-$\mu$K energies, depending on the reduced mass. In current setups based on Paul traps typically tens of partial waves contribute to the scattering cross sections and resonant features are washed out. Observation of ion-atom Feshbach resonances thus still represents a major experimental challenge.

From the theoretical side the most difficult task is to predict the positions of the resonances. This requires the knowledge of the scattering lengths of different channels, which cannot be accurately calculated without additional {\it a priori} knowledge. However, it is possible to obtain a lot of general properties of specific systems with just the knowledge about the long range interaction and hyperfine structure.

Two basic ways of studying Feshbach resonances include a numerical coupled-channel calculation and a quantum defect treatment based on the frame transformation concept. In the latter method one parametrizes the problem at short range with scattering lengths belonging to different hyperfine states in the molecular basis (the one corresponding to the total nuclear and electronic angular momentum), where the quantum defect matrix $\mathbf{Y}^{IS}$ is diagonal (see Section~\ref{subsec:mqdt}). The transformation back to the asymptotic scattering channels reads $\mathbf{Y}=\mathbf{Z}(B)\mathbf{UY}^{IS}\mathbf{U}^\dagger\mathbf{Z}^\dagger (B)$. Here $\mathbf{U}$ is the frame transformation between the molecular and asymptotic long-range basis and $\mathbf{Z}$ is the transformation to magnetic-field-dressed states~\cite{IdziaszekNJP11}, which for single valence electron (alkali atoms) is described by the Breit-Rabi formula.

Let us consider as an example the collision between an alkaline earth ion with no nuclear spin, e.g.~$^{40}$Ca$^+$, with an alkali atom such as $^{87}$Rb. This configuration is currently the most popular experimental choice. The ionic Zeeman states are $\left|f_\mathrm{i}=1/2,m_{f_\mathrm{i}}=\pm 1/2\right>$, while for the atom we have hyperfine structure with possible $f_\mathrm{a}=1$ and $f_\mathrm{a}=2$. We neglect weak couplings between different partial waves so that the total angular momentum projection $M_F$ is conserved ($F=f_\mathrm{i}+f_\mathrm{a}$). In the molecular basis the proper quantum numbers are $I=i_\mathrm{i}+i_\mathrm{a}$ and $S=s_\mathrm{i}+s_\mathrm{a}$. As $s_\mathrm{i}=s_\mathrm{a}=1/2$, the channels can be divided into singlet ($S=0$) and triplet ($S=1$) ones. The quantum defect matrix is diagonal in the molecular basis, while in the asymptotic channel representation the couplings between the channels depend on the parameter $1/a_\mathrm{c}=1/a_\mathrm{s}-1/a_\mathrm{t}$, where $a_\mathrm{s}$ is the singlet and $a_\mathrm{t}$ the triplet scattering length. We notice that for similar values of the two scattering lengths the couplings will be weak and the resulting resonances should be narrow. It is possible to manipulate the density of resonances to some extent by selecting different hyperfine levels or switching between isotopes.

\begin{figure}[tb!]
\includegraphics[width=0.9\linewidth]{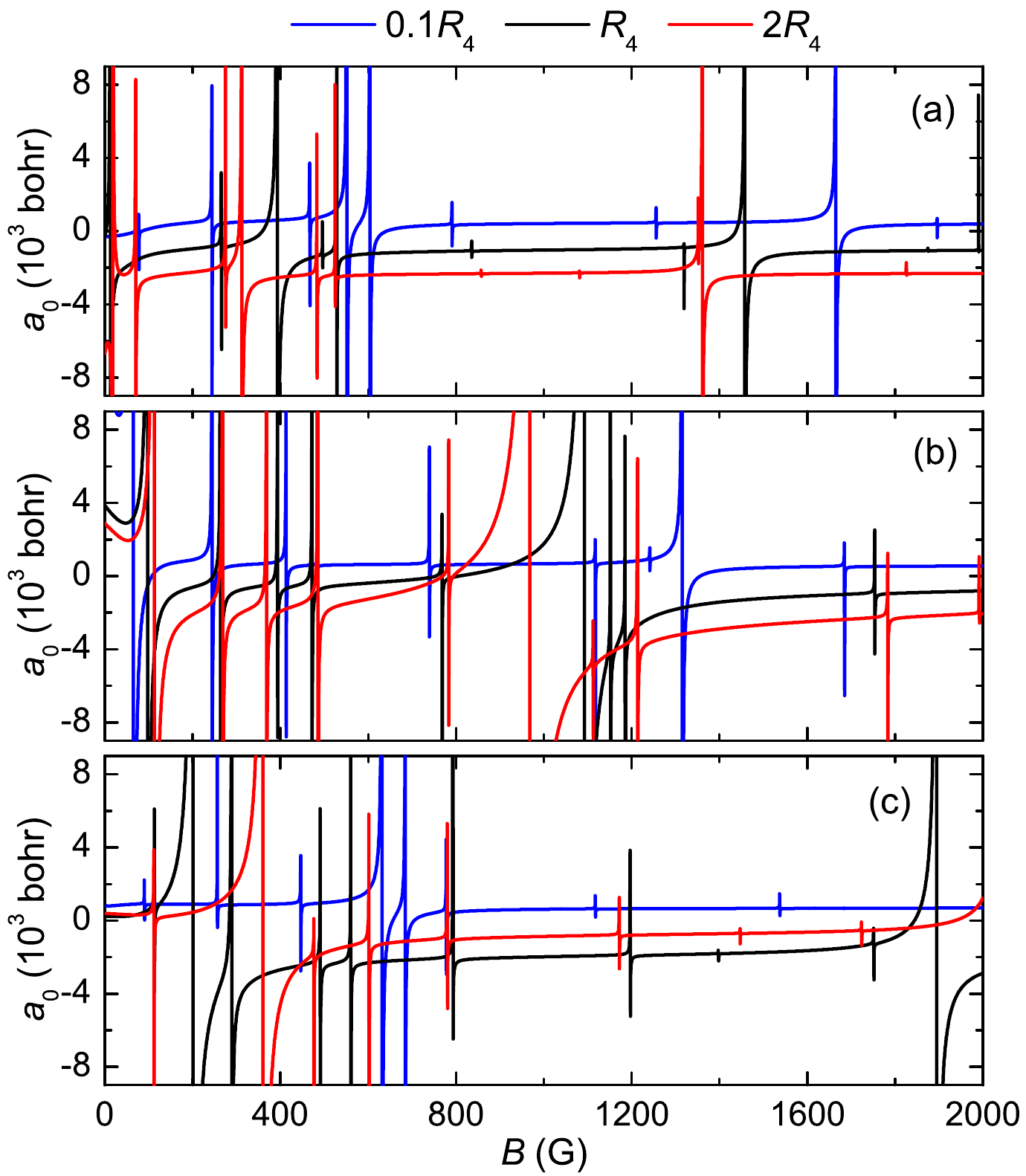}
\caption{The $s$-wave scattering length as a function of magnetic field for collisions of a Cr atom with a Sr$^+$~(a), Ba$^+$~(b), and Yb$^+$~(c) ion, obtained from coupled-channel calculations. Typical scattering lengths of $0.1R^\star$, $R^\star$, and $2R^\star$ for the $X^6\Sigma^+$ electronic state and $-0.1R^\star$, $-R^\star$, and $-2R^\star$ for the  $a^8\Sigma^+$ electronic state are assumed, respectively. Adapted from~\cite{TomzaPRA15b}.}
\label{fig7_TomzaPRA15b}
\end{figure}

\begin{figure}[tb!]
\includegraphics[width=0.9\linewidth]{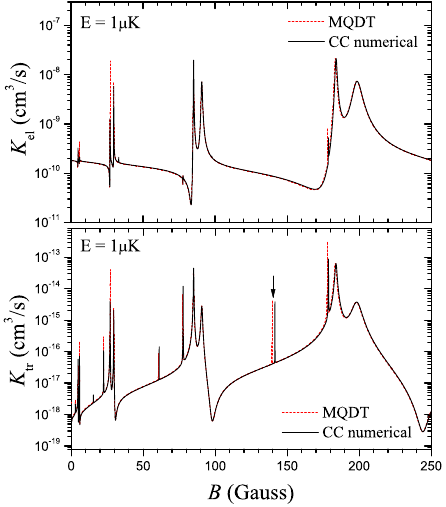}
\caption{Magnetic field dependence of the elastic and reactive (charge exchange and radiative transfer) collision rate constants for the Ca$^+$+Na system calculated with MQDT and close-coupled model with the same scattering lengths at collision energy $E/k_B=1\mu$K. Scattering lenghts of $R^\star$ and -$R^\star$ are assumed for the $X^1\Sigma^+$ and $a^1\Sigma^+$ electronic states, respectively. From~\cite{IdziaszekNJP11}.}
\label{fig:rescomp}
\end{figure}

Numerical calculations of Feshbach resonance spectra have been performed by~\cite{IdziaszekNJP11,TomzaPRA15a,TomzaPRA15b,GacesaPRA17} for a number of ion-atom systems including Yb$^+$+Li, which is the most promising in this respect from the experimental point of view due to the smallest reduced mass. Feshbach resonances were also investigated for collisions involving Cr atoms. Results are presented in Fig.~\ref{fig7_TomzaPRA15b}.  As the values of the relevant scattering lengths could not be calculated, these calculations were parametrized by changing the values of scattering lengths associated with different electronic symmetries. This can be done by scaling the potential energy curves. The calculations took into account the full structure of short range couplings, revealing the presence of narrow resonances originating from the hyperfine interaction in the Cr atom in addition to broad resonances coming from the hyperfine structure of the ions, which have much larger hyperfine coupling constant than~Cr.

At finite temperatures one should include contributions from higher partial waves and it is convenient to study the total collision rate constants instead of the scattering length. Already at $\mu$K temperatures one can notice a much higher density of narrow resonances with partial wave $l$ up to around $10$. This makes the resonance spectrum much harder to interpret and suggests that in current setups it is not possible to efficiently control the interactions. On the other hand, charge transfer processes are dominated by low-$l$ contributions even at relatively high energies due to the necessary tunnelling through the centrifugal barrier, which is exponentially suppressed. This makes them potentially promising for measuring the values of singlet and triplet scattering lengths experimentally. An exemplary dependence of collision rate constants on the magnetic field at finite temperature is shown in Fig.~\ref{fig:rescomp} where predictions of the MQDT model assuming the same short-range phase for each partial wave are compared to numerical calculations performed for Ca$^+$+Na system. As it is shown, Feshbach resonances can enhance the rate constants by several orders of magnitude. We note that for $l=4$ there is a small discrepancy between the two models (pointed by the black arrow), indicating that angular momentum-dependent corrections to the quantum defect matrix are needed to fully reproduce the spectrum. 

In all the considerations above we assumed that the collision partners are not subject to any external potential. In realistic experimental conditions the ion is placed in a tight external trapping potential, which can contain time-dependent terms when using the Paul trap. This introduces coupling between the center of mass and relative motion and modifies the asymptotic behaviour of the wave function. This can strongly affect the properties of resonances.

An additional effect which has to be considered is the coupling of the ion charge to the magnetic field. This leads to quantized cyclotron orbits for the ion movement, which can be neglected only if the orbit radius is much smaller than $R^\star$. This issue has been investigated by~\cite{SimoniJPB11}. As the ion motion is confined in directions perpendicular to magnetic field axis, the problem qualitatively resembles quasi-1D scattering. In the low field limit the effective confinement is weak and the scattering can be well described by the energy-dependent pseudopotential. However, for strong magnetic fields the effective 1D coupling constant exhibits multiple resonances with contributions from many partial waves. This should be included in the interpretation of future experimental results performed at high magnetic fields.

\subsection{Collisions in traps}
\label{subsec:colltrap}

As we have discussed in the previous subsection, magnetic Feshbach resonances provide a unique tool to control the coupling between a scattering and a molecular state. This controllability can be also attained by manipulating the external confinement of both the atom and the ion. For instance, by controlling the separation between the traps of the two interacting particles, trap-induced shape resonances (TIR) between the molecular bound states and the trap-extended states can be observed~\cite{StockPRL03}. In reduced-dimensional setups, where the confinement in one or two spatial directions is sufficiently tight, the effective scattering amplitude exhibits so-called confinement-induced resonances (CIR)~\cite{OlshaniiPRL98}. Such types of resonances have been also predicted for ion-atom systems~\cite{IdziaszekPRA07,MelezhikPRA16}.

We begin the discussion with the TIRs and we consider an ion-atom system that is described by the Hamiltonian:
\begin{align}
\label{H3D}
\hat H = & \sum_{\nu = \mathrm{i},\mathrm{a}} \left[ \frac{\mbf{\hat p}_\nu^2}{2 m_\nu} +
\frac{1}{2} m_\nu \omega_\nu^2 (z_\nu - d_\nu)^2
+ \frac{1}{2} m_\nu \omega_{\perp \nu}^2 \rho_\nu^2
\right]\nonumber \\
& + V(|\mbf{\hat r}_\mathrm{i}-\mbf{\hat r}_\mathrm{a}|).
\end{align}
Here the label i (a) refers to the ion (atom), respectively, $d_\nu$ denotes the positions of the centers of atom and ion traps, $\rho^2 = x^2 + y^2$ and $V(R)$ denotes the ion-atom interaction potential. The trapping potentials are axially symmetric and displaced along the $z$ axis. The coexistence of radio-frequency (rf) and dipole traps in the same spatial region is nontrivial and crucially depends on the separation of time scales of the rf and laser fields~\cite{IdziaszekPRA07}. 

The short-distance behavior of the wave function can be described e.g in the spirit of the quantum-defect theory (see~\ref{subsec:mqdt}). At distances $R \ll R^{\star}$
the relative and center-of-mass (COM) degrees of freedom are decoupled, and the relative motion is governed by the Hamiltonian $\hat H_0 = \mbf{\hat p}^2/2 \mu + V(R)$, since for $R \ll R^{\star}$ the interaction potential is much larger than typical trapping potentials and heights of the angular-momentum barrier. The short-distance behavior of the radial wave functions reads
\begin{equation}
\label{Sol1}
R_l(R,E) \sim \sin\left(R^\star/R + \varphi_l(E)\right), \quad R_0 \ll R \ll \sqrt{R^{\star}/k},
\end{equation}
where $R_0$ marks the distance at which the interaction potential deviates from the asymptotic long-range $1/R^4$ behavior and $k$ is the wave vector related to the asymptotic kinetic energy, defined in terms of the total energy as $E=\hbar^2 k^2/(2 \mu)$, and $\varphi_l(E)$ are quantum-defect parameters (i.e., short-range phases), which we assume to be independent of the energy and angular momentum, i.e. $\varphi_l(E) \equiv \varphi$. Hence, in the numerical calculations we can replace $V(R)$ by its asymptotic $1/R^{4}$ behavior and impose on the wave function the boundary condition stated by Eq.~\eqref{Sol1} with the single parameter $\varphi$. 

\begin{figure}
\includegraphics[width=0.9\linewidth]{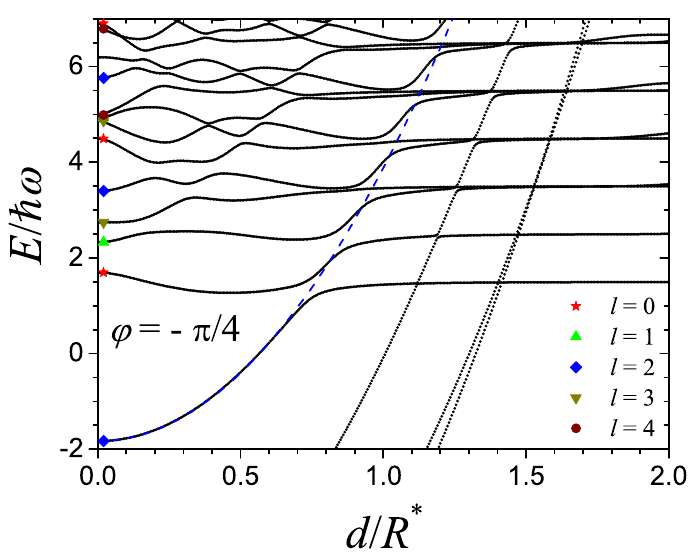}
\caption{Energy spectrum of the relative motion for an atom and an ion confined in identical spherically symmetric harmonic traps with frequency $\omega$ versus the distance $d$ between traps, calculated for $\varphi = - \pi/4$, and  $R^{\star} = 3.48\, a_\mrm{HO}$, where $a_\mrm{HO} = \sqrt{\hbar/(\mu \omega)}$. The blue dashed line shows the approximate dependence of the molecular state energy on $d$, while symbols at $d=0$ depict the angular momentum of the molecular states. From~\cite{IdziaszekPRA07}.}
\label{Fig:spectrum3D}
\end{figure}

An example of the energy spectrum of the Hamiltonian~\eqref{H3D} as a function of the distance between the traps $d=|d_\mathrm{i}-d_\mathrm{a}|$ is shown in Fig.~\ref{Fig:spectrum3D} 
with $\omega_\mathrm{i} = \omega_\mathrm{a} = \omega$. In this case the COM and relative motion can be separated, and therefore we can focus on the relative motion of the two particles only. At $d=0$ the angular momentum $l$ is a good quantum number and the states have definite angular symmetry, which is depicted by the appropriate symbols in Fig.~\ref{Fig:spectrum3D}. At intermediate distances, the energy curves exhibit avoided or diabatic crossings, depending on the symmetry of eigenstates and on the strength of the coupling term in the Hamiltonian. In the ion-atom system, avoided crossings can be attributed to resonances between molecular and vibrational states that appear when the energy of a vibrational level coincides with the energy of a molecular state shifted by the external trapping. The latter, to a good approximation, is given by $\mu \omega^2 d^2 /2$. This behavior can be explained by noting that the bound states $|\Psi_\mrm{mol}\rangle$ are localized around $R=0$, and thus $\langle \Psi_\mrm{mol}(d)| \hat H_\mrm{rel}(d) |\Psi_\mrm{mol}(d)\rangle \approx E_\mrm{b}+\frac{1}{2} \mu \omega^2 d^2$. Here $E_{\mrm{b}}$ is the binding energy at $d=0$ and $\hat H_\mrm{rel}$ is the relative-motion part of the Hamiltonian. The location of the avoided crossing depends on the short-range phases.

\begin{figure}
\includegraphics[width=\linewidth]{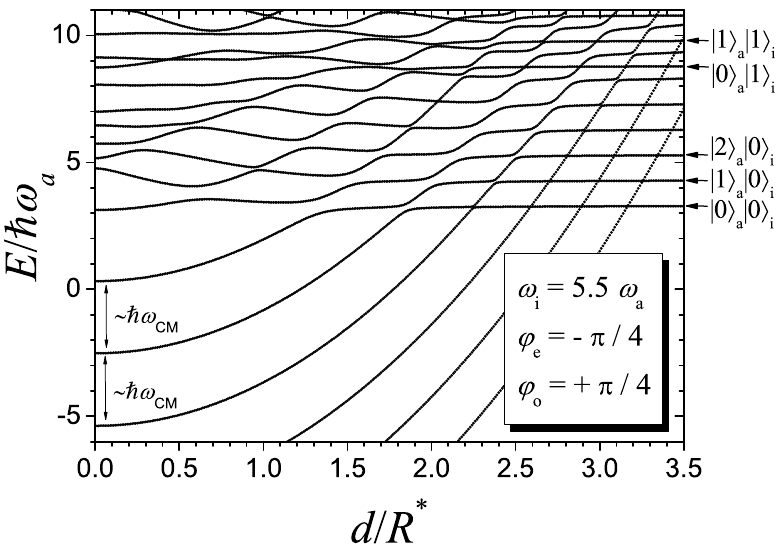}
\caption{Energy spectrum for an atom and an ion confined in harmonic traps with $\omega_\mathrm{i} = 5.5\, \omega_\mathrm{a}$ as a function of the distance $d$. Calculations are performed for $\varphi_\mathrm{e} = - \pi/4$, $\varphi_\mathrm{o} = \pi/4$ and $l_\mathrm{a} = 0.9 R^{\star}$. From~\cite{IdziaszekPRA07}.}
\label{Fig:SpectrFull}
\end{figure}

In the general case, when the trapping frequencies for the atom and ion are not equal, the relative and COM degrees of freedom are coupled and the energy spectrum exhibits a richer structure. This is illustrated in Fig.~\ref{Fig:SpectrFull}, but in one dimension (1D), showing the adiabatic energy levels as a function of $d$ for some example parameters: $\omega_\mathrm{i} = 5.5\, \omega_{\rm a}$ and $l_\mathrm{a} = 0.9 \,R^{\star}$. This choice corresponds to the interaction of $^{40}$Ca$^{+}$ and $^{87}$Rb in the traps with $\omega_\mathrm{a} = 2 \pi \times 10$ kHz and $\omega_\mathrm{i} = 2 \pi \times 55$ kHz. The figure indicates the presence of several avoided crossings between vibrational and molecular states, which are a manifestation of TIRs. In comparison to the case of identical trapping frequencies, we observe that the molecular part of the energy spectrum contains states with different numbers of excitations in the COM degree of freedom. They can be easily identified at $d=0$, when the molecular levels are then equally separated by $\hbar \omega_\mathrm{CM}$ with $\omega_\mathrm{CM}^2 = (m_\mathrm{a}\omega_\mathrm{a}^2+ m_\mathrm{i} \omega_\mathrm{i}^2)/M$. The arrows on the right-hand side of Fig.~\ref{Fig:SpectrFull} indicate the asymptotic states for large separations, which can be labeled by the number of excitation quanta for the atom and ion trap, respectively.

Let us now discuss the regime of quasi-one-dimensional dynamics, where the transverse confinement for both the atom and the ion is much stronger than the confinement in the longitudinal direction. In such a case, for energies smaller than the excitation energy in the transverse direction, the motion can be effectively assumed to be frozen to the ground state of the transverse trap. Nevertheless, the transverse motion plays an important role at short distances, since the true ion-atom interaction is of three-dimensional nature. For simplicity, hereafter we assume $\omega_{\perp \mathrm{i}}$ = $\omega_{\perp \mathrm{a}} = \omega_{\perp}$, although TIRs and CIRs appear also for different trapping frequencies. At large distances, the total wavefunction of the relative motion can be decomposed into a product of longitudinal and transverse components, enabling therefore to obtain an effective 1D Hamiltonian, which reads
\begin{align}
\label{eq:H1D}
\hat H_{\mathrm{1D}} = \sum_{\nu=\mathrm{i,a}} \left[
\frac{\hat p^2_{\nu}}{2 m_{\nu}} +
\frac{1}{2} m_\nu \omega_\nu^2 (z_\nu - d_\nu)^2
\right]
+V_{\mathrm{1D}}(\vert z_\mathrm{i} - z_\mathrm{a}\vert),
\end{align}
with
\begin{align}
\label{eq:int1D}
V_{\mathrm{1D}}(\vert z_\mathrm{i} - z_\mathrm{a}\vert)  = \int\int d\rho_\mathrm{i} d\rho_\mathrm{a} \vert\psi_0(\rho_\mathrm{i},\rho_\mathrm{a})\vert^2 V(\vert\mathbf{r}_\mathrm{i} - \mathbf{r}_\mathrm{a}\vert).
\end{align}
Here $\psi_0(\rho_\mathrm{i},\rho_\mathrm{a})$ is the (Gaussian) ion-atom wavefunction of the transverse harmonic confinement. The asymptotic behavior of the interaction for $\vert z \vert$ much larger than the trap length scales is $V_{\mathrm{1D}}(\vert z\vert) = - C_4/z^4$. 

\begin{figure}[tb!]
\includegraphics[width=\linewidth]{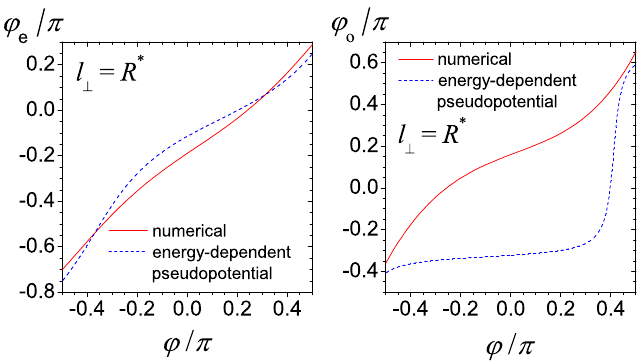}
\caption{Even and odd short-range phases $\varphi_\mathrm{e}$ and $\varphi_\mathrm{o}$ calculated for $a_\perp = R^{\star}$. Numerical results (red solid lines) are compared with predictions of the model replacing the ion-atom interaction with the energy-dependent pseudopotential. Adapted from~\cite{IdziaszekPRA07}.}
\label{Fig:phi1}
\end{figure}

In one dimension, the wavefunction can be split into the even and odd part, similarly to the partial wave expansion in 3D~\cite{OlshaniiPRL98}, which obey the following boundary conditions
\begin{align}
\label{eq:asym1D}
\Psi_{\mathrm{rel}}^\mathrm{e}(z,k)&\sim \vert z\vert \sin[R^\star/\vert z\vert + \varphi_\mathrm{e}(k)]\quad z \ll \sqrt{R^\star/k},\\
\Psi_{\mathrm{rel}}^\mathrm{o}(z,k)&\sim z \sin[R^\star/\vert z\vert + \varphi_\mathrm{o}(k)]\quad  z\ll \sqrt{R^\star/k}.\nonumber
\end{align}
Here the labels $\mathrm{e}$ and $\mathrm{o}$ indicate the even and odd solutions of the Schr\"odinger equation, respectively. Thus, in quasi-1D ion-atom systems two quantum defect parameters are needed: the even, $\varphi_\mathrm{e}(k)$, and odd, $\varphi_\mathrm{o}(k)$, short-range phases. This distinction is due to the fact that the general solution for two distinguishable particles can be any superposition of even and odd functions, contrarily to indistinguishable particles for which parity has to be conserved. The relationship between the short-range phases $\varphi_\mathrm{e,o}$ and the three-dimensional short-range phase $\varphi$ can be found numerically~\cite{IdziaszekPRA07}. Figure~\ref{Fig:phi1} shows an example of the dependence of $\varphi_\mathrm{e}$ and $\varphi_\mathrm{o}$ on $\varphi$ for $a_\perp = R^{\star}$. It compares the results of numerical calculations with predictions based on the pseudopotential method. In the regime $a_{\perp} \ll R^\star$ (strong ion-atom intercation or tight transverse confinement), where the pseudopotential approach is not applicable at all, the short-range phases $\varphi_\mathrm{e,o}$ vary even more rapidly as a function of $\varphi$~\cite{IdziaszekPRA07}, indicating the possibility of the occurrence of a CIR.

\begin{figure}[tb!]
\includegraphics[width=0.85\linewidth]{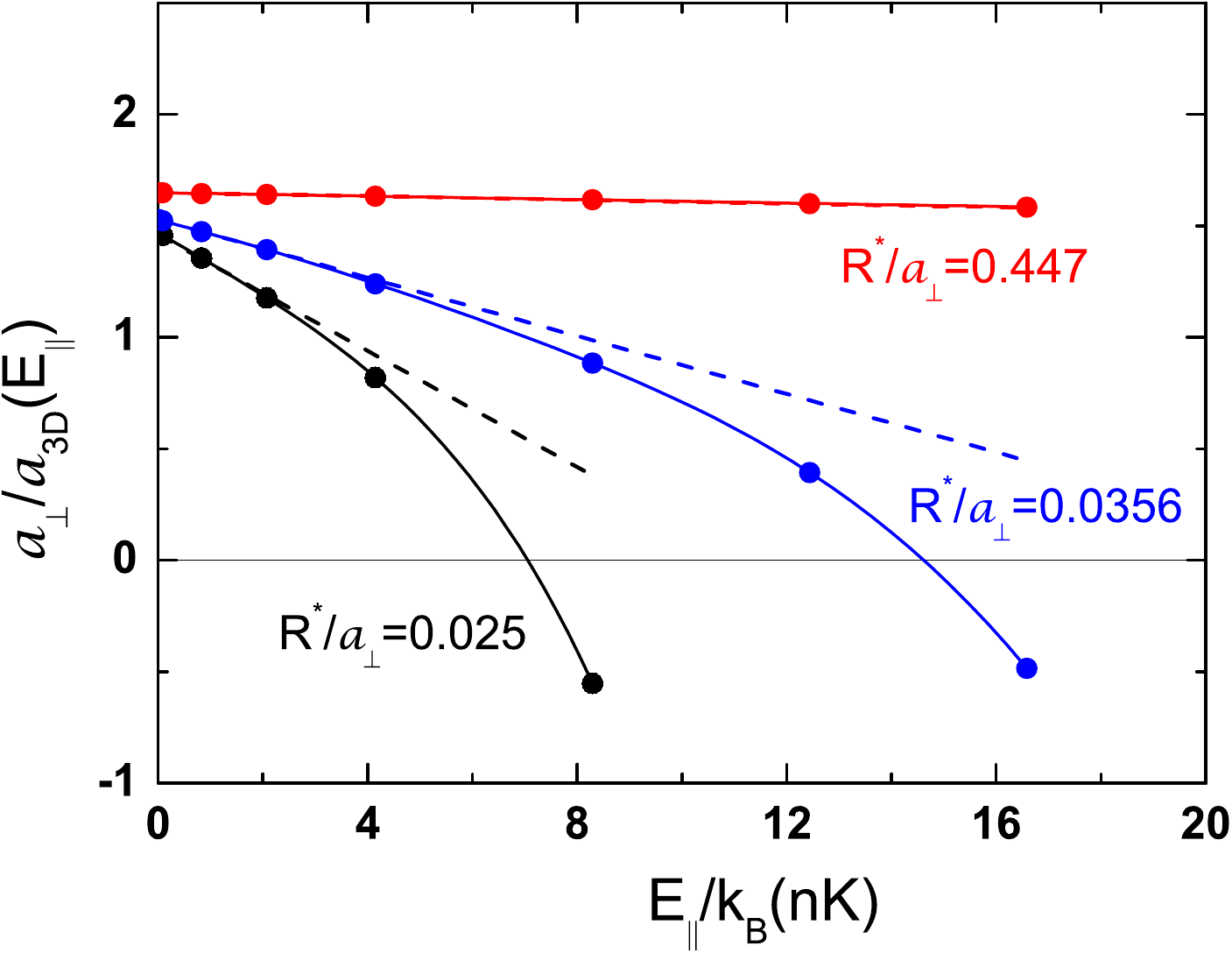}
\caption{Ratio $a_{\perp}/\as(E_{\parallel})$ calculated at the location of the ion-atom CIR position as a function of the longitudinal collision energy $E_{\parallel}$ for three different values of the ratio $R^\star/a_{\perp}$ for the ion-atom pair $^{171}$Yb$^+$+$^{6}$Li. The circles represent the calculated values of $a_{\perp}/\as(k)$ via the integration of Eq.~(\ref{2Dhamiltonian}) with the boundary condition~(\ref{asymptotics}). The solid curves correspond to $a_{\perp}/\as(k)  = 1.4603 - 0.6531 (m_{\rm a}/\mu) (E_{\parallel}/\hbar\omega_{\perp})$ at $R^\star/a_{\perp}=0.025$, where the
effective-range approximation for the energy-dependent scattering length $\as(k)$ has been applied, and Eq.~(\ref{CIRaic}) at higher
values of $R^\star/a_{\perp}$. From~\cite{MelezhikPRA16}.}
\label{fig3_MelezhikPRA16}
\end{figure}

Confinement-induced resonances have been proved an important tool to control the interaction in ultracold ensembles in reduced spatial dimensions. A similar controllability can be expected in ion-atom systems as well. \cite{MelezhikPRA16} studied this problem within the static ion approximation and solved numerically the 3D time-independent Schr\"odinger equation
\begin{equation}
\left(-\frac{\hbar^2}{2 m_\mathrm{a}}\nabla^2_R
+\frac{m_\mathrm{a}}{2} \omega_{\perp}^2\rho^2 + \frac{C_{12}}{R^{12}}-\frac{C_4}{R^4}\right)\psi({\bf
r}) = E \psi({\bf r}).
\label{2Dhamiltonian}
\end{equation}
Here the static ion is located at ${\bf r}_\mathrm{i}=0$ and the short-range repulsive term $C_{12}/R^{12}$ has been used to simplify the short-range dynamics while maintaining a reasonable number of bound states. The solution has been obtained with the following boundary condition for $z \rightarrow \pm\infty$
\begin{equation}
\psi(z,\rho) = \left[\exp(ikz) + f^{\pm}(k,a_{\perp},\as)\exp(i k\vert z
\vert)\right]\varphi_0(\rho),
\label{asymptotics}
\end{equation}
where in the transverse direction the atomic wavefunction has been assumed to be asymptotically in the ground state $\varphi_0(\rho)$ of the two-dimensional  harmonic oscillator. Here $f^{\pm}$ denote the forward and backward scattering amplitudes, and ${\bf r} \equiv (x,y,z)$, $R=\vert\mathbf{r}\vert$, ${\bf \rho} = (x,y)$. 

\begin{figure}[tb!]
\includegraphics[width=0.85\linewidth]{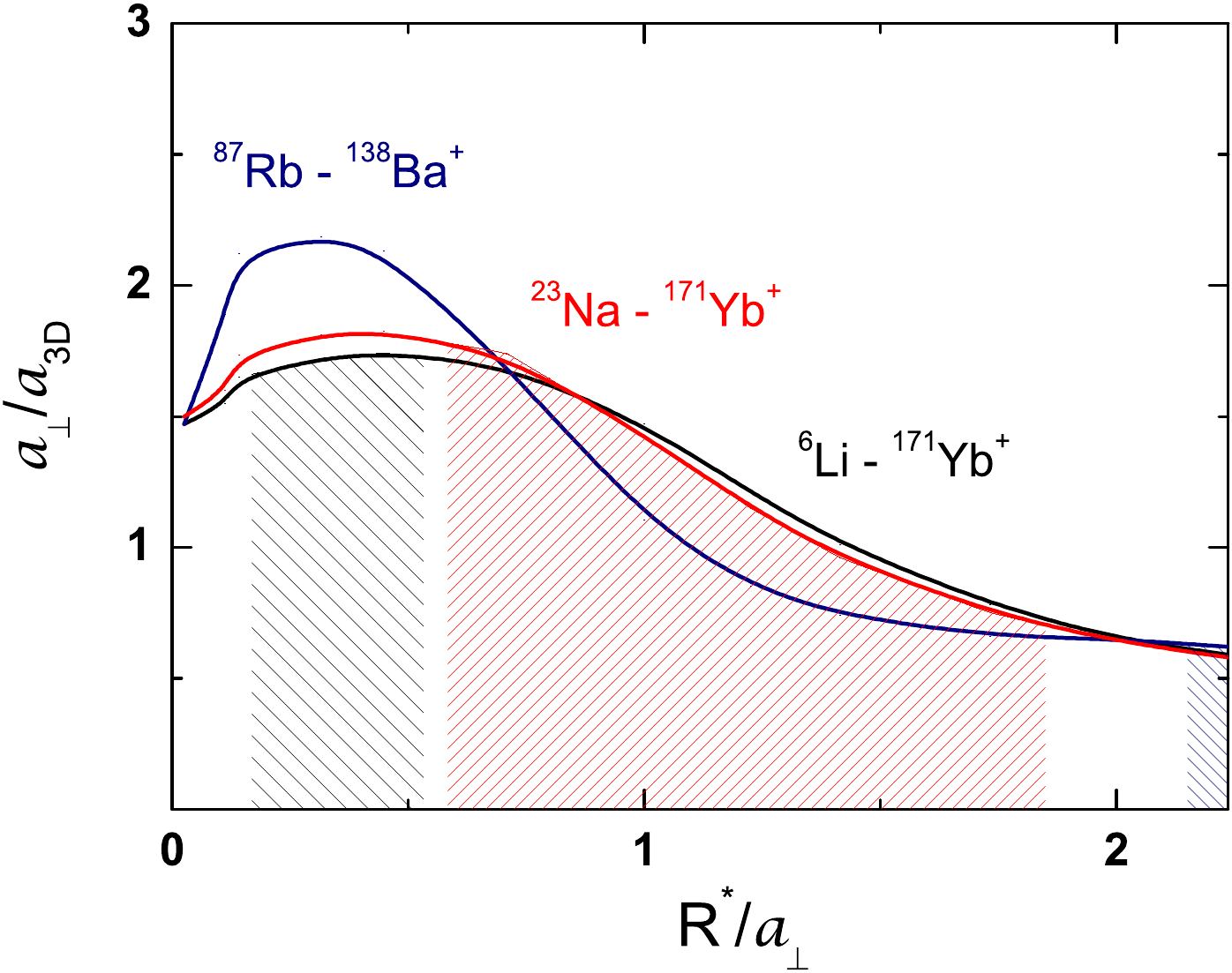}
\caption{Ratio $a_{\perp}/\as$ in which the CIR occurs as a function of $R^\star/a_{\perp}$ calculated for different
ion-atom pairs in the zero-energy limit $E_{\parallel}/E^\star=10^{-6}$. Shaded areas indicate the frequency range 
$\omega_{\perp}=2\pi\times (10-100)$kHz of the atomic transverse trap for each of the ion-atom pairs. Since $R^\star$ relies on the reduced ion-atom mass as well as the trap width $a_\perp$, the frequency ranges (i.e., the shaded areas) depend on the specific pair, too. Adapted from~\cite{MelezhikPRA16}.}
\label{fig4_MelezhikPRA16}
\end{figure}

In the framework of the outlined formalism, two opposite regimes have been investigated by~\cite{MelezhikPRA16}: the long wavelength limit, where $R^\star\ll a_{\perp}$, and the limit $R^\star\gtrsim a_{\perp}$. In the former case, it has been found that the pseudopotential approximation with energy-dependent scattering lengths describes accurately the position of the CIR and that in the zero-energy limit the atom-atom resonance condition $a_{\perp}/a_{\mathrm{3D}}= 1.4603 ...$ is retrieved. For moderate energies (a few nK), the position of the CIR is well described by the semi-analytical formula
\begin{align}
\frac{a_{\perp}}{\as(k)} =
1.4603 +\Delta\left(\frac{R^\star}{a_{\perp}}\right)-0.6531 \left(a_{\perp}k\right)^2=\nonumber \\
=1.4603 +\Delta\left(\frac{R^\star}{a_{\perp}}\right)-0.3266 \left(\frac{m_a}{\mu}\right)\left(\frac{a_{\perp}}{R^\star}\right)^2\left(\frac{E_{\parallel}}{E^*}\right).
\label{CIRaic}
\end{align}
Here $\Delta(R^\star/a_{\perp})$ denotes the shift of the CIR from the zero-energy value, and $E_{\parallel}$ denotes the longitudinal collision energy of the atom. Figure~\ref{fig3_MelezhikPRA16} displays the range of applicability of Eq.~(\ref{CIRaic}) for the Yb$^+$+Li ion-atom pair. We note that the case $R^\star/a_{\perp}=0.447$ ($\omega_{\perp}\simeq 2\pi\times 71$ kHz) falls already into the region of experimentally reachable values of atom traps~\cite{HarterCP14,CetinaPRL12}. This demonstrates that already moderate trapping frequencies allow the exploration of a broad range of values of $R^\star/a_{\perp}$, which is not the case for atom-atom collisions in a waveguide.

In the limit $R^\star\gtrsim a_{\perp}$, that is, when the effective spatial range of the ion-atom interaction is comparable or larger than the transverse width of the atom waveguide, the pseudopotential approximation cannot be applied. Even though in this limit semi-analytical expressions for the CIR position are not derivable, an interesting ``isotope-like'' effect, that is, a strong dependence of the CIR position on the ratio $R^\star/a_{\perp}$ (i.e., on the atomic mass), can be potentially observed in experiments. In Fig.~\ref{fig4_MelezhikPRA16}, such a reliance is shown in the zero-energy limit, where the CIR position for different ion-atom pairs in different regions of the ratio $R^\star/a_{\perp}$ (shaded areas) is illustrated. For instance, for $a_{\perp}\simeq R^\star$ one needs $\omega_{\perp} = 2\pi\times 2.2$ kHz for the ion-atom pair $^{138}$Ba$^+$+$^{87}$Rb, while for $^{171}$Yb$^+$+$^{23}$Na we have $\omega_{\perp} = 2\pi\times 30$ kHz. Hence, compared to atom-atom CIRs, in ion-atom systems one has more flexibility in the tunability of the ion-atom interaction. Besides, the study indicates that the CIR width increases by enhancing the ratio $R^\star/a_{\perp}$. This effect is also persistent at finite collision energies~\cite{MelezhikPRA16}.

\section{Experimental realizations of hybrid systems}
\label{sec:Exper}

In this section, we describe the experimental techniques used to create and study ion-atom systems, their advantages and shortcomings. In particular, we discuss in detail the micromotion of charged particles in time-dependent potentials and its consequences, including the limits on sympathetic cooling of ions and non-thermal energy distributions. We also discuss some of the schemes that have been proposed or are currently pursued to avoid the micromotion problem.

\subsection{Basic experimental techniques}

Ions are typically created by isotope selective photo-ionization of atoms originating from an atomic oven placed some distance away from the trap, and subsequently Doppler cooled to the center of the trap. When more than one ion is loaded into the trap, the ions crystallize to form linear strings, planar or even three-dimensional structures~\cite{Birkl1992}. By shuttering the photo-ionization beam while observing the ion fluorescence signal, any desired number of ions can be loaded.

\begin{figure}[tb!]
\includegraphics[width=0.9\linewidth]{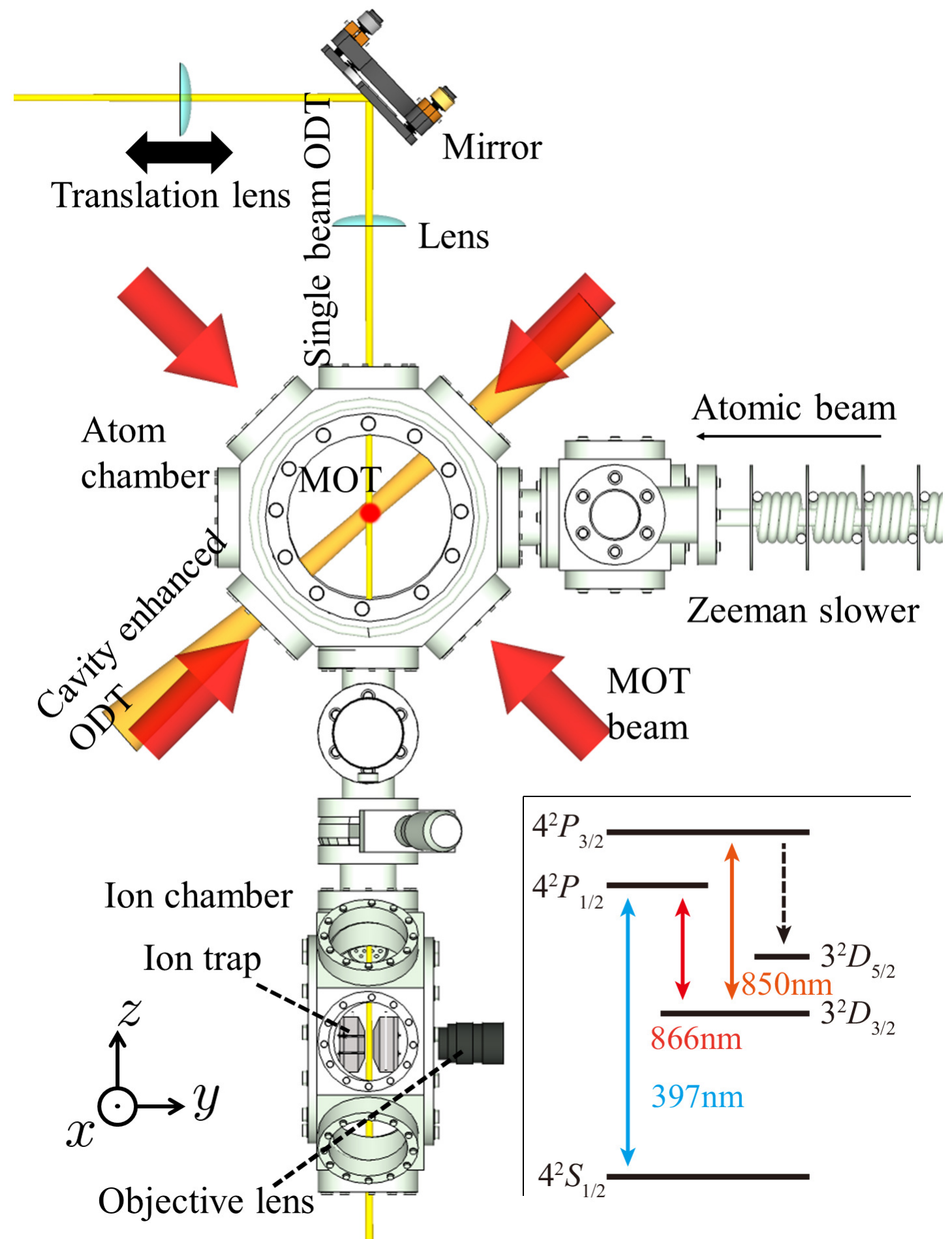}
\caption{Schematic drawing of the experimental setup used by~\cite{SaitoPRA17}. The upper part shows the atom chamber, whereas the lower part presents the ion chamber. $^6$Li atoms are supplied from the right side of the atom chamber and are decelerated by a Zeeman slower. First, $^6$Li atoms are trapped by a conventional MOT, and then, they are trapped by an optical dipole trap at the center of the atom chamber. The atoms are transported to the position of the ion trap by moving one lens placed on a translation stage. Ca$^+$ ions are trapped at the center of the ion-trap electrode placed inside the ion chamber. The 397-nm cooling laser for the ions is incident along both the radial and the axial directions of the ion trap, and the 866-nm repump laser is incident along the axial direction. Fluorescence of the ions is collected by an objective lens and detected by a photomultiplier tube and an EMCCD camera. The energy-level diagram of the $^{40}$Ca$^+$ ion is shown in the inset. From~\cite{SaitoPRA17}.
}
\label{fig:Saito_PRA_setup}
\end{figure}

The earliest experiments combining cold trapped atoms and ions employed Paul traps in combination with magneto-optical trapping of atoms~\cite{SmithJMO05,GrierPRL09,RellergertPRL11,HallPRL11,RaviAPB12}. With such experimental setups, temperatures around the Doppler limit for the atoms (several 100~$\mu$K) are within reach. To go to lower atomic temperatures, more elaborate subsequent experimental techniques are needed such as evaporative cooling and magnetic or optical trapping of atoms. Combining the Paul trap with ultracold atom technology poses particular challenges. For instance, optical trapping requires optical access for high power lasers that may damage the ion trap. Magnetic trapping, on the other hand, requires nearby electromagnets capable of supplying sufficient magnetic field gradients for trapping atoms. The radio frequency Paul trapping field may also cause atomic spin flips, leading to losses. Since the initial magneto-optical trap is loaded from an atomic beam or an otherwise increased background pressure of atoms, care needs to be taken in protecting the electrodes of the ion trap from contamination with atoms. Although the atoms used are generally conducting, oxidation can occur over long periods in the vacuum system and charge contamination or the formation of dipoles on the electrode's surface may compromise the ionic trapping field~\cite{BrownutRMP15}. For this reason the atomic cloud is sometimes prepared some distance away from the ion trap and subsequently transported to the ions either by magnetic~\cite{ZipkesNature10} or optical~\cite{SchmidRSI12,HazePRA13,MeirPRL16} fields. An example of such a setup, from the Mukaiyama group, is presented in Fig.~\ref{fig:Saito_PRA_setup}.

Ions are generally detected using fluorescence imaging either by collecting light on a photomultiplier tube or a CCD camera. Since the ions strongly repel each other via the Coulomb force, it is possible to resolve individual ions in most experiments, with typical inter-ion distances lying in the 1-10~$\mu$m range~\cite{JamesAPB1998}. The internal atomic structure of many of the used ion species also allows state detection using fluorescence imaging. Typically, one or more states correspond to a bright or fluorescing ion, while other states correspond to no detected photons (dark ions). Subsequent optical pumping to the bright states allows for distinguishing dark ions from impurity ions or reaction products that are always dark. Ion loss can be detected by structural changes in the ion crystal. Atomic ensembles can be detected using absorption imaging, which gives access to the atom number and density, as well as the atomic temperature via time-of-flight analysis after switching off the atom trap. In contrast to ion imaging, atom imaging typically leads to atom loss, such that a new atomic ensemble has to be prepared after detection.

Novel tools from quantum information science also allow for detecting quantities related to the motion of the ions~\cite{LeibfriedRMP03}. For instance, it is possible to map average numbers of motional quanta onto the internal state of the ion, giving access to the ion's temperature. In a recent work this technique was even used to probe the distribution of motional quanta corresponding to non-thermal states~\cite{MeirPRL16} for ultra cold ion-atom mixtures. At higher temperatures, ion-photon scattering rates give information about ionic energies during Doppler re-cooling after interacting with atomic ensembles~\cite{WesenbergPRA07,ZipkesNature10,MeirPRL16}. Recently, this technique has been extended to much larger temperatures, allowing e.g. measuring large energy transfers after electronic state quenching in collisions between ions and atoms \cite{SikorskyPRA17,MeirARX17b}.

Since the atomic traps are much less deep than ion traps, it can happen that collisions between atoms and energetic ions lead to atom loss~\cite{ZipkesNature10}. This effect has been used to detect the atomic density profile by looking at atom loss induced by trapped ion~\cite{SchmidPRL10}. Atom loss can also be employed to detect and minimize excess micromotion~\cite{ZipkesNature10,ZipkesPRL10,SchmidPRL10,HarterAPL13,Mohammadi2018minimizing}. Inelastic collisions leading to ion loss due to association or charge transfer can also be detected~\cite{GrierPRL09,ZipkesNature10,SchmidPRL10}. The mass of reaction products may be obtained by ion mass spectroscopy~\cite{Welling1998,Baba2001,Drewsen2004}. Here the reaction product is sympathetically cooled by co-trapped parent ions and the mass is inferred by obtaining the mass-dependent trap frequencies of ion crystals in the electric field of the Paul trap. This is usually done by parametric excitation using an oscillating electric field and observing changes in ion fluorescence. State-dependent fluorescence imaging has been used to detect spin relaxation of a single $^{171}$Yb$^+$ ion in a cloud of spin polarized Rb atoms~\cite{RatschbacherPRL13}. These tools also allow to infer the coherence time of the ionic spin in the atomic bath using Ramsey experiments~\cite{RatschbacherPRL13}.

\subsection{Electrical ion trapping}
\label{subsec:eltrap}

A common method of trapping ions is using a linear Paul trap~\cite{PaulRMP90,LeibfriedRMP03}. The Paul trap employs an oscillating electric field and offers tight confinement to ions such that they are well localized and can have very long lifetimes in the trap. A linear Paul trap is generated by the static and radio-frequency electric fields $\mathbf{E}_{\rm{PT}}(\mathbf{r},t)=\mathbf{E}_{\rm s}(\mathbf{r})+\mathbf{E}_{\rm rf}(\mathbf{r},t)$ with
\begin{equation}\label{eqS}
\begin{split}
\mathbf{E}_{\rm s}(x,y,z)&=E'_{\rm s}\left(\frac{x}{2},\frac{y}{2},-z\right)\,,\\
\mathbf{E}_{\rm rf}(x,y,z,t)&=E'_{\rm rf}\cos (\Omega_{\rm rf} t)\,\left(x,-y,0\right)\,.
\end{split}
\end{equation}
Here, $E'_{\rm s}$ and $E'_{\rm rf}$ are the electric field gradients of the static and radio-frequency fields, while $\Omega_{\rm rf}$ is the trap drive frequency. Assuming an ion of mass $m_{\rm i}$ and charge $e$ we introduce the stability parameters $a=2eE'_{\rm s}/(m_\mathrm{i}\Omega_{\rm rf}^2)$ and $q=2eE'_{\rm rf}/(m_\mathrm{i}\Omega_{\rm rf}^2)$ and obtain the following classical equations of motion
\begin{equation}
\begin{split}
\ddot{x} (\xi)- \left[a + 2q \cos(2\xi) \right]x(\xi) & = 0\,,\\
\ddot{y} (\xi)- \left[a - 2q \cos(2\xi) \right]y(\xi) & = 0\,,\\
\ddot{z} (\xi) + 2a z(\xi) & = 0\,,
\end{split}
\end{equation}
with $\xi=\Omega_{\rm rf}t/2$. The equations for the transverse motion take the canonical form of Mathieu equations, which can be solved via the Floquet theorem. This yields the following general solutions for the coordinates to the lowest order in the stability parameter $q$
\begin{equation}\label{eq_orbits}
\begin{split}
x (t) & \approx \left[ C_x \cos(\omega_x t) + S_x \sin(\omega_x t) \right] \left[ 1 + \frac{q}{2} \cos(\Omega_{\rm rf} t ) \right]\,,\\
y (t) & \approx \left[ C_y \cos(\omega_y t) + S_y \sin(\omega_y t) \right] \left[ 1 - \frac{q}{2} \cos(\Omega_{\rm rf} t ) \right]\,,\\
z (t) & = \left[ C_z \cos(\omega_z t) + S_z \sin(\omega_z t) \right] \,, 
\end{split}
\end{equation}
with the constants of integration denoted by $S_j$ and $C_j$, $j=\{x,y,z\}$. The motion of the ion in the transverse $xy$-plane is given by a slow secular motion of frequency $\omega_{x,y}\approx \frac{\Omega_{\rm rf}}{2}\sqrt{a+q^2/2}$, and a fast micromotion of frequency $\Omega_{\rm rf}$. Stable ion trapping requires $q < 0.9$ for $a\rightarrow 0$, with typical experimental values for $q$ being 0.1-0.4 with $a\ll q$. This means that the micromotion is generally a factor 10-50 faster than the secular motion, has an amplitude that is smaller than that of the secular motion. The motion of the ion in the $z$ direction is purely harmonic since there is no radio-frequency field in this direction. Additional voltages on the ion trap electrodes can be used to break the symmetry of the electric fields. In this case, stability parameters can be defined in all directions, i.e.~$a_x+a_y+a_z=0$ and $q_x+q_y+q_z=0$ as described by~\cite{LeibfriedRMP03}.

\subsection{Micromotion in ion-atom systems}
\label{sec:micromotion}

\subsubsection{Excess micromotion and its fundamental limits}

Since the collision energy of atoms and ions is in most experiments limited by the micromotion of the ions, it is important to minimize its effects in order to achieve the lowest temperatures in ion-atom systems. As can be seen from the approximate solutions of the Mathieu equations~ (\ref{eq_orbits}), each ionic orbit has associated with it intrinsic micromotion, that cannot be reduced by any experimental means for given stability parameters $q$ and $a$. However, experimental imperfections arrise in real Paul trap that can cause {\it excess} micromotion. Here, we review some of the techniques developed for detecting and minimizing this excess micromotion~\cite{BerkelandJAP98}.

{\it The quantum limit} -- Let us first examine the minimum micromotion energy that we could hope to reach under perfect experimental conditions, i.e. we only consider intrinsic micromotion. The description of the ion trapped in the time-dependent electric fields of the Paul trap in the previous subsection was classical. In this case, an ion cooled to zero kinetic energy would sit exactly in the center of trap and all micromotion would vanish. A quantum mechanical analysis~\cite{LeibfriedRMP03} shows that the dispersion of the ionic wavepacket leads to a minimum micromotion amplitude of $x_{\rm mm}^{\rm min}=l_{\rm ho}q/2$, corresponding to a "breathing" motion of the ionic ground state with size $l_{\rm ho}=\sqrt{\hbar/(2m\omega_x)}$, and similarly for the $y$ direction. The energy associated with this motion amounts to an average kinetic energy of $\mathcal{E}_{\rm mm}^{\rm min}=m\Omega_{\rm rf}^2\left(x_{\rm mm}^{\rm min}\right)^2/4$. If we assume that the static stability parameter $a \rightarrow 0$, we have that $\omega_x\approx \Omega_{\rm rf} q/\sqrt{8}$, such that $\mathcal{E}_{\rm mm}^{\rm min}\approx\hbar\omega_x/4$, that is, the minimum micromotion kinetic energy is of the same order as the zero-point energy of the trap. Obviously, the present discusion is limited to small systems of trapped ions, in which the ions are trapped in a position where the oscillating electric fields vanish. The Coulomb repulsion between the ions will prevent large ion densities from being achieved and outer ions will have micromotion energies that are much larger than this quantum limit. This is particularly true for ion crystals that are two- or three-dimensional.

{\it In-phase excess micromotion} -- An important modification to the Paul trap occurs when the static and radio-frequency fields do not disappear at the same point~\cite{BerkelandJAP98}. An additional uncontrolled static field, for instance in the $x$ direction, $\mathbf{E}_{\rm emm}=(E_{\rm emm},0,0)$, displaces the ion's equilibrium position by $x_{\rm emm}=eE_{\rm emm}/(m_{\rm i}\omega^2_x)$. In this situation, the ion undergoes excess micromotion of amplitude $r_{\rm emm} =qx_{\rm emm}/2$ corresponding to an average kinetic energy of $\mathcal{E}_{\rm emm}=m_{\rm i}q^2\Omega_{\rm rf}^2x_{\rm emm}^2/16$. In a collision with an atom, this kinetic energy can be transferred to the ion's secular motion, causing ion heating as described below. Additional static electric fields can be applied to compensate the micromotion in this situation, as described in the next subsection.

{\it Axial micromotion} -- It can also happen that $q_z\neq 0$ due to asymmetries in the construction of the linear Paul trap. In this case, it may happen that the ion cannot be trapped in the center of the radio-frequency field in the $z$ direction, leading to axial micromotion of amplitude $q_z z_0/2$, with $z_0$ the distance from the center of the radio frequency field. This micromotion also poses a challenge when working with linear chains of ions. For instance, two ions sitting in the center of a linear Paul trap will align along the $z$ axis at a distance $l_z=\pm (e^2/2\pi\epsilon_0m_\mathrm{i}\omega_z^2)^{1/3}/2$ from the trap center, owing to their Coulomb repulsion. This causes micromotion of amplitude $l_zq_z/2$.

Asymmetries in the Paul trap geometry or electronics caused by e.g. inaccurate construction or by poor filtering of the electrode feeds can lead to an oscillating homogeneous field in the axial direction. In this case, there is no point along the $z$ axis where the oscillating field vanishes. This type of micromotion may be compensated by feeding a  radio-frequency voltage with appropriate phase and amplitude to one of the Paul trap endcaps.

{\it Out-of-phase excess micromotion} -- Finally, excess micromotion can occur because the radio frequency null is time-dependent. This originates from differences in impedance of the electrodes and their electric connections to the rf source, or from length differences in the wiring between the radio-frequency electrodes. This type of micromtion, which is sometimes called quadrature micromotion, leads to an additional oscillating electric field at the position of the ion of magnitude~\cite{BerkelandJAP98}
\begin{equation}
E_{\rm qmm}=E_{\rm electrode}(0)\sin\varphi_{\rm ac}\sin(\Omega_{\rm rf} t)\,.
\end{equation}
Here, $E_{\rm electrode}(0)$ is the electric field amplitude at the ion's position due to the oscillating voltage on a single electrode and $\varphi_{\rm ac}$ is the phase difference between the two electrodes. This leads to a micromotion amplitude of $x_{\rm qmm}=eE_{\rm qmm}\sin\varphi_{\rm ac}/(m\Omega_{\rm rf}^2)$. Since the voltages applied in most Paul traps are very high, this type of micromotion can lead to large kinetic energies in the ion, even for small phase differences $\varphi_\mathrm{ac}$.

\subsubsection{Micromotion detection and compensation}\label{sec_compmm}

Since attainable temperatures in hybrid ion-atom experiments are usually limited by excess micromotion, it is of key importance to be able to compensate it. The extent to which the micromotion can be compensated depends on the accuracy with which compensating electric fields can be applied to the ion, on drifts due to thermal effects, and on the accuracy with which the micromotion can be detected in an experiment.

{\it Micromotion detection by monitoring ion positions} -- Radial excess micromotion induced by stray DC electric fields shift the ions out of the node of the rf-quadrupole field. These stray fields can be measured by monitoring the ion positions with a camera while varying the radial confinement by tuning the radio frequency amplitude~\cite{GlogerPRA15}. The ion's average position in the presence of a field $E_{\rm emm}$ is given by $x_{eq}=eE_{\rm emm}/(m\omega_x^2)$. In experiments where high-numerical aperture imaging is possible, average ion positions can be distinguished at the $\sim$~200~nm scale. Assuming $^{171}$Yb$^+$ with the lowest trap frequency of $\omega_x=2\pi\times 25\,$kHz, it would allow for detecting and compensating stray electric fields down to about $E_{\rm emm}\leq$~0.01~V/m, corresponding to a micromotion energy of $\sim$~0.3~mK, if we assume $\Omega_{\rm rf}=2\pi\times\,2$MHz and $q=0.2$.

We can improve on this method by using detection schemes that probe the ion position but do not rely on the resolution of the imaging system. This can be done, e.g.~by monitoring the fluorescence of the ion in an inhomogeneous laser field~\cite{BerkelandJAP98}, or by inducing a spatially
dependent AC Stark shift by a tightly focused laser beam to observe its influence on the fluorescence~\cite{HuberNatCom14}. We can also introduce an inhomogeneous magnetic field and probe the transition frequency of two Zeeman or hyperfine states with differential magnetic field susceptibilities as a function of the radial confinement. For instance, consider two Zeeman states with a transition frequency of $\hbar\omega_{\rm trans}=\hbar\omega_0+\mu_\mathrm{B} B'x$, with $\omega_0$ the bare transition frequency, $\mu_\mathrm{B}$ the Bohr magneton, $B'$ the magnetic field gradient, and $x$ the position of the ion. The transition can be measured with an accuracy of $\sim 2\pi\times100\,$Hz, using e.g. Ramsey spectroscopy. Assuming $B'=0.1$~T/m, which is not too demanding experimentally, this would allow to determine $x$ down to about $70$~nm and to compensate the excess micromotion energy to about 40~$\mu$K by applying external electric fields, as described below.

{\it Micromotion detection by monitoring spectral lines} -- A much employed technique to detect micromotion is by observing the spectral composition of atomic transitions as a result of the modulation of the excitation laser in the rest frame of the ion. In the following, we follow the derivation of~\cite{BerkelandJAP98}. Consider the electric field $\mathbf{E}_{\rm L}(t)$ of a laser interacting with an ion undergoing excess micromotion. In the ion's rest frame, the electric field can be written as
\begin{equation}\label{eq_EL_restframe}
\mathbf{E}_{\rm L}(t)={\rm Re}\left(E_0 e^{i(\mathbf{k}\cdot(\mathbf{r}_0+\mathbf{r}_{\rm emm})-i\omega_Lt)}\right),
\end{equation}
with $\mathbf{k}$ the wavevector of the laser, $\mathbf{r}_0$ representing the secular motion and $\mathbf{r}_{\rm emm}$ the excess micromotion, $E_0$ the electric field amplitude of the laser and $\omega_\mathrm{L}$ its frequency. We can rewrite Eq.~\eqref{eq_EL_restframe} in terms of the Bessel functions $J_n(\beta)$~\cite{BerkelandJAP98}
\begin{equation}\label{eq_Bessel}
\mathbf{E}_{\rm L}(t)={\rm Re}\left(E_0e^{i\mathbf{k}\cdot \mathbf{r}_0}\sum_n J_n(\beta)e^{in(\Omega_{\rm rf} t+\delta+\pi/2)-i\omega_\mathrm{L}t}\right),
\end{equation}
with $\mathbf{r}_\mathrm{emm}=\mathbf{r}^{\parallel}\cos(\Omega t)+\mathbf{r}^{\perp}\sin(\Omega t)$ and modulation index $\beta^2=\left(\sum_n k_nr_n^{\parallel}\right)^2+\left(\sum_n k_nr_n^{\perp}\right)^2$. The phase difference $\delta$ is given by $\tan(\delta+\pi/2)=(\sum_n k_nr_n^{\perp})/(\sum_n k_nr_n^{\parallel})$.

From Eq.~\eqref{eq_Bessel}, we get the resonance condition $\omega_\mathrm{L}=\omega_0+n\Omega_{\rm rf}$, with $\omega_0$ the bare transition frequency. Therefore, the absorption line acquires a set of sidebands at frequencies $n\Omega$, with integer $n$. From the relative strength of these sidebands we can infer the modulation index $\beta$ and thereby the amplitude of the excess micromotion.

In the situation where the transition linewidth $\Gamma_0$ is much smaller than the trap drive frequency, $\Gamma_0\ll \Omega$, the sidebands can be resolved and their relative strengths can be probed immediately. The comparison in Rabi frequency between the zero-order sideband (carrier) and the first order sideband carrier gives access to the modulation index: $\Omega_1/\Omega_2=J_0(\beta)/J_1(\beta)$. Care needs to be taken to eliminate systematic errors in this approach. \cite{MeirARX17}, show that a source of such errors can be oscillating magnetic fields at the position of the ion which are caused by induced rf currents that modulate the transition frequency. In this case, the apparent compensation electric field depends on the magnetic susceptibility of the optical transition used. Thermal motion of the ion may also cause systematic errors as second-order coupling between inherent micromotion and thermal harmonic motion sidebands can overlap with the excess micromotion sidebands~\cite{MeirARX17}.

In the limit where $\Gamma_0\gg \Omega$, the sidebands cannot be resolved. In this case, the excess micromotion becomes apparent from the line broadening of the transition. The spectral decomposition, Eq.~\eqref{eq_Bessel}, can be convoluted with a spectral distribution representing broadening due to e.g.~natural linewidth, magnetic fields and atomic substructure, and it can be fit to the observed spectrum. Obviously, for vanishingly small excess micromotion, this method becomes unreliable. However, one can artificially increase the excess micromotion by applying a large offset field. Inverting the offset field allows accurate determination of the zero-crossing of the modulation index. Alternatively - and more accurately - the fluorescence of the ion interacting with a low-intensity laser beam, which is modulated by the Doppler shift, can be correlated with the trap drive frequency $\Omega_{\rm rf}$. This method is known as the photon-correlation method and has the advantage that it is sensitive to both excess and quadrature micromotion as described in detail by~\cite{BerkelandJAP98}.

Since the method described above only works for micromotion that is parallel to the wavevector of the laser, each direction of micromotion can be independently minimized by using three laser beams with projections onto three independent directions. In case the compensating fields cannot be applied in the same directions, iterations may be necessary to accurately minimize micromotion in all directions.

{\it Micromotion detection by motional excitation} -- Excess micromotion can also be detected by motional excitation of the secular ion motion~\cite{Narayanan2011,Ibaraki2011}. If the ion is displaced by a stray electric field $\mathbf{E}_{\rm emm}$, it will experience an oscillating field from the radio-frequency electrodes. Introducing an oscillating field on these electrodes, that is close to the secular frequency of the ion motion, will result in motional excitation. Interplay between this excitation and the cooling and detection laser will result in a dip in ion fluorescence. If the ion is in the center of the trap, no excitation can occur, since the oscillating field vanishes such that the ion is compensated in that direction.

{\it Micromotion detection by monitoring atom loss} -- In an ion-atom mixture, the atomic loss rate depends on the energy of the trapped ion. This is because more energetic ion-atom collisions result in a higher probability for atom to subsequently leave the shallow atom trap.  \cite{ZipkesNature10,ZipkesPRL10,SchmidPRL10,HarterAPL13} used the observed atom loss to probe ionic micromotion energy. Offset fields were used to compensate this micromotion as described below. This type of micromotion detection has the benefit that it also works for ions that do not allow for fluorescence imaging, because they lack closed transitions or only have transitions at inaccessible wavelengths.

{\it Compensating micromotion by external electric fields} -- To compensate excess micromotion, usually Paul traps are equipped with a number of electrodes for applying compensating electric fields. The use of ultra-stable voltage sources and proper filtering prevents problems with drifting micromotion compensation and ion heating. Since charge distributions sticking to electrodes after oxidation can cause time-varying stray fields, micromotion compensation has to be repeated regularly. The precision with which excess micromotion can be compensated varies between experimental realisations~\cite{HarterAPL13}. Typically residual static electric fields are in the range of about 1~V/m, with~\cite{HarterAPL13} quoting the lowest value of 0.02~V/m, corresponding to a residual excess micromotion temperature of 0.5~$\mu$K for Rb$^+$. In these experiments, excess micromotion sets the temperature scale in ion-atom interactions, opening the way towards studying ion-atom mixtures in the quantum regime. Compensating out-of-phase micromotion requires the application of time-dependent radio-frequency offset fields with appropriate phase.

{\it Preventing excess micromotion} -- Although it should not be recommended to rely on prevention alone, some care can indeed be taken in reducing the amount of excess micromotion. In particular, axial and out-of-phase micromotion may be mitigated by proper mechanical and electronic construction. For instance, care needs to be taken that the electrodes of the Paul trap are exactly parallel and the connectors to the radio-frequency electrodes are made of equal length. Proper filtering and impedance tuning of the electrodes is also useful. Stray fields due to charges sticking on nearby non-conductive surfaces can be prevented by avoiding the use of insulating materials close to the ions. Unfortunately, many ion species require ultraviolet lasers for cooling and detecting, which can easily extract electrons from trapping electrodes. They are then carried around the Paul trap due to the large electric fields present. Oxidized material on electrode surfaces, for instance coming from atomic ovens used to load ions or atoms, can also pose a problem as the charges accumulate there. Heating up the ion trap permanently or periodically can reduce their effects, as can smart design of laser beam paths and ovens to prevent charge extraction and contamination of surfaces. Useful information on the prevention and drifts of stray electric fields can be found e.g. in Refs.~\cite{HarlanderNJP10,HarterAPL14}

\subsubsection{Effects of micromotion on hybrid ion-atom systems}

At first sight it is somewhat surprising that a system of ions trapped in a Paul trap does not thermalize with a buffer gas. This was first addressed by~\cite{MajorPR68} as early as 1968. A useful way of shedding light on the problem is by considering hard sphere collisions in one dimension between an ion and atoms that are stationary (i.e.~assuming classically zero atomic temperature). Assuming that there is no excess micromotion, the ion follows an orbit $x(t)$ given by Eq.~\eqref{eq_orbits}. At $t=t_{\rm col}$, a hard-sphere collision occurs with a stationary atom such that the new position and velocity are given by $x'(t_{\rm col})=x(t_{\rm col})$ and $v'(t_{\rm col})=Av(t_{\rm col})$, with $A=(m_{\rm i}-m_{\rm a})/(m_{\rm i}+m_{\rm a})$ according to energy and momentum conservation. Since we can write the velocity of the ion as a sum of a secular and a micromotion part, $v(t)=v_{\rm sec}(t)+v_{\rm mm}(t)$, it is obvious that $v'(t)$ could have a higher or lower secular velocity than $v(t)$ depending on the value of $v_{\rm mm}(t_{\rm col})$, i.e.~heating or cooling can occur. Whether heating or cooling is more likely on average, depends solely on the mass ratio via $A$, in this simple model. One useful way to picture the situation is that the collision with an atom disrupts the micromotion of the ion, which transfers energy from the micromotion to the secular motion.

\begin{figure}[tb!]
\includegraphics*[width=0.9\linewidth]{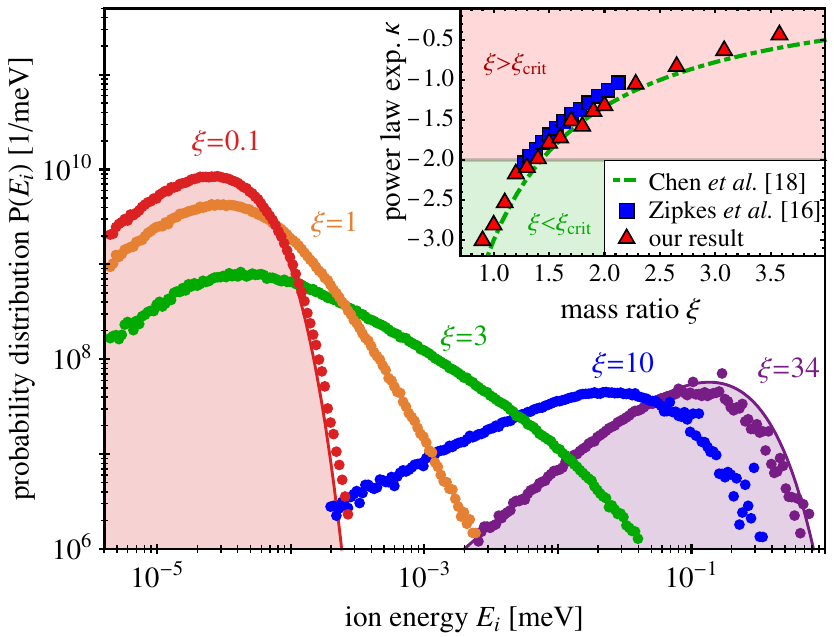}
\caption{Normalized equilibrium energy distributions for different mass ratios $\xi=m_\mathrm{a}/m_\mathrm{i}$ in a Paul trap. The buffer gas cloud distribution is given by a Gaussian of size $\sigma_\mathrm{a}=R_0/100$ and temperature of $T_\mathrm{a}=200 \mu$K. Also shown is the energy distribution in the Boltzmann regime (red curve) and the energy distribution for $\xi = 34$ (purple curve), according to the expressions in Table~\ref{table:regimes}. The inset compares the exponents $\kappa$ of the power-law in the energy distribution (as defined in Table~\ref{table:regimes}), for different models. The condition $\kappa = -2$ separates the regimes of stable from unstable ion motion. From~\cite{HoltkemeierPRL16}.
}
\label{fig:fig2_HoltkemeierPRL16}
\end{figure}

\begin{figure*}[tb!]
\includegraphics[width=0.7\linewidth]{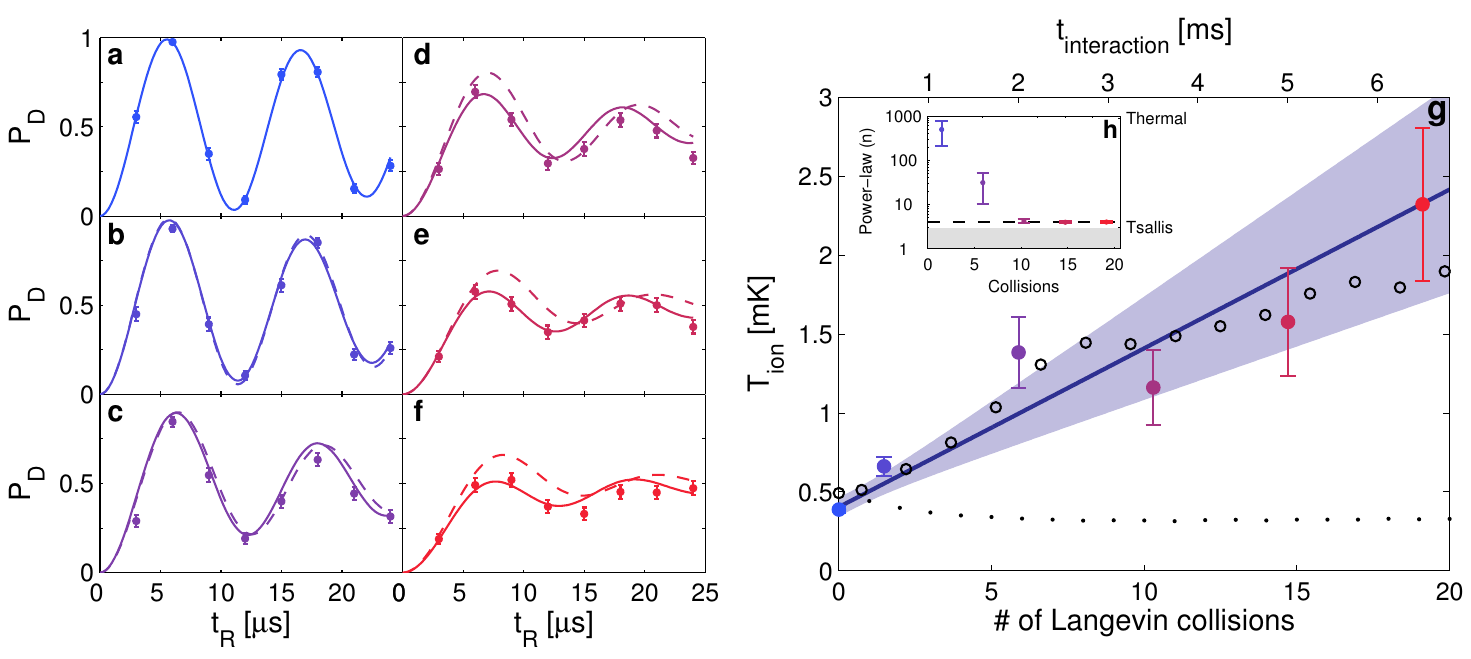}
\caption{Carrier Rabi spectroscopy between the $S_{1/2}$ ground state and the metastable $D_{5/2}$ state in a Sr$^+$ ion. (a)-(f) Each graph corresponds to a different interaction time between the ion and ultracold Rb atoms (0, 0.5, 2, 3.5, 5, 6.5 ms, respectively). The solid lines indicate a fit using a Tsallis distribution, while the dashed lines show the fit of the data to a thermal distribution. (g) The ion’s temperature (filled colored circles) as a function of the interaction time (top x-axis) and the average number of Langevin collisions (bottom x-axis). Open circles come from a simulation that takes into account the polarization potential. Black dots are simulation results taking into account only hard-sphere collisions. (h) Ion’s power-law parameter $n$, which measures to what extent thermal a distribution can be used. The ion’s energy distribution starts with $n\gg 1$, consistent with a Maxwell-Boltzman distribution, and converges to $n=4.0(2)$ after $\sim$~10 collisions. For $n>10$, thermal and Tsallis distributions are almost indistinguishable. From~\cite{MeirPRL16}.
}
\label{fig2_MeirPRL16}
\end{figure*}

\begin{table}[b]
\begin{center}
\caption{Analytical expressions for the ion's kinetic energy distribution in three different regimes for a Paul trap, as obtained from fitting the numerical results shown in Fig.~\ref{fig:fig2_HoltkemeierPRL16}. From~\cite{HoltkemeierPRL16}.}
\label{table:regimes}
\begin{tabular}{lll}
\hline \hline
Boltzmann regime ($\xi\ll\xi_{\mathrm{crit}}$) & $P(E_{\mathrm{i}})\propto E_{\mathrm{i}}^{3/2} \, \exp{(-\frac{E_{\mathrm{i}}}{k_\mathrm{B} T_\mathrm{a}})}$ \\
Power-Law regime ($\xi\sim\xi_{\mathrm{crit}}$) & $P(E_{\mathrm{i}})\propto \left\{
  \begin{array}{l l}
    E_{\mathrm{i}}^{3/2}, & \  E_{\mathrm{i}}\ll k_\mathrm{B} T_\mathrm{a} \\
    E_{\mathrm{i}}^{\kappa}, & \ E_{\mathrm{i}}\gg k_\mathrm{B} T_\mathrm{a}
  \end{array} \right. $ \\
	\vspace{0.1cm}
Localization regime ($\xi\gg\xi_{\mathrm{crit}}$) & $P(E_{\mathrm{i}})\propto E_{\mathrm{i}}^{\kappa} \, \exp{(-\frac{E_{\mathrm{i}}}{E_\mathrm{a}^{\star}})}$  \\
\hline \hline
\end{tabular}
\end{center}
\end{table}

Although the above oversimplified model gives some intuition for the underlying physics, a complete description of the problem of a buffer gas interacting with ions trapped in a Paul trap has proven to be remarkably complex. \cite{DeVoePRL09} considered the case in which trapped ions were interacting with a buffer gas of finite temperature via hard-sphere collisions and calculated numerically that the ions should develop non-Gaussian energy distributions with power-law tails after undergoing hard-sphere collisions with the atoms. These distributions, which are shown in Fig.~\ref{fig:fig2_HoltkemeierPRL16}, fit well to a Tsallis distribution~\cite{Tsallis1988}, which is a generalization to the Maxwell-Boltzmann distribution. This long-tailed energy distribution reaches temperatures that are much higher than that of the buffer gas. It was found by \cite{DeVoePRL09} that the mass ratio parametrizes the deviation from the normal Maxwell-Boltzmann distribution, with the smaller ion-atom mass ratios corresponding more closely to normal thermal distributions. Similar results were obtained by~\cite{ZipkesNJP11}, who studied the kinematics of trapped ions for a buffer gas at $T=0$ in relation to ion-atom mass ratio, trap geometry, differential cross section, and non-uniform neutral atom density distribution, and identified excess micromotion as the main limit in attainable temperatures.

\cite{ChenPRL14} developed an analytical model that can predict the steady-state and dynamics of the kinetic energy of a single ion in the buffer gas as well as the transition from sympathetic cooling to heating, and its dependence on trap parameters and masses of the particles. These results indicated that the observation of non-Maxwellian statistics could indeed be attributed to random heating collisions. \cite{RousePRL17,RousePRA18} derived the analytical form of the ionic energy distribution showing that it indeed follows a Tsallis distribution. \cite{HoltkemeierPRL16,HoltkemeierPRA16} have extended the considerations to the case of higher order radio-frequency traps and inhomogeneous buffer gases. We discuss their results below.

In a remarkable experiment, \cite{MeirPRL16,MeirARX17} directly probed the energy distribution of a trapped Sr$^+$ ion immersed in a cloud of Rb atoms at a temperature of 5~$\mu$K using methods developed for ion trap quantum information processing~\cite{LeibfriedRMP03}. The ion was prepared in the ground state of secular motion and all excess micromotion energy was compensated to values below 0.5~mK. The ion was prepared in a pure electronic ground state after interacting with the atoms for some time. A narrow linewidth laser coupled this ground state to the meta-stable $D_{5/2}$ state, which resulted in Rabi oscillations. The Rabi frequency $\Omega_{n_x,n_y,n_z}$ of these oscillations depends on the amount of motional quanta $n_j$ present in the secular motion of the ion in each direction $j=x,y,z$~\cite{MeirPRL16}
\begin{equation}
\Omega_{n_x,n_y,n_z}=\Omega_0 \prod e^{-\frac{\eta_i}{2}}L_{n_i}\left(\eta_i^2\right),
\end{equation}
with the Lamb-Dicke parameters $\eta_j=k_jl_j^{\rm ho}$, and $k_j$ the wavevectors of the light and $l_j^{\rm ho}=\sqrt{\hbar/(2m_{\rm i}\omega_j)}$ the size of the ionic groundstate wavepacket in direction $j$. The function $L_n(x)$ is the Laguerre polynomial of degree $n$, and $\Omega_0$ is the bare Rabi frequency. Thermal occupation of excited states of the secular harmonic oscillator states results in mixing of various frequency components in the Rabi oscillation leading to damping. State-selective fluorescence detection allows probing the Rabi oscillations. A maximum-likelihood fit of the Rabi oscillations averaged over many experimental runs allows an identification of the relative population of each of the harmonic oscillator states. In this way, the authors were able to show that indeed a Maxwell-Boltzmann distribution does not describe the data as well as a Tsallis distribution, as shown in Fig.~\ref{fig2_MeirPRL16}.

\begin{figure}[tb!]
\begin{centering}
\begin{overpic}[trim=0 0 0 -40, scale=0.60, unit=1mm]{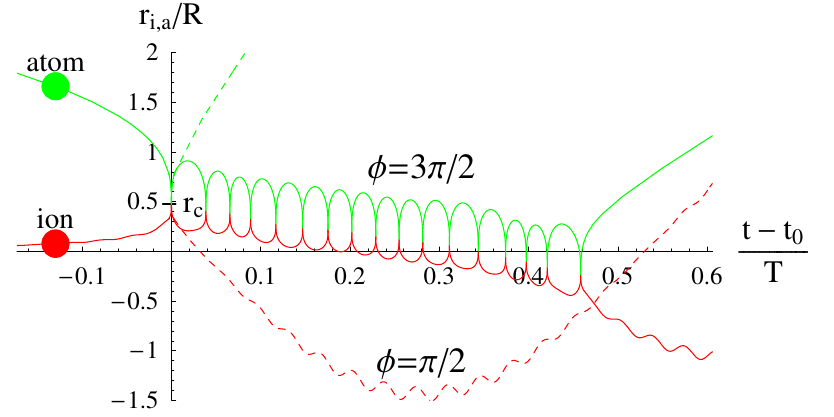}
\put(35,29){\includegraphics[scale=0.5]{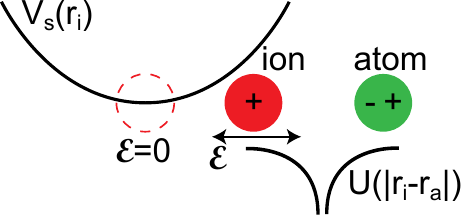}}
\end{overpic}
\par\end{centering}

\caption{Trajectories of an ion $r_\mathrm{i}\left(t\right)$ and
an atom $r_\mathrm{a}\left(t\right)$ during a classical one-dimensional
low-energy collision. The atom of mass $m_\mathrm{a}$ approaches the ion
of mass $m_\mathrm{i}=2m_\mathrm{a}$ held in the center of a rf trap with secular
frequency $\omega=2\pi/T$ and Mathieu parameter $q=0.1$, leading
to a hard-sphere collision at $r_\mathrm{i}=r_\mathrm{a}=r_\mathrm{c}$, $t=0$ and rf phase
$\phi$. For $\phi=\pi/2$ (dotted lines), the trap field adds energy
to the system, causing heating. For $\phi=3\pi/2$ (solid lines),
the rf field removes energy, binding the atom to the ion and causing
further collisions at various rf phases until enough energy is accumulated
to eject the atom. From~\cite{CetinaPRL12}.}
\label{fig:trajs}
\end{figure}

The experiment of~\cite{MeirPRL16} also sheds light on another important question: what sets the energy scale in the situation where there is no excess micromotion while the buffer gas is close to $T=0$? \cite{CetinaPRL12} considered the case of an ion that is cooled to the center of the Paul trap and does not undergo micromotion classically. When an atom with negligible initial temperature collides with this ion, we would not expect much to happen when we consider classical hard-sphere collisions. However, \cite{CetinaPRL12} discovered that the picture changes significantly when it is taken into account that the ion-atom interaction is indeed of a long-range $1/R^4$ nature. As the atom approaches the stationary ion, the ion starts to move at a velocity that is no longer adiabatic with respect to the Paul trap's electric fields and is pulled away from the trap center. In this situation, energy is transferred from the time-dependent trapping fields into the ion-atom system and rapid heating occurs. As before, it was found that the mass ratio $m_\mathrm{a}\ll m_\mathrm{i}$ can be used to mitigate this heating effect. \cite{MeirPRL16} recreated in an experiment the situation considered theoretically by~\cite{CetinaPRL12}. As the ion was cooled to the ground state of motion in all directions and micromotion was compensated close to the quantum limit, the initial heating of the ion was caused by the long-range character of the ion-atom interaction. The work by~\cite{CetinaPRL12} thus provides an important limitation to achievable temperatures in sympathetic cooling of ions by atoms. In Fig.~\ref{fig:trajs}, some of the ion-atom trajectories calculated by~\cite{CetinaPRL12} are shown.

In the presence of multiple trapped ions, one has to include the inevitable micromotion-induced heating, as it is no longer possible to keep all ions in the center of the trap. Indeed, collisions amongst ions themselves in dense ion systems can lead to heating in the presence of micromotion. \cite{ChenPRL13} have experimentally analysed these effects by studying $^{174}$Yb$^+$ ions that were heated due to micromotion interruption during ion-ion collisions. The observed time evolution of the ion temperature was compared to a theoretical model for ion-ion heating which allowed the authors to directly measure the Coulomb logarithm. This provides a simple analytical description of ion cloud density, temperature, and structural phase. These results have been extended to the case where large ion crystals were interacting with cold atoms by \cite{SchowalterNatCom16} in an experiment supported by molecular dynamics simulations. Interestingly, a bifurcation of steady-state ion energies has been found, showing that buffer-gas cooling has strong limitations for larger ion crystals.
\cite{Meir2018} used a single-shot Doppler-cooling thermometry with single-event and energy-mode resolution to study the sympathetic cooling dynamics of an energetic ion immersed in an ultracold bath of neutral atoms. They demonstrated the capabilities of the new method to detect a single collision and the direction of ion motion following this collision. They used this capability to observe a deviation of the scattering angle distribution from the Langevin model predictions manifested by a forward-scattering peak. They also directly observed the nonequilibrium dynamics of atom-ion collisions in a Paul trap with single-collision resolution.

\subsubsection{Quantum description of micromotion in ion-atom systems}

Up until now, we have only discussed the classical description of micromotion in ion-atom systems. \cite{LeibfriedRMP03} have provided a quantum description of single ions in Paul traps and~\cite{NguyenPRA12} were the first to extend this description to the situation where an atom was also present based on Floquet theory. Below, we give a brief description of this procedure.

The Hamiltonian of an ion of mass $m_\mathrm{i}$ trapped in a Paul trap with a trap drive frequency $\Omega$ and stability parameters $q$ and $a$, is given by
\begin{equation}\label{H_ion}
H_\mathrm{ion}(t)=\frac{p_{\rm i}^2}{2m_{\rm i}}+\frac{1}{8}m_{\rm i}\Omega^2z_{\rm i}^2\left[a+2q\cos(\Omega t)\right]\,,
\end{equation}
where we only considered one direction $z_{\rm i}$ of motion with momentum $p_{\rm i}$ for simplicity. Following~\cite{CookPRA85,NguyenPRA12} we write the ion wavefunction as
\begin{equation}
\Psi(z_\mathrm{i},t)={\rm exp}\left(-\frac{i}{4\hbar}m_{\rm i} q \Omega z_{\rm i}^2\sin (\Omega t)\right)w (z_{\rm i},t)\,.
\end{equation}
Using this Ansatz in the Schr\"odinger equation, the following effective Hamiltonian for the wavefunction $w (z_i,t)$ is obtained
\begin{equation}
H_{\rm eff}(t)=H_{\rm sec}+H_\mathrm{mm}(t)=\frac{p_{\rm i}^2}{2m_{\rm i}}+\frac{1}{2}m_\mathrm{i}\omega_{\rm i}^2z_{\rm i}^2+H_\mathrm{mm}(t)\,,
\end{equation}
with the micromotion term given by
\begin{equation}
H_\mathrm{mm}(t)=-m_{\rm i}g^2\omega_{\rm i}^2z_{\rm i}^2\cos(2\Omega t)-g\omega_\mathrm{i} \{z_{\rm i},p_{\rm i}\}\sin(\Omega t)\,.
\end{equation}
Here, $g=[2(1+2a/q^2)]^{-1/2}$. Neglecting this micromotion term forms the basis of the so-called secular approximation, in which we are left with the Hamiltonian of a time-independent harmonic oscillator with trap frequency $\omega_{\rm i}=\frac{\Omega}{2}\sqrt{a+\frac{q^2}{2}}$.

In case there is also an atom present described by the Hamiltonian $H_\mathrm{a}$, we can write the combined Hamiltonian as $H_\mathrm{tot}=H_\mathrm{static}+H_\mathrm{mm}(t)=H_\mathrm{a}+H_{\rm sec}+H_\mathrm{ai}+H_\mathrm{mm}(t)$, with $H_\mathrm{ai}$ the interaction term between the atoms and ions. The static part of this Hamiltonian can be solved by transforming to relative $r$ and center-of-mass $R$ ion-atom coordinates. The center-of-mass wave functions can be expanded onto a suitably chosen basis of orthogonal states, e.g.~Fock states $|\Phi_n(R)\rangle$, whereas the relative-coordinate wave functions $|\phi_k(r)\rangle$ can be obtained using e.g. quantum defect theory as explained in Sec.~\ref{subsec:mqdt}. Any remaining interactions between the relative and center-of-mass coordinates in the static Hamiltonian can be easily taken into account by computing appropriate matrix elements, such that the eigenenergies $E_l$ and eigenstates $|\psi_l(R,r)\rangle$ of the static Hamiltonian can be obtained by diagonalisation. Here $n$, $k$ and $l$ denote the quantum numbers labeling the states. The micromotion Hamiltonian in the center-of-mass and relative coordinates is given by
\begin{eqnarray}
H_\mathrm{mm}(t)&=&-m_{\rm i}g^2\omega_{\rm i}^2\left(R^2+\frac{\mu^2}{m_{\rm i}^2}r^2+\frac{2\mu}{m_{\rm i}}rR\right)\cos (2\Omega t)\nonumber\\
&-&g\omega_{\rm i}\left(\{R,p\}+\frac{\mu}{m_{\rm i}}\{r,p\}+\frac{m_{\rm i}}{M}\{R,P\}\right.\nonumber\\
&+&\left.\frac{\mu}{M}\{r,P\}\right) \sin (\Omega t)\,.
\end{eqnarray}
Here $\{.,.\}$ denotes the anti-commutator and $\mu=m_{\rm i}m_{\rm a}/M$ denotes the reduced mass with $M=m_{\rm a}+m_{\rm i}$ the total mass. The relative and center-of-mass momenta are denoted by $p$ and $P$, respectively.

Next, we use Floquet theory to obtain the energies and eigenstates in terms of the unperturbed eigenstates and we have to diagonalise the Hamiltonian~\cite{NguyenPRA12}
\begin{equation}
H_\mathrm{F}=H_{\rm static}+H_\mathrm{mm}(t)-i\hbar\frac{\partial}{\partial t}\,.
\end{equation}
We use the unperturbed Floquet eigenstates  $\vert u_{jl}\rangle=e^{ij\Omega t}|\psi_l(R,r)\rangle$ of $H_\mathrm{F}-H_\mathrm{mm}(t)$ as our basis with Floquet energies $\epsilon_{jl}=E_l+j\hbar\Omega$. Here the integer $j$ denotes the class of the Floquet state. Then, we introduce the generalised matrix elements
\begin{equation}
\label{eq:genmatrxel}
 \langle \langle u^*_{j'l'}|H_\mathrm{mm}(t)|u_{jl} \rangle
 \rangle=\frac{1}{T}\int_0^T dt \langle u^*_{j'l'}|H_{mm}(t)|u_{jl}\rangle\,,
\end{equation}
where now $T$ indicates the period of the micromotion. To gain further insight into the micromotion effect we write the matrix elements as follows~\cite{NguyenPRA12,JogerPRA14}
\begin{widetext}
\begin{equation}\label{eq_Vmm}
\begin{split}
 \langle u^*_{j'l'}|H_\mathrm{mm}(t)|u_{jl} \rangle
 &=\langle \psi_{l'} |\left[V_1\cos (2\Omega t)+ V_2 \sin (\Omega t)  \right]e^{i(j-j')\Omega t}|\psi_{l}\rangle\,,\\
V_1 &= -m_{\rm i} g^2\omega_{\rm i}^2\left(R^2+\frac{m_{\rm a}^2}{M^2}r^2+\frac{2m_{\rm a}}{M}Rr\right)\,,\\
V_2 &= -\frac{i g \omega_{\rm i} m_{\rm i}}{\hbar}\left(E_{l'}-E_{l}\right)\left(R^2+\frac{m_{\rm a}^2}{M^2}r^2+\frac{2m_{\rm a}}{M}Rr\right)\,.
\end{split}
\end{equation}
\end{widetext}

Using these matrix elements, the total ion-atom problem may be solved by diagonalization taking an appropriate number of Floquet classes into account. Although the exact form of the solution will obviously depend on the problem at hand, a few general remarks can already be made by taking a closer look at Eq.~\eqref{eq_Vmm}. First of all, we note that selection rules exist for resonances between the Floquet classes when $\epsilon_{j'l'}=\epsilon_{jl}$, which are given by $j'=j\pm 2$ for $V_1$ and $j'=j\pm 1$ for $V_2$. In the situation studied by~\cite{JogerPRA14}, the largest effects were due to the matrix elements containing the relative coordinate $r$ in $V_2$. This is because for the scattering wavefunctions $|\phi_n(r)\rangle$, no selection rules exist for $\langle \phi_n(r) | r | \phi_{n'}(r) \rangle$ and $\langle \phi_n(r) | r^2 | \phi_{n'}(r) \rangle$, such that two states $n,n'$ that are separated very far in energy can have significant coupling. Such combinations of states can cause resonances between closeby Floquet classes, which significantly alter the solutions as compared to the secular approximation. We note that the prefactors to the terms containing $r$ and $r^2$ in Eq.~\eqref{eq_Vmm} are given by $m_{\rm a}/M$ and $m_\mathrm{a}^2/M^2$, respectively. Therefore, we find again that adverse micromotion effects may be reduced by choosing $m_{\rm i} \gg m_{\rm a}$.

The above calculation was used by~\cite{NguyenPRA12} to study the effect of micromotion on the controlled collision between a single atom and ion. This setup was proposed by~\cite{DoerkPRA10} as a means to implement a quantum gate between an ion and an atom and is discussed in Sec.~\ref{subsec:gate}. Micromotion would put serious restraints on the proposed scheme, as the atom and ion could couple to higher energy states, leading to significant heating. \cite{JogerPRA14} studied the effect of micromotion on a double-well system with a single ion trapped in the middle. The idea here is that the ion can control the atomic tunneling between the wells via its internal spin state or motion, leading to ion-atom entanglement. They found that this system is affected in a similar way as the situation considered by~\cite{NguyenPRA12}, although the use of an ion-atom combination with a suitable mass-ratio should allow for the implementation of the proposed scheme.

\begin{figure}[tb!]
\includegraphics*[width=0.85\linewidth]{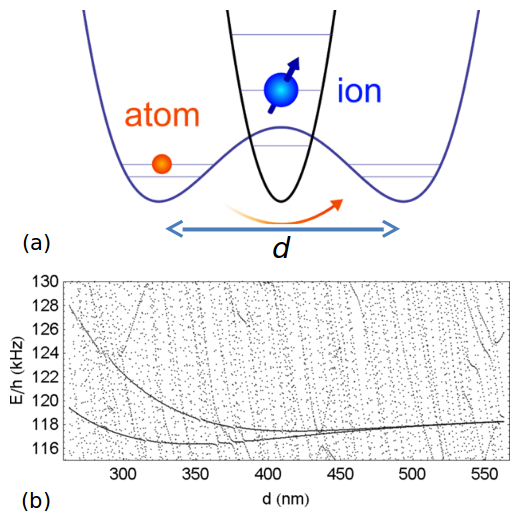}
\caption{Effect of micromotion on the quantum dynamics of an atomic double-well system interacting with a single ion shown in panel (a). The quasi-energy spectrum is shown as a function of inter-well separation $d$ (panel (b)). The spectrum shows the symmetric and antisymmetric ground states of the double-well system, and its energy difference is proportional to the atomic tunneling rate. It can be seen that for large $d$, there is almost no tunneling, whereas as $d$ gets smaller tunneling (i.e.~energy splitting) occurs. Since the atom needs to pass the ion during tunneling, the rate depends on the internal state of the ion and can thus be used to entangle the ion and atom~\cite{GerritsmaPRL12}. Taking micromotion into account leads to many additional quasienergies that run almost vertically up as a function of $d$. As $d$ gets smaller, the coupling between the two ground states and the high energy Floquet states gets larger, until for very small $d$, the two ground states disintegrate which signals the breakdown of double-well system. For the calculation, $^7$Li and $^{171}$Yb$^+$ were assumed with atomic trap frequency $\omega_\mathrm{a}=2\pi\times98\,$kHz and $\Omega_{\rm rf}=2\pi\times 967\,$kHz with $q=0.4$ and $a=0$ for the ion such that $\omega_\mathrm{i}=2\pi\times 137\,$kHz. For the calculation the Floquet classes $j=−2,\dots,2$ were taken into account. Adapted from~\cite{JogerPRA14}.
}
\label{fig:fig_DW_mm}
\end{figure}

The Floquet formalism presented above is limited in applicability to very small systems, i.e.~systems with only two particles in one dimension, or systems with exceptional symmetry. In particular, for most practical cases an enormous amount of basis states and Floquet classes needs to be taken into account to reach convergence. This is compounded by the large energy separation usually encountered in ion-atom systems. The trapping frequencies of the atoms usually lies in the kHz range, whereas the trap drive frequency of the Paul trap lies in the MHz range, such that each Floquet class contains an enormous amount of states. An example of the Floquet quasienergies for the double-well system considered by~\cite{JogerPRA14} as a function of the inter-well separation $d$ is shown in Fig.~\ref{fig:fig_DW_mm}. We can see that although many energy levels are present in the spectrum, the coupling to the states of interest (the symmetric and antisymmetric ground states of the double-well system) are indeed very small, so that the micromotion should not pose a problem for the parameters considered here.

\begin{figure}
\includegraphics*[width=0.85\linewidth]{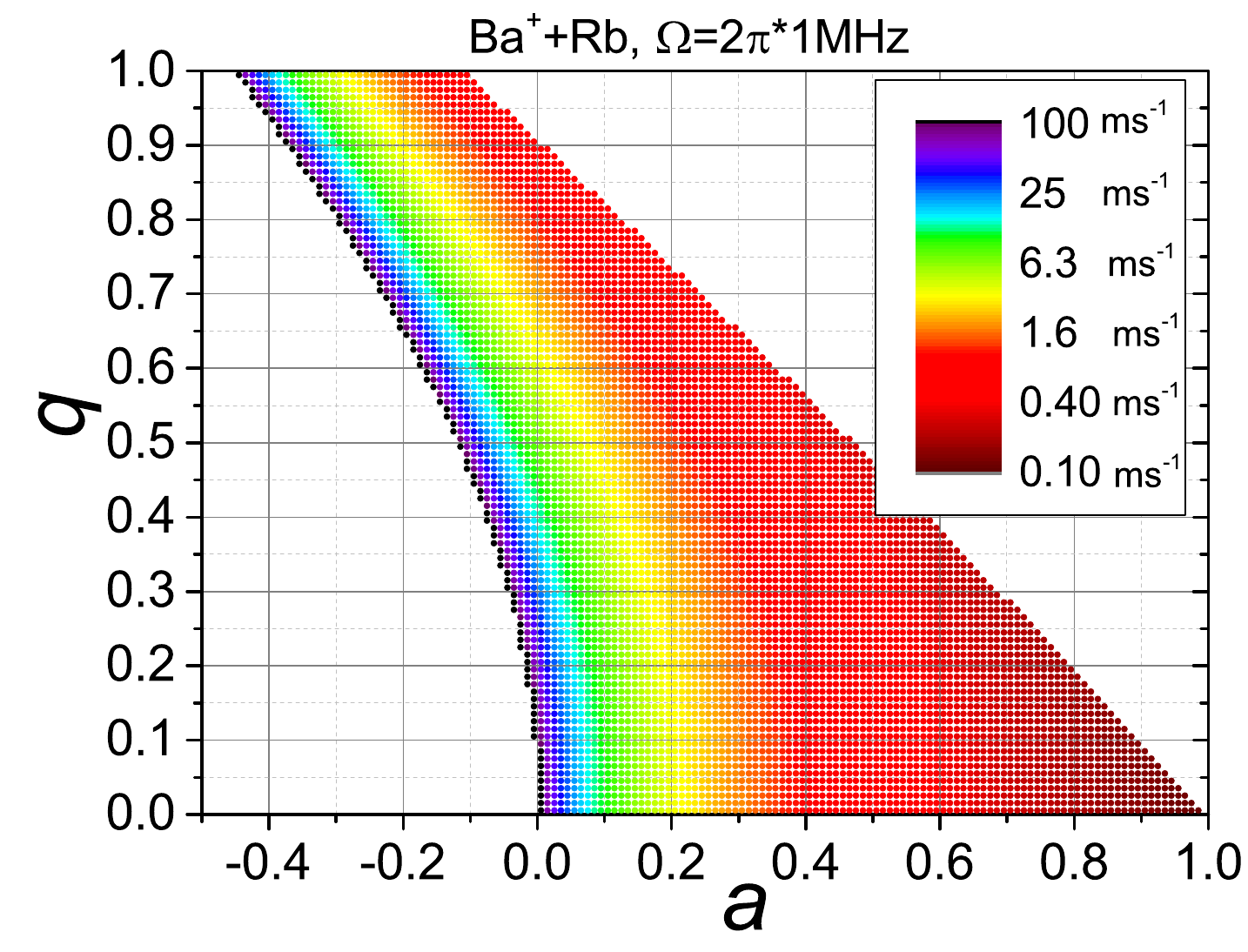}
\caption{Cooling rate in one spatial direction for an $^{138}$Ba$^+$ immersed in $^{87}$Rb gas with density $10^{12}$cm$^{-3}$ and temperature of 200$\mu$K calculated using master equation formalism as a function of Paul trap parameters $a$, $q$ with $\Omega=2\pi\times 1$MHz. Adapted from~\cite{KrychPRA15}.
}
\label{fig2_KrychPRA15}
\end{figure}

{\it Master equation description} -- Previously discussed limits on sympathetic cooling were based on classical or semiclassical approaches. In contrast,~\cite{KrychPRA15} aimed to study this problem quantum mechanically. In their approach, the ion is treated as an open quantum system and the atomic gas provides a reservoir. Using a regularized version of the ion-atom potential and implying Born and Markov approximations,~\cite{KrychPRA15} derived the master equation for the reduced density matrix of the ion $\rho$
\begin{equation}
\begin{split}
\dot{\rho}=\frac{1}{i\hbar}\left[H_S,\rho\right]-\sum_{\mathbf{k},\mathbf{k^\prime}}{\bar{n}_\mathbf{k}\left(\bar{n}_{\mathbf{k}^\prime}+1\right)c_{\mathbf{kk^\prime}}c_{\mathbf{k^\prime k}}/\hbar^2} \\
\times \int{d\tau\left( e^{i\tau(\omega_k-\omega_{k^\prime})} \left[e^{i(\mathbf{k}-\mathbf{k^\prime})\hat{\mathbf{r}}},e^{-i(\mathbf{k}-\mathbf{k^\prime})\hat{\mathbf{r}}(t,-\tau)}\hat{\rho}\right]\right.} \\
+\left. e^{-i\tau(\omega_k-\omega_{k^\prime})}\left[\hat{\rho}e^{i(\mathbf{k}-\mathbf{k^\prime})\hat{\mathbf{r}}(t,-\tau)},e^{-i(\mathbf{k}-\mathbf{k^\prime})\hat{\mathbf{r}}}\right] \right),
\end{split}
\end{equation}
where $\bar{n}_\mathbf{k}$ are the occupation numbers of the gas modes with energy $\hbar\omega_k$, $c_{kk^\prime}$ is the Fourier transform of the interaction potential and $\hat{\mathbf{r}}$ is the ion position operator. The problem can be simplified in the Lamb-Dicke regime where the length scale of the ion secular motion is much smaller than de Broglie wavelength of the reservoir, which allows for expanding the exponents into power series, as the expectation value of the operator
$(\mathbf{k}-\mathbf{k^\prime})\hat{\mathbf{r}}$ is much smaller than one. Up to the third order, different directions are not coupled and one arrives at simple equations of motion for the expectation values of combinations of ionic operators. In particular, the expectation value of the position and momentum operators is described by
\begin{align}
\dot{\bar{r}}_j & =\bar{p}_j/M,\\
\dot{\bar{p}}_j & =-M\tilde{\omega}_j^2\bar{r}_j-\bar{p}_j\sum_{n,m}{C_n^j C_m^j \tilde{\eta}_{nj}\cos\left((n-m)\Omega t\right)},
\end{align}
where $\omega$ and $C$ coefficients characterize the motion of the bare ion in the Paul trap, and $\eta$ is set by the interaction with the reservoir. The second term in momentum equation describes a time-dependent friction force which can lead to ion cooling or heating. Similar equations can be derived for operators quadratic in $r$, $p$. Based on these solutions,~\cite{KrychPRA15} analyzed the cooling rates for several systems and different trap parameters. We show as an example the stability diagram for $^{138}$Ba$^+$+$^{87}$Rb in Fig.~\ref{fig2_KrychPRA15}. Within this formalism it is also possible to derive the minimal achievable ion energies. The most problematic assumption of this treatment is neglecting the back-action of the ion on the gas. In experiments, the ion is able to remove a significant fraction of atoms from the condensate, which potentially can change the cooling dynamics.

\subsubsection{Octupole and higher-order ion traps}
\begin{figure}[tb!]
\includegraphics*[width=0.8\linewidth]{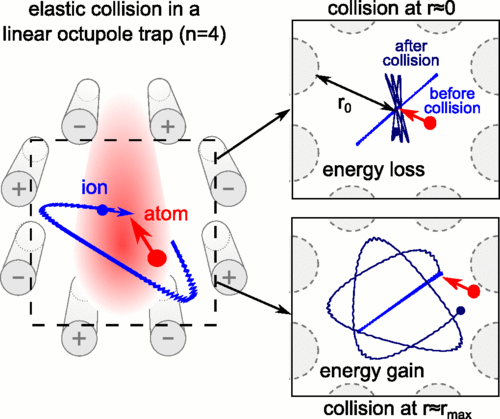}
\caption{Schematic of a linear rf trap with $n=4$. The atoms are indicated in red; the ion is shown in blue. The eight radio frequency electrodes are indicated by the gray cylinders. The two insets on the right illustrate two elastic collisions, one close to the trap center and the other close to the ion's turning point. The trap radius $r_0$ is indicated by the arrow. From~\cite{HoltkemeierPRL16}.
}
\label{fig:Holtkemeier_octupole}
\end{figure}

Another way to mitigate the effects of micromotion-induced heating is to use octupole or higher order ion traps~\cite{WesterJPB09,DeiglmayrPRA12,HoltkemeierPRL16,HoltkemeierPRA16}, in which the electric field increase as $r^{n-1}$, with $r$ the distance from the center of the trap and $n\geq 2$ the multi-pole order of the trap. Such traps can be created by employing more than 4 radio-frequency electrodes for the ion trap, as shown in Fig.~\ref{fig:Holtkemeier_octupole}. In this situation, the center of the trap features nearly field-free regions~\cite{DeiglmayrPRA12} such that atoms trapped in this region should cause more efficient sympathetic cooling than is possible in Paul traps. \cite{HoltkemeierPRA16} showed in theory that the critical mass ratio - at which efficient sympathetic cooling is still possible - depends on the multipole order $n$ and the spatial extent of the atomic cloud in the center of the ion trap. In this way, a broad range of ions could be sympathetically cooled, and the higher order traps up to $n=22$ find application in studying molecular ions and anions via cryogenic~\cite{GerlichPS95,WesterJPB09} and ultracold gases~\cite{DeiglmayrPRA12}.

A disadvantage of the use of higher order traps for some of the applications envisioned in hybrid ion-atom systems is that they do not allow for tight localization of individual ions. This seriously reduces some of the merits of the trapped ion platform such as long lifetimes and the ability to ground state cool the ions with lasers.

\subsection{Alternative techniques}
\label{subsec:alt}

\subsubsection{Optical ion trapping}

Since the radio frequency trap limits attainable temperatures due to micromotion, a promising new route towards colder ion-atom mixture is to use alternative trapping methods for the ion. In recent years, optical trapping of ions in optical tweezers and optical lattices has been demonstrated~\cite{SchneiderNatPhot10,SchneiderPRA12,EnderleinPRL12,HuberNatCom14,SchaetzJPB17}. Obviously the optical trapping potential, which interacts with the dipole polarizability of the ion, is much shallower than that of the radio frequency trap, which interact with the charge of the ion. Therefore, trapping lifetimes have for a long time been limited to milliseconds, with recent improvements providing 3$\,$s~lifetime~\cite{Lambrecht2016}, comparable to atoms under similar trapping conditions. Technical improvements should lead to further lifetime enhancements in the future. 

It is important to note that, although the trap depth is much smaller than for the Paul trap, the confinement can be made equally tight using focused lasers or optical lattices, where trapping frequencies in the MHz regime have been achieved in 1D~\cite{EnderleinPRL12} and are within reach for higher dimension. Preparation and cooling of ions can be performed in a Paul trap that is then adiabatically switched off and the ions are loaded into an optical trap. Using such an approach, trapping of multiple ions forming a one-dimensional Coulomb crystal in a single-beam optical dipole trap have been demonstrated~\cite{SchmidtPRX18}. Anharmonicities in the trapping potential in combination with the radio frequency drive may result in parametric excitation and of the ions leading to ion loss. This may be mitigated by more accurate trapping design and deeper optical potential, i.e.~by using higher power lasers.  Alternatively, loading in a deep dipole trap of low trapping frequencies and loading out of a MOT via photoionization will avoid the rf-overlap in first place. An interesting benefit to the optical dipole trap approach is that the atoms may be straightforwardly trapped in the same potential, leading to very accurate overlap between the two species. Since the laser field also constitutes an oscillating electric field, it was found that ions in optical traps also display micromotion, albeit at a very low amplitude (and at optical frequencies)~\cite{CormickNJP11}.

\subsubsection{Schemes involving Rydberg excitations}

\begin{figure}[tb!]
\includegraphics*[width=0.9\linewidth]{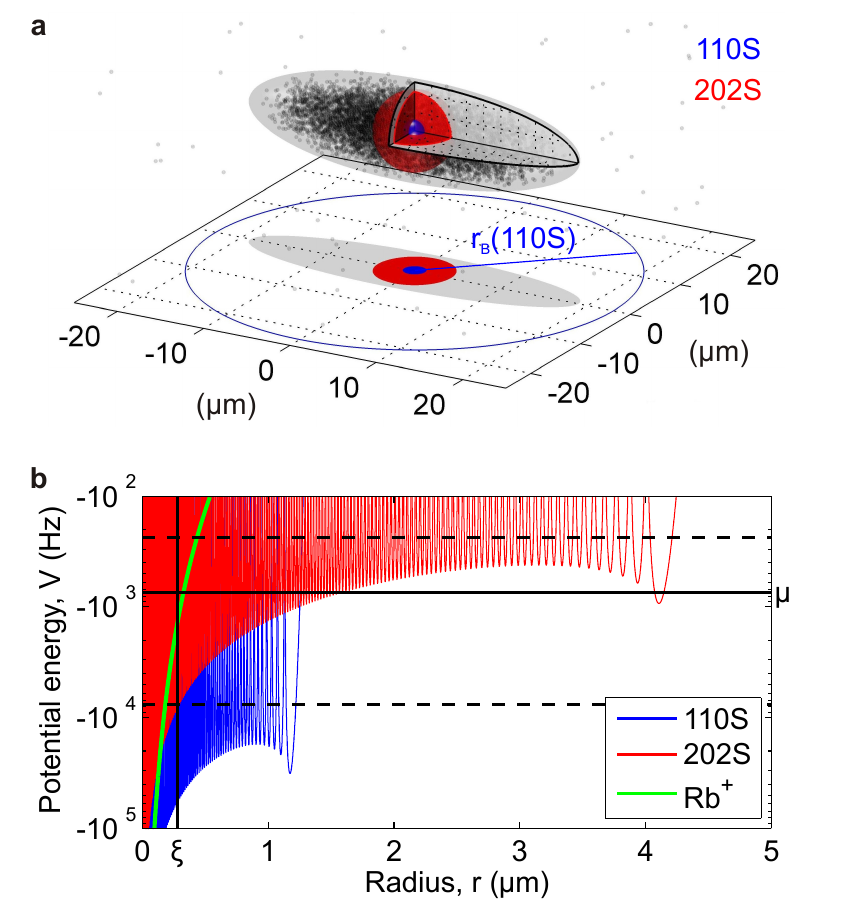}
\caption{Single Rydberg electron immersed in a BEC: (a) comparison of the size of the Rydberg states and BEC cloud, (b) corresponding interaction potentials. From~\cite{BalewskiNature13}.}
\label{fig1_BalewskiNature13}
\end{figure}

A very different idea of realizing a hybrid ion-atom system is to produce the charges directly from the ultracold gas. The simplest way to do it without introducing additional traps for the charged particles is via Rydberg excitation. In this way, the Rydberg electron and the ionic core remain bound but the typical distance between them is large enough to allow for observing interaction effects with neutral atoms in the cloud. The electron plays typically a much more important role in the dynamics of the system than the ion. This is caused by the vast difference in reduced masses in electron-atom and ion-atom collisions. Low reduced mass makes the centrifugal barrier huge and prevents high partial wave scattering. At the same time, the $s$-wave interaction can be well described by the zero range interaction proportional to $4\pi\hbar a/\mu$. Within the mean-field approximation, the potential experienced by the atoms due to the presence of a Rydberg electron can then be expressed as $V(r)=\frac{4\pi\hbar^2 a}{\mu}\left|\psi(r)\right|^2$, where $\psi(r)$ is the Rydberg orbital. \cite{BalewskiNature13} demonstrated that the Rydberg electron can indeed impact the whole condensate and lead to phonon excitations. The size of the orbital, which determines how many atoms on average will interact with the electron, can be adjusted by changing the principal quantum number $n$ and the density of the cloud. Studying the line shapes of the excitation spectra revealed sharp lines caused by the formation of molecules in the few-atom regime, which turn into a broad shift in the many-atom limit~\cite{GajNatComm14}, where a Rydberg polaron can be formed~\cite{CamargoPRL18}. The theoretical description of Rydberg polarons has been provided~by~\cite{SchmidtPRL16,SchmidtPRA18} in terms of the functional determinant approach.

The role of the atomic ion in such systems has so far been rather marginal. This is due to the fact that the interaction with the electron is orders of magnitude stronger and spread over a large number of atoms, while the ion acts only locally, meaning that in a dilute gas the effects of the ion are not detectable on the time scale of the experiment which is limited to a few $\mu$s by the decay of the Rydberg atom. Inelastic processes leading to the formation of molecular Rb$_2 ^+$ ion have been observed~\cite{SchlagmullerPRX16}, but the role of the Rydberg electron was crucial for the process. In order to probe the ion dynamics in the Rydberg setup, it would be beneficial to diminish the role of the electron. The possible strategies to achieve this include i) working with large Rydberg orbits and small atomic clouds, so that the electron wave function does not overlap strongly with the atoms (see Fig.~\ref{fig1_BalewskiNature13}), ii) increasing the density of the gas to make ion-atom collisions more frequent and be able to work at shorter time scales, iii) confining the atomic gas in a reduced dimensional setup to further decrease its overlap with the electron. Going along these lines, recent experiment~\cite{Kleinbach2018} working with up to 190S Rydberg state and a very tight, anisotropic optical tweezer trap for the atoms detected the presence of the ion by showing that it contributes to broadening of the Rydberg excitation line. In order to be able to probe the dynamics of the ion, the electron would need to be removed from the gas completely, e.g. by exciting it to a circular state with high angular momentum. This would potentially allow to study the ion remaining in the cold gas for long times and at sub-microkelvin temperatures, close to the $s$-wave collisional limit. Next, \cite{Engel2018} manged to observe Rydberg blockade on a highly excited ultracold Rydberg atom induced by a single ion mediated over tens of micrometer distances. The ion was produced from an ultracold atomic ensemble via near-threshold photoionization of a single Rydberg excitation, employing a two-photon scheme which was specifically suited for generating a very low-energy ion.

Another strategy involving the Rydberg excitation has been proposed by~\cite{Schmid2017}. In this scheme, one starts from an ultralong-range Rydberg molecule composed of a neutral atom bound in the outer well of the Rydberg potential. This molecule is then photoionized, which initializes the ion-atom scattering event with the initial wavepacket formed from the Rydberg molecule wavefunction with the mean kinetic energy as low as few $\mu$K and a strongly non-thermal profile. Importantly, as the molecule is spherically symmetric, the collision can only happen in the $s$-wave (although with considerably higher energy) if stray electric fields which would induce partial wave mixing are compensated well enough. Dynamics of the wavepacket strongly depends on the sign of the ion-atom scattering length $a$. For large and positive $a$, there is a weakly bound molecular state which can have large overlap with the initial state. As a result, two separate shells (molecular and dispersive one) are formed during the collision. Measuring the bound fraction of the wavepacket as well as its expansion velocity allows then to obtain information about the scattering length.

\begin{figure}[tb!]
\centering
\includegraphics[width=8cm]{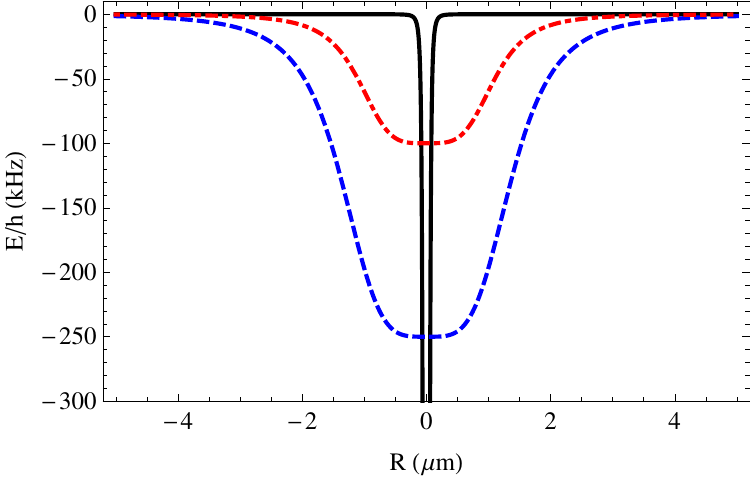}
\caption{Adiabatic potentials for a ground state atom and an ion (solid black), for a dressed atom with $\Omega=2\pi$~10~MHz and $\Delta=2\pi$~1~GHz (red dash-dotted) and $2\pi$~0.4~GHz (blue dashed) assuming coupling to the $\mid 30S_{1/2}\rangle$ state of lithium. Figure taken from~\cite{SeckerPRA16}.}
\label{fig_dress}
\end{figure}

A new direction of research is to study trapped ions interacting with ultracold atoms that are coupled to Rydberg states~\cite{HahnPRA00}. The strong polarizabilities of the Rydberg atoms would increase the interaction strength between atoms and ions by many orders of magnitude, as compared to the case of ground state atoms. Such interactions may be mediated over much larger length scales as well. In particular, an atom that is dressed with a Rydberg state $|nS_{1/2}\rangle$, with a (effective) Rabi frequency $\Omega$ and detuning $\Delta_0$, experiences an adiabatic potential~\cite{SeckerPRA16}
\begin{equation}\label{eqVad}
V(R)=-\frac{A R_w^4}{ R^4+ R_w^4}
\end{equation}
\noindent when it is a distance $R$ away from a trapped ion.  The depth of this potential is given by $A=\hbar\Omega^2/\Delta_0$ and its width by $R_w=(C_4^{|nS\rangle} \hbar\Delta_0)^{1/4}$. Here, $n$ denotes the principal quantum number of the Rydberg state and $C_4^{|nS\rangle}$ is proportional to the polarizability of the Rydberg state. When we consider a lithium atom with $n=30$, $\Omega = 2\pi$~10~MHz, $\Delta_0=2\pi$~1~GHz, we find $A/h=$~100~kHz and $R_w=1$~$\mu$m, such that $R_w \gg R^*$ (e.g., assuming Yb$^+$). For these numbers, the lifetime of the dressed atom is enhanced by a factor 10$^4$ as compared to the Rydberg state, such that coherent experiments on the 100~ms timescale seem possible. In Fig.~\ref{fig_dress} we show the resulting adiabatic potential.

Possible applications envisioned for the Rydberg-ion system would include ion-atom quantum gates and interfaces~\cite{SeckerPRA16} and novel (repulsive) ion-atom interactions~\cite{SeckerarXiv16,Wang2019controlling} obtained by Rydberg dressing on a dipole-forbidden transition. Such repulsive ion-atom interactions could allow reaching ultracold temperatures despite micromotion and for any mass-ratio.  The feasibility of such a setup, in which Rydberg excitation occurs within a Paul trap, have been established by two recent experiments that demonstrated Rydberg excitation of trapped {\it ions}~\cite{FelkderPRL2015,HigginsPRL2015}. Recently, \cite{Ewald2018} reported on the observation of interactions between ions in a Paul trap overlapped with ultracold Rydberg atoms. They observed inelastic collisions, manifested in charge transfer between the Rydberg atoms and ions, exceed Langevin collisions for ground state atoms by almost three orders of magnitude in rate, which indicates a huge increase in interaction strength. Furthermore,~\cite{Haze2019stark} demonstrated Stark spectroscopy of Rydberg levels in the trap for measuring electric fields. These results pave the way towards tuning interactions between ultracold atoms and ions by laser coupling to Rydberg states for future studies e.g. of charge transport~\cite{Mukherjee2019}.

\subsubsection{Photoionization}

Creation of charges in an ultracold cloud can also be accomplished using other methods than Rydberg excitation. Direct photoionization of atoms using a femtosecond laser has been studied by~\cite{Wessels2017}. So far, the experiment only focused on the process of strong-field ionization itself and not on the ionic products.

\section{Applications of cold ion-atom systems}
\label{sec:app}

In this section we review experimental studies of ion-atom dynamics, in particular charge exchange and other inelastic processes. Then we address the theoretical proposals for studying ions immersed in ultracold atomic systems in the context of future applications such as quantum simulations, computations, and probing. We discuss what unique features the ion-atom systems can add to the field of quantum technologies as compared to other existing systems.

\begin{table*}[tb!]
\caption{Summary of cold ion-atom systems investigated experimentally.} 
\begin{ruledtabular}
\begin{tabular}{lll}
Ion & Atom & References   \\
\hline
 Yb$^+$ & Yb &\cite{GrierPRL09}\\
 Yb$^+$ & Rb &\cite{ZipkesNature10,ZipkesPRL10,RatschbacherNatPhys12,RatschbacherPRL13}\\ 
 Rb$^+$ & Rb &\cite{SchmidPRL10,HartePRL12,RaviAPB12,RaviNatCommun12,SchmidRSI12,LeePRA13,RayAPB14} \\
        &    &\cite{Engel2018,Kleinbach2018,Haze2019stark} \\
 Ba$^+$ & Rb &\cite{SchmidPRL10,SchmidRSI12,HallMP13b,KrukowPRA16,KrukowPRL16}\\
 Ca$^+$ & Rb &\cite{HallPRL11,HallMP13a,EberleCPC16}\\
 Yb$^+$ & Ca &\cite{RellergertPRL11}\\
 Ba$^+$ & Ca &\cite{SullivanPRL12}\\
 Na$^+$ & Na &\cite{SivarajahPRA12,GoodmanPRA15}\\
 Ca$^+$ & Li &\cite{HazePRA13,HazePRA15,SaitoPRA17,HazePRL18}\\
 Ca$^+$ & Na &\cite{SmithAPB14} \\
 Sr$^+$ & Rb &\cite{MeirPRL16,MeirARX17,SikorskyNatCom18,Meir2018}\\
 K$^+$ & Rb &\cite{DuttaPRL17} \\
 Rb$^+$ & Cs &\cite{DuttaPRL17}\\
 Cs$^+$ & Cs &\cite{Dutta2017} \\
 Cs$^+$ & Rb &\cite{Dutta2017}\\
 Yb$^+$ & Li &\cite{Joger2017,Furst2017,Ewald2018}\\
\end{tabular}
\label{tab:exp}
\end{ruledtabular}
\end{table*}
\subsection{Cold collisional studies}
\label{sec:cold_col}

A good understanding of the collisional properties of hybrid ion-atom systems is crucial for further applications. This requires not only high performance theoretical calculations, but also taking into account potentially very complex dynamics resulting from the specific features of the experimental setup. The most notable example here is the micromotion of the ion. 

It is experimentally challenging to separate different processes taking place in the system, such as inelastic two- and three-body collisions, possibly involving photons coming from the trapping lasers. Inhomogeneous atomic density and nonthermal kinetic energy distribution of the ion have to be taken into account. Ion micromotion typically sets the lower limit for the achievable mean ion energy, but also provides a tool for controlling it by introducing excess micromotion on purpose.
In this subsection we focus on the collisional ion-atom physics in realistic environments. So far, all experimental results have been obtained using hybrid setups involving the Paul trap except for a single experiment in which the ion is created by ionization of a Rydberg atom~\cite{Engel2018}. The studied systems are listed in Table~\ref{tab:exp}. We start the discussion with experiments probing the dynamics resulting mainly from two-body elastic and charge transfer collisions, and then move to spin relaxation and three-body effects.

\begin{figure}[b]
 \includegraphics[width=1.0\linewidth]{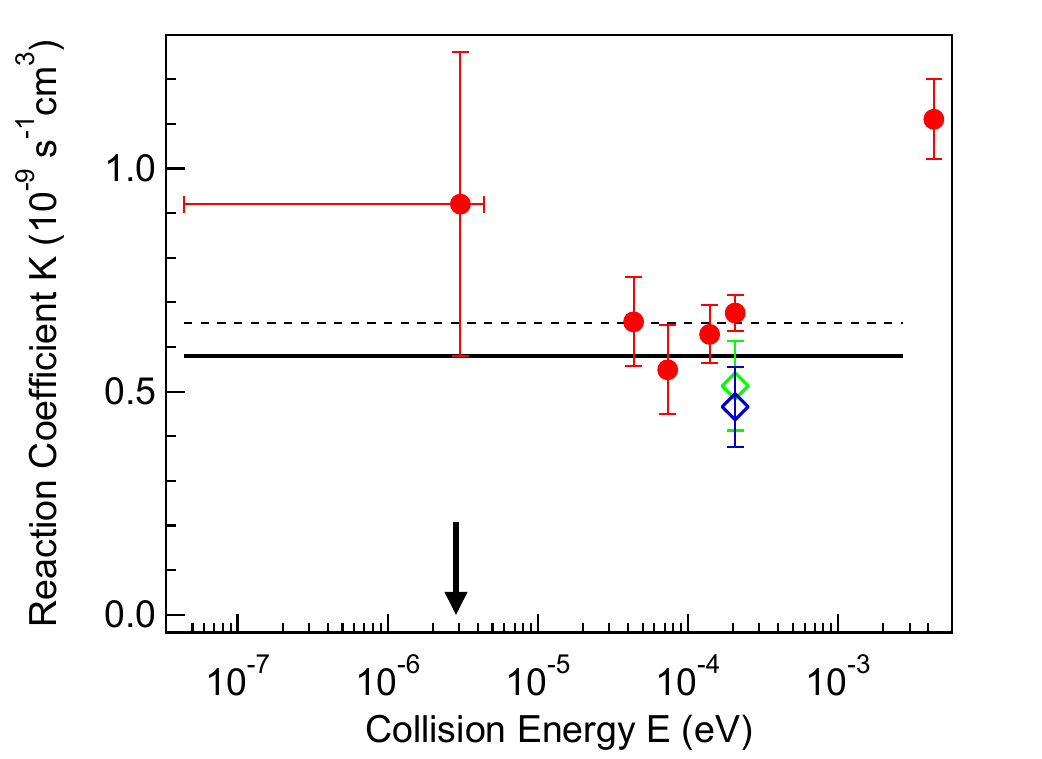}
 \caption{\label{fig:secV1_1}Charge exchange rate coefficient measured for Yb$^+$+Yb system as a function of average collision energy. Circles, green and blue diamonds represent $^{172}$Yb$^+$+$^{174}$Yb, $^{172}$Yb$^+$+$^{171}$Yb and $^{174}$Yb$^+$+$^{172}$Yb, respectively. The solid line gives the Langevin rate, while the dashed line accounts for the contribution of the electronically excited Yb present due to MOT lasers. The black arrow indicates the ionization isotope shift between $^{174}$Yb and $^{172}$Yb. From~\cite{GrierPRL09}.}
 \end{figure}

In a proof-of-principle experiment,~\cite{SmithJMO05} successfully built a hybrid ion-atom trap designed to co-trap laser-cooled Ca$^+$ ion together with cold Na atoms. They showed that the rf fields generated by the linear Paul trap do not destroy the magneto-optical trap used for the atoms. The first observations of cold ion-atom collisions was made by~\cite{GrierPRL09}, studying charge exchange in different isotopic combinations of Yb$^+$+Yb at energies ranging from 35$\,$mK to 45$\,$K. Even at the lowest achievable energies the collisions involved over 40 partial waves. Detection of charge exchange events was possible with isotope-selective fluorescence imaging. The measured rates, shown in Fig.~\ref{fig:secV1_1}, were consistent with the predictions of the Langevin model. In a subsequent theoretical analysis,~\cite{ZhangPRA09} calculated the differential cross sections and showed that at experimentally probed collision energies the elastic collision leads to a peak at small scattering angles, whereas charge transfer is characterized by enhanced backscattering.

\begin{figure}
\includegraphics[width=1.0\linewidth]{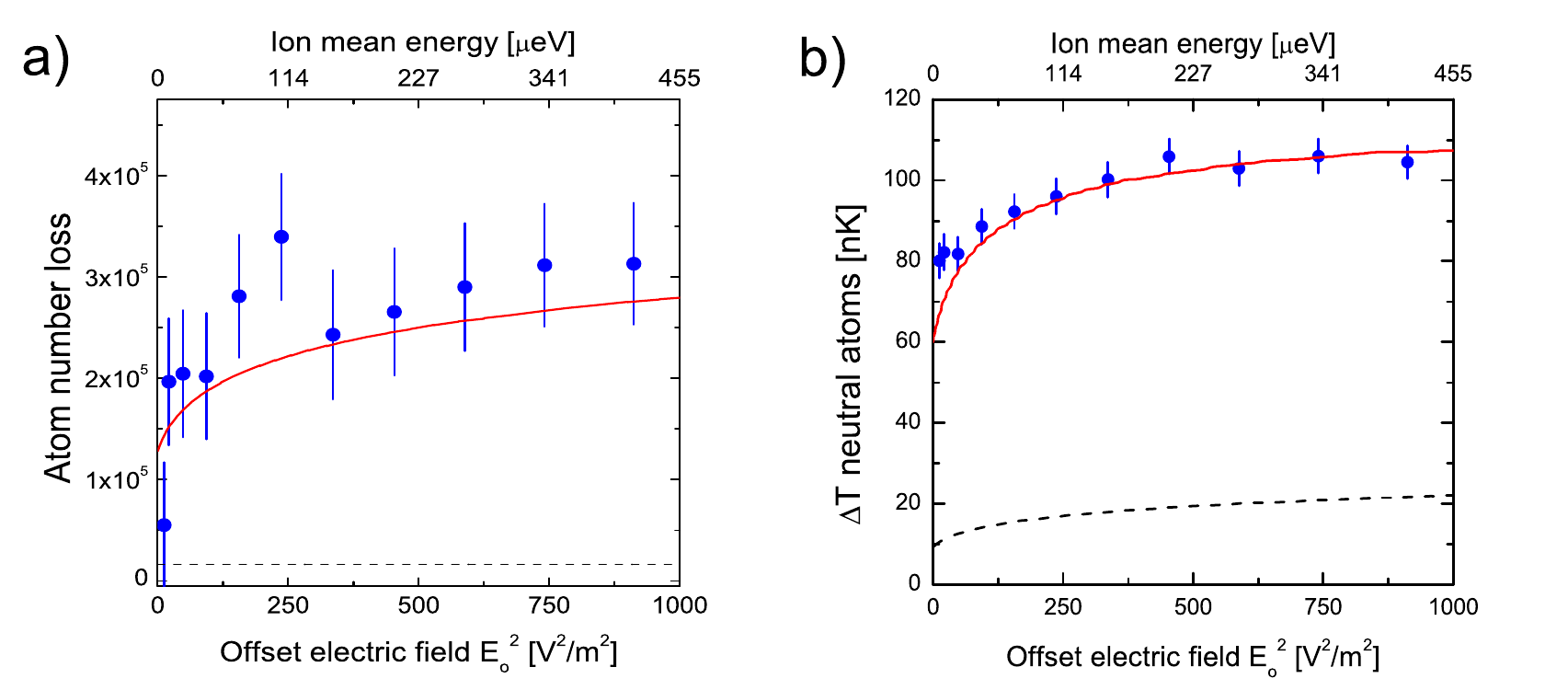}
\caption{(a) Atom loss from the magnetic trap as a function of the ion energy in the $^{172}$Yb$^+$+Rb experiment. Solid line results from the numerical simulation basing on the full collision cross section, while the dashed line gives the predictions of the Langevin model. Points are experimental results. (b) Temperature increase of the atomic cloud vs. mean ion energy. From~\cite{ZipkesPRL10}.}
\label{fig2_ZipkesPRL10}
\end{figure}

Heteronuclear Yb$^+$+Rb collisions were investigated by~\cite{ZipkesNature10,ZipkesPRL10}. Collisional processes were probed at energies corresponding to 0.2-5$\,$K. Elastic collisions with the ion typically lead to loss of atoms from the trap if the momentum transfer exceeds the trap depth, so measuring the number of atoms via absorption imaging can be used to infer the elastic collision cross section. Furthermore, for small energy transfers the atom stays in the trap and thermalizes with the rest of the cloud, which results in heating the whole system. Another effect which has to be included is the ''evaporative heating'' of the atoms, as the ion preferentially removes them from the center of the trap where the density is the highest and typical momenta are the lowest. Finally, one should keep in mind that the steady state of the ion is influenced by the Paul trap. This complicated dynamics could be modelled using Monte Carlo calculations described in detail by~\cite{ZipkesNJP11}. Figure~\ref{fig2_ZipkesPRL10} shows the experimental results which are well reproduced by the simulations, provided that the  differential cross section calculated from the quantum scattering model is used instead of the classical approximation~\cite{CotePRA00}. Reactive collisions were also studied in the same experiment by means of fluorescence imaging. In contrast to the homonuclear case, the observed reaction rates were five orders of magnitude smaller than the characteristic Langevin rate. Measurements of the density dependence of the reaction rates showed that the observed loss of Yb$^+$ is due to two-body collisions. Surprisingly, no molecular ions were detected although one would expect radiative association to be important. Similar results, with low reaction rates and no production of molecular ions, were obtained by~\cite{SchmidPRL10} with Ba$^+$+Rb system. However, the lack of molecular ions may result from multiple processes such as secondary collisions with the atoms or photodissociation caused by the trapping laser fields. 

\begin{figure*}[tb!]
\includegraphics[width=1.0\linewidth]{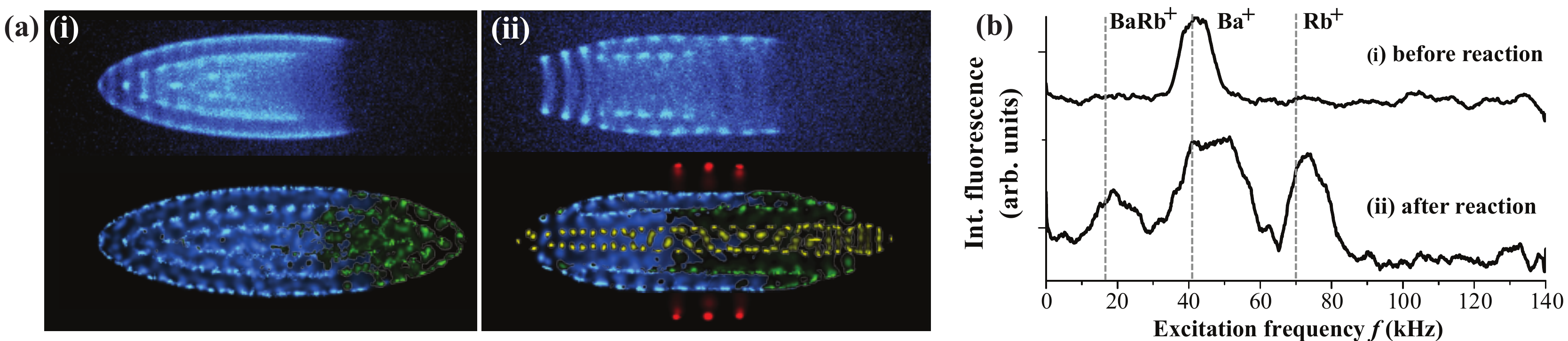}
\caption{(a) False-color fluorescence images of a Ba$^+$ Coulomb crystal (upper panels) and their molecular dynamics simulation (lower panel) before (left) and after (right) reaction with Rb atoms. Product ions, dark in the experiment, are depicted (blue: $^{138}$Ba$^+$, green: lighter isotopes of Ba$^+$, yellow: $^{87}$Rb$^+$, red: BaRb$^+$). (b) Resonance excitation mass spectra before and after the reaction. From~\cite{HallMP13b}.}
\label{fig3_HallMP13b}
\end{figure*}

In the next step, \cite{RatschbacherNatPhys12} studied controlled chemical reactions by tuning the quantum states of the reactants. Such full control over the quantum mechanical degrees of freedom in a chemical reaction is of key interest for quantum-contolled chemistry and allows the identification of fundamental interaction processes and even the steering of chemical reactions on the single particle level. Charge-exchange reaction rates of electronically excited ions showed that the $^2D_{3/2}$ state of Yb reacts with Rb at the Langevin rate, while for the $^2F_{7/2}$ state the rate was two orders of magnitude lower. Changing the hyperfine state of the atoms from $\left|2,2\right>$ to $\left|1,1\right>$ while keeping the ion in the ground electronic state lead to enhancement of the reaction rate by a factor of 35, demonstrating the important role of the hyperfine interaction. By monitoring the ion fluorescence, it was shown that in about 4\% of the events the electronically excited ion stayed in the trap after the inelastic process and was recooled again after some time, indicating radiative decay back to the ground state. 

\cite{HallPRL11,HallMP13a} studied cold reactive collisions between laser-cooled Ca$^+$ ions forming a crystal and Rb atoms. They observed rich chemical dynamics and interpreted it in terms of nonadiabatic and radiative charge exchange as well as radiative molecule formation. They studied the role of light-assisted processes and showed that the efficiency of the dominant chemical pathways is considerably enhanced in excited reaction channels. 
Next, \cite{HallMP13b} studied cold chemical reactions between laser-cooled Ba$^+$ ions and Rb atoms. A number of ions forming a Coulomb crystal was overlapped with an atomic cloud. Monitoring the number of remaining ions enabled the measurement of the reaction rate (see Fig.~\ref{fig3_HallMP13b}(a)). Molecular dynamics simulations were used to characterize the ion energy distribution. The average energy could be controlled by changing the number of ions, as for large crystals more ions were far from the trap center where micromotion is larger. Different reaction products could be detected via resonant excitation mass spectrometry as presented in Fig.~\ref{fig3_HallMP13b}(b). Molecular ions were clearly detected. However, multiple secondary processes could not be excluded so a direct comparison to theoretical branching ratios was not possible. Different electronic states of the ion could be probed by adjusting the ion cooling laser. The  energy dependence of the reaction rates was consistent with the Langevin predictions. 

In order to improve the energy resolution of the measurements,~\cite{EberleCPC16} developed a novel dynamic hybrid trap. Their approach is based on pushing a cloud of laser-cooled Rb atoms through a stationary Coulomb crystal of cold ions by using precisely controlled, tunable radiation pressure forces. The atom kinetic energies can be controlled over an interval ranging from 30$\,$mK up to 350$\,$mK with energy spreads as low as 24 mK. In an alternative approach, \cite{PuriRSI18} demonstrated the high-resolution collision energy control through ion position modulation in atom-ion hybrid systems by translating an ion held within a radio-frequency trap through a magneto-optical atom trap. The technique allows one to control ion kinetic energies from 0.05 to 1 K with a fractional resolution of $\sim$10.

A different heteronuclear system, namely Yb$^+$ colliding with Ca atoms, was studied by~\cite{RellergertPRL11}. In this case the observed chemical reaction rate was close to the predictions of the Langevin model due to strong charge transfer processes. This was explained by theoretical calculations of the transition dipole moment and Franck-Condon factors, which turned out to be large in this system due to an avoided crossing between the $X^2\Sigma^+$ and $A^2\Sigma^+$ states. The branching ratio between the radiative association and charge transfer process was measured to be below 2\%, while theoretical calculation predicted a much higher value. 

\cite{SullivanPRL12} investigated the case of Ba$^+$+Ca system in which the charge exchange process is energetically forbidden unless Ca is electronically excited by the cooling laser. However, the electric field of the ion shifts the atoms away from resonance at short distances, suppressing the reaction in this channel. On the other hand, excitation of the ion changes only the dispersion interaction with the atoms, and the ion is not shifted from the resonance, making inelastic processes possible. By estimating the population of different electronic states in the trap using the Liouville equation it was possible to extract the reaction rate $k_p$ for the Ba$^+(^2P_{1/2})$+Ca$(^1 S)$ inelastic collisions, finding $k_p=4.2(1.9)\times 10^{-10}\,$cm$^3$/s.

\cite{RaviNatCommun12} studied cooling of an Rb$^+$ ion by Rb atoms. It has been shown that placement of the dense atomic cloud in the center of the Paul trap is beneficial for cooling efficiency, as collisions occur in the region where the ion macromotion is classically the fastest. An additional cooling mechanism characteristic for collisions of the ion with its parent neutral particle has been suggested: cooling by resonant charge exchange. The electron can be transferred between the particles without any other change in the internal states, producing a translationally cold ion and a very hot atom which leaves the trap. Further experimental evidence that resonant charge exchange can be important for the cooling was given by~\cite{DuttaPRL17}, where the dynamics of Cs$^+$ ions in Cs and Rb gases was compared and the cooling was more efficient when using Cs.

Collisions between Rb$^+$ ions and laser-cooled Rb atoms were also investigated by~\cite{LeePRA13}. Here the ions were produced directly from the atomic cloud by two-photon ionization which results in high average kinetic energy. The measured quantity was atomic fluorescence from the MOT. Analysis of rate equations provided an estimate for the total two-body collision rate (including elastic and resonant charge exchange) $k=1.23(42)\times10^{-7}$cm$^3$/s. This large value results from the fact that 28\% of the atomic population is estimated to be in the electronically excited state due to using the magnetooptical trap.

The study of the Ca$^+$+Na system started by~\cite{SmithJMO05} enabled the measurement of the charge transfer rates reported by~\cite{SmithAPB14}. In this case they loaded the Paul trap with Ca$^+$ ions in the presence of a Na MOT and found a strong loss of Ca$^+$ ions. Here radiative charge transfer was predicted to be very slow. However, the measured loss rate constant was estimated to be $2\times 10^{-11}$cm$^3$/s. It was conjectured that a non-radiative process from the Ca$^+ (^2S)$+Na$(^2P)$ entrance channel is responsible for the observed loss. As in the case of~\cite{LeePRA13}, excited state population of Na results from the MOT lasers. It is noteworthy that the rate coefficient, while unexpectedly large, is still two orders of magnitude smaller than the Langevin rate. The same experimental setup allowed for measurements of Na$^+$+Na collision rates~\cite{GoodmanPRA15} by investigating both the atomic population in the MOT and the number of ions in the Paul trap. As the Na$^+$ ions are optically dark, they were detected directly with an electron multiplier. Using an improved version of the rate equation model proposed by~\cite{LeePRA13}, the total collision rates for both ground- and excited-state Na atoms were determined and turned out to agree with theoretical predictions.

The trap geometry and electric fields present in the system are important for the efficiency of the sympathetic cooling of single or multiple ions by the atomic cloud. For this reason~\cite{GoodmanPRA12} performed a realistic classical trajectory simulation of ion trajectories in a hybrid trap for experimentally realistic parameters. They concluded that it should be possible to simultaneously cool several ions at the same time, provided that the atomic density is sufficiently high to combat the effect of rf heating. In a subsequent experiment,~\cite{SivarajahPRA12} have demonstrated sympathetic cooling of hot Na$^+$ ions by Na atoms trapped in a magneto-optical trap. 

The Smith group also investigated the dynamics of loading ions into a linear Paul trap by photoionization of atoms from a magneto-optical trap. In combined theoretical and experimental studies, \cite{BlumelPRA15,WellsPRA17} shown the universal, nonlinear, non-monotonic behavior of the saturation curves of magneto-optical-trap-loaded Na$^+$ ions stored in an ion-neutral hybrid trap as a function of loading rate.

Sympathetic cooling of the ion by the atomic cloud is expected to be most efficient for large ion/atom mass ratios. For this reason, the Mukaiyama group focused on the Ca$^+$+Li system. Observation of elastic collisions via decay of the atomic cloud was reported by~\cite{HazePRA13}. The energy was controlled by changing the loading time of the ions and covered the range 100$\,$mK to 3$\,$K. The measured rates were in agreement with the semiclassical model~\cite{CotePRA00}. Recently, \cite{HazePRL18} demonstrated sympathetic cooling in this system, when efficient collisional cooling was realized by suppressing collision-induced heating. Charge-exchange collisions were investigated in the next step by~\cite{HazePRA15} in the mK energy regime. While for Ca$^+$ in the $S_{1/2}$ state the inelastic processes are improbable,  excited electronic states could be populated in the experiment by the cooling lasers using the optical pumping technique. Collisional quenching from the $D_{3/2}$ state did not contribute to the losses, as the ion would stay in the trap and become quickly recooled again. Only Li$^+$ ions were detected as reaction products. The measured rate coefficients were $k_{D_{3/2}}=8.2(6)\times 10^{-11}$cm$^3/$s and $k_{P_{1/2}}=3.0(5)\times 10^{-10}$cm$^3/$s, while $k_{S_{1/2}}<7\times 10^{-13}$cm$^3/$s (only the upper bound could be given in this case). These values are again much lower than the Langevin rate, which for Ca$^+$+Li is $k_\mathrm{L}=5\times 10^{-9}$cm$^3/$s.
Further investigation of the charge exchange process for this system has been given by~\cite{SaitoPRA17}. This time the energy could be manipulated from 1$\,$mK to 1$\,$K by controlling the excess micromotion. The energy dependence of inelastic collision cross sections was in agreement with the Langevin predictions.

\cite{Joger2017} investigated the inelastic collision rates for the Yb$^+$+Li system which is even more favorable for reaching the quantum limit in hybrid ion-atom systems due to large ion/atom mass ratio. For the ground state collisions they found the charge transfer and association rate to be at least $10^3$ times smaller than the Langevin collision rate, in agreement with theory~\cite{TomzaPRA15a}. For ions prepared in the excited electronic states, they found that the inelastic collision rate is dominated by charge transfer and does not depend on the ionic isotope nor the collision energy in the range 1-120$\,$mK. Interestingly, they found the loss rates for the Yb$^+$ ion in the $^2F_{7/2}$ state an order of magnitude larger than for the $^2D_{3/2}$ state. By comparing these measurements with the results presented by~\cite{RatschbacherNatPhys12} it can be conjectured that the electronic configuration of atomic thresholds and related molecular states surrounding the ion-atom entrance channel is more important than the strength of spin-orbit coupling to determine the short-range probability of non-radiative charge transfer for collisions involving Yb$^+$ ions.

\begin{figure}[tb!]
\includegraphics*[width=0.95\linewidth]{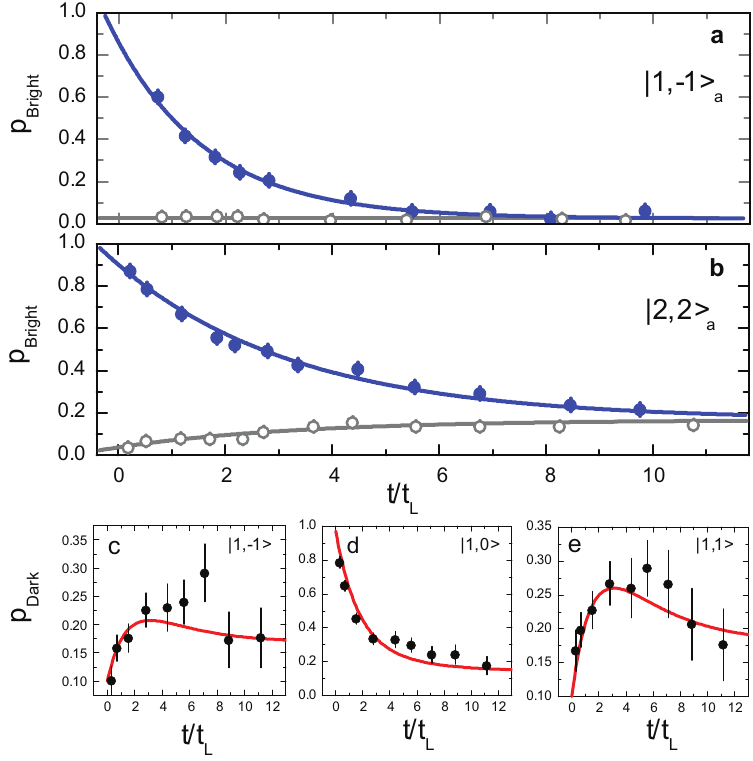}
\caption{Hyperfine spin relaxation in $^{171}$Yb$^+$ interacting with ultracold Rb atoms. (a) The probability $p_\mathrm{Bright}$ after preparation of the ion in the $|F=1,m_F=0\rangle_i$ (full symbols) or the $|F=0,m_F=0\rangle_i$ (open symbols) state vs the interaction time for atoms in the $|F=1,m_F=-1\rangle_\mathrm{a}$ state. Error bars denote 1 standard deviation uncertainty intervals resulting from a total of 5000 measurements. The fit values at t=0 are limited by detection errors. (b) Similar data for collisions with $^{87}$Rb atoms in the $|F=2,m_F=2\rangle_\mathrm{a}$ state and a total of 19 000 measurements. (c)-(e) Zeeman-resolved detection within the $F=1$ manifold after preparation in $|F=1,m_F=0\rangle_i$ (atoms in $|F=2,m_F=2\rangle_\mathrm{a}$) with 1200 measurements per Zeeman state. The measurements are performed by applying resonant $\pi$-pulses, exchanging the population of the dark state $|F=0,m_F=0\rangle_i$ with $|F=1,m_F=-1\rangle_i$, $|F=1,m_F=0\rangle_i$ or $|F=1,m_F=1\rangle_i$ immediately before the detection of the probability of the ion being in the dark state $p_\mathrm{Dark}$. From~\cite{RatschbacherPRL13}.}
\label{fig3_RatschbacherPRL13}
\end{figure}

Apart from measuring the collision rates for different hyperfine levels, it is an intriguing idea to treat the ion as a qubit and measure its decoherence due to the atomic bath. This was achieved by~\cite{RatschbacherPRL13} using $^{174}$Yb$^+$, which is an ideal two-level system due to its vanishing nuclear spin. It turned out that an initially polarized ion decayed into a mixed state within only a few Langevin collisions, both for the case when spin-exchange collisions were allowed (Rb was prepared in $F=1,m_F=1$ state) and when they were forbidden (which is the case for the fully stretched state $\left|F=2,m_F=2\right>_{\rm Rb}$ and $m_f=1/2$ for Yb$^+$). This indicates the prominent role of the second-order spin-orbit coupling in the Yb$^+$+Rb system and was confirmed theoretically by~\cite{TscherbulPRL15}. Such spin-orbit interactions seriously impede on some of the applications envisioned for ion-atom systems. This is particularly true for applications in quantum information science as the quantum information is usually stored in the internal spin states of the ions and atoms. \cite{RatschbacherPRL13} studied also the more complicated case of $^{171}$Yb$^+$ which has $F=0$ and $F=1$ states. Their results are presented in Fig.~\ref{fig3_RatschbacherPRL13}. Here the initially populated $\left|F=1,m_F=0\right>_{\rm Yb}$ state decayed exponentially to the ground $F=0$ state if the atoms were in $F_{\rm Rb}=1$ manifold, but for $F_{\rm Rb}=2$ case a nonzero steady-state population of the initial state remained. This could be understood as the effect of large energy release during hyperfine changing collisions which could be transferred both to kinetic and spin degrees of freedom.

In more recent work, spin changing collisions were studied in Sr$^+$ interacting with ultracold $^{87}$Rb atoms~\cite{SikorskyNatCom18} and Yb$^+$ interacting with cold $^6$Li~\cite{Furst2017}. For these systems spin-exchange rates were also found to be on the order of the Langevin collision rates. Spin-relaxation rates were however found  to be significantly smaller than in Yb$^+$+Rb especially for the combination Sr$^+$+Rb, where this rate was found to be 48(7) times slower than the Langevin collision rates~\cite{SikorskyARX18}. Theoretical studies of the spin-exchange processes~\cite{Furst2017,SikorskyARX18,CoteARX18} have revealed the surprising result that spin-exchange rates strongly depend on the difference between the singlet and triplet scattering length even in the mK regime, where many partial waves contribute and thermal averaging has to be applied. By measuring the cross section at relatively high temperatures, one can thus gain information about the $s$-wave regime.
This phase-locking between different partial waves in spin-exchange collisions has important consequences. For Yb$^+$+$^6$Li, for instance, the observed large spin-exchange rate in the mK regime indicates a large difference between singlet and triplet scattering lengths. These results suggest that broad magnetic Feshbach resonances can be expected when the $s$-wave regime of collisions is reached~\cite{TomzaPRA15a,ChinRMP10}. Since it is expected that Yb$^+$+$^6$Li can reach this regime in a Paul trap~\cite{Furst2018}, this result is very encouraging in the quest for observing ion-atom Feshbach resonances. In Sr$^+$/Rb, this effect may be experimentally confirmed in the near future by comparing spin-exchange rates of different Sr isotopes~\cite{SikorskyARX18}. The origin of the phase-locking lies in the short-range nature of the spin-exchange interaction, which makes it insensitive to the shape of the centrifugal barrier~\cite{SikorskyARX18,CoteARX18}, in a similar spirit as setting constant short-range phases in the quantum defect approach. 
 
\begin{figure}
\includegraphics[width=0.95\linewidth]{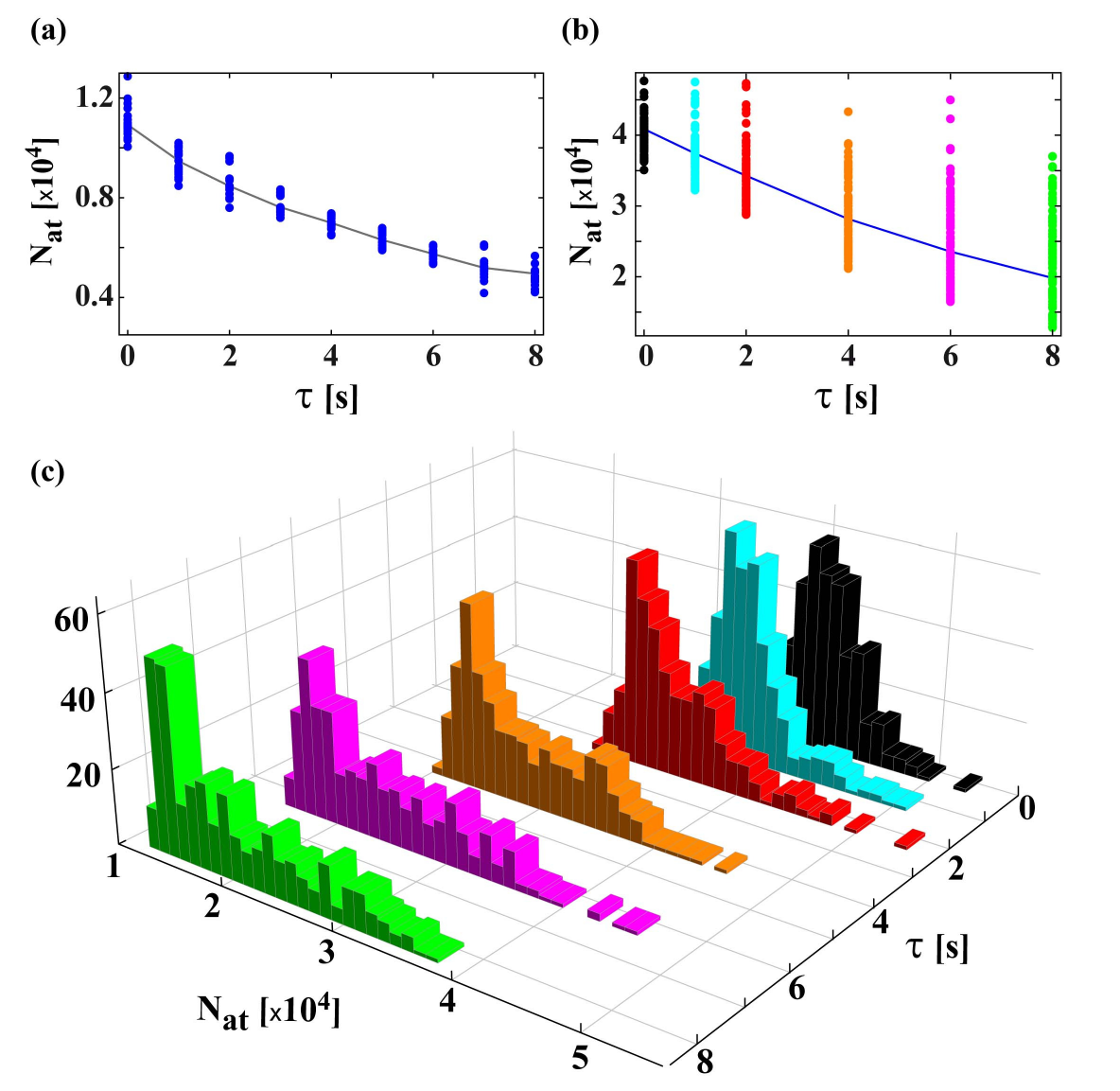}
\caption{\label{fig:secV1_3} Dynamics of the decay of the atomic cloud under the influence of the ion observed in the Rb$^+$+Rb experiment. (a) Remaining atom number as a function of time for ion energy $E\approx 35$mK and initial atomic density $n_{\rm at}=10^{11}\,$cm$^{-3}$. (b) Same but for $E\approx0.5\,$mK and $n_{\rm at}\approx1.1\times 10^{12}\,$cm$^{-3}$. (c) Histograms containing the data from (b). From~\cite{HartePRL12}.}
\label{fig1_HarterPRL12}
\end{figure}

If the atomic cloud has high enough density, three-body processes which happen at the rate $\Gamma=K_3 n^2$ can become important. Three-body collisions involving the ion were first investigated by~\cite{HartePRL12} for the Rb$^+$+Rb+Rb system. Additionally, the ion constantly undergoes two-body collisions which lead to atom loss. Figure~\ref{fig:secV1_3} shows the atomic decay for two different atomic densities. For the dilute cloud with $n\approx 10^{11}\,$cm$^{-3}$ three-body collisions happen at very low rate. At higher densities, the spread in the data becomes much greater. This can be explained by three-body collision which leads to association of Rb$_2$ molecule and large energy transfer to the ion. The ion is then ejected onto a highly excited orbit of the Paul trap, where the atomic density is much lower and slowly recooled back via elastic collisions with the atoms. At the time spent in the low density regions, the ion cannot efficiently remove the atoms. As a result, the atom number distribution develops a tail towards large values which is not seen at low densities. The estimated rate constant for the three-body process is $k_3=3.3(3)\times10^{-25}\,$cm$^6$/s while for two-body collisions the rate coefficient is $k_2=5.0(5)\times10^{-9}\,$cm$^3$/s. Assuming $10^{12}\,$cm$^{-3}$ density gives $\Gamma_3\approx 10^{-1}$s$^{-1}$ while $\Gamma_2\approx 5\times 10^3$s$^{-1}$. Another valuable information is the binding energy of the Rb$_2$ molecule produced in the recombination process. This was estimated by lowering the ion trap depth and checking if the ion is lost or not, giving $\Delta E=0.4(1)$eV.

The techniques described above allowed~\cite{HarterNP13} to perform an experiment, in which spectroscopy was performed on Rb$_2$ molecules that were created by 3-body recombination, thus allowing to probe the population distribution over the molecular states. For this, the molecules were ionized to Rb$_2^+$ followed by ion detection, either via a co-trapped Ba$^+$ ion, or by observing the atom loss as a result of the presence of the ions. This pioneering work in state-to-state product detection  for a neutral reactive process has been extended by~\cite{WolfScience17}, who introduced a two-color ionization scheme that permitted them to measure quantitative branching fractions for producing various rotational levels of the last five most weakly bound states of the Rb$_2$ molecules. Numerical 3-body calculations based on the long-range potentials provided semi-quantitative agreement and aided the development of propensity rules for the distribution of recombination products in this ultracold state-to-state chemistry experiment.

Three-body recombination involving an ion in the mK regime was further investigated using Ba$^+$ ion in Rb gas by~\cite{KrukowPRL16}. By introducing excess micromotion in a controllable fashion, it was possible to measure the energy dependence of the recombination rate $k_3$. Here the measurement relied on fluorescence imaging of Ba$^+$ shortly after placing the ion in the atomic cloud. The total loss rate $\Gamma=k_2 n+k_3(E) n^2$ includes two-body charge transfer for which the rate is independent of energy as well as three-body recombination. Agreement with theoretical predictions of~\cite{PerezJCP15} that $k_3\propto E^{-3/4}$ could only be reached by including the nonthermal energy distribution of the ion in the analysis. In this experiment the peak atomic density $n=1.9\times 10^{12}\,$cm$^{-3}$ was high enough to make three-body loss of the ion the dominant process compared to two-body charge transfer rate. By studying the dependence of the ion loss rate on atomic density~\cite{KrukowPRA16}, it was possible to separate the two- and three-body processes, providing the measurement of the charge transfer rate $k_2=3.1(6)(6)\times10^{-13}\,$cm$^3/$s. At the same time, at the mean ion energy of $2.2(9)\,$mK the three-body rate coefficient was $k_3=1.04(4)(45)\times 10^{-24}\,$cm$^6$/s. These values could only be precisely extracted if the loss of atom number due to elastic collisions during the measurement was taken into account. This analysis shows that three-body loss processes can be non-negligible even at comparatively low atomic densities.

\subsection{Quantum simulation}
\label{seq:quant_sim}

Analog quantum simulation concerns the emulation of the physics of a system of interest, which is typically hard to control, with another system (i.e., the simulator), which is easier to control~\cite{CiracNP12}. An example is to experimentally determine the evolution of a system which is intractable to numerical analysis. The ultimate goal is to possibly identify new states of matter that could then be useful for technological applications. A prototypical example is the long-standing search for high-temperature superconductivity. Contrary to universal quantum computation, which aims at developing a device capable to implement any unitary operation on a large number of quantum bits (qubits), and therefore to perform any many-body quantum Hamiltonian evolution, analog quantum simulation is by definition not universal and aims at specific many-body problems. 
 
\begin{figure*}[tb!]
\includegraphics[width=180mm]{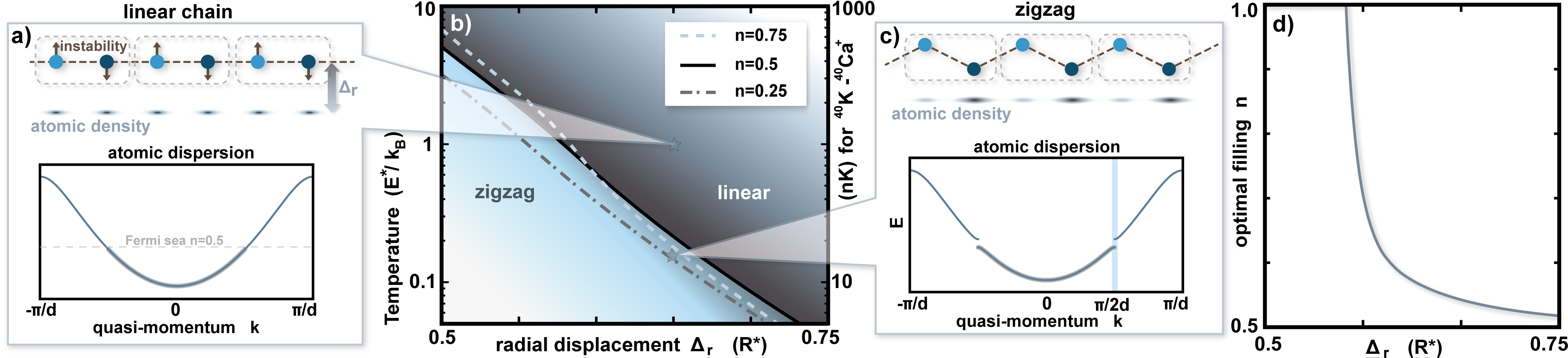}
\caption  {\label{FIG:zigzag} Experimentally tuning the ion trapping allows for the observation of a Peierls-like phase transition when atoms are confined in an effective 1D dipole trap at a transverse separation $\Delta_r$ from the ion string. 
The discrete translational symmetry of the linear chain~(a) is spontaneously broken towards a zigzag pattern of the ions~(c) and a band gap opens at $k=\pi/2d$ for the atoms. At different atomic fillings $n$ (per ion) and below a critical temperature $T_c$, a phase transition occurs~(b). From~\cite{BissbortPRL13}.}
\end{figure*}

Separately, both trapped ions and cold atoms have proven to be excellent platforms for quantum simulations of condensed-matter physics~\cite{BlattNP12,BlochNP12}. Nonetheless, both systems have their advantages and disadvantages, which limit their applicability to investigate a broader range of models. For instance, cold atoms in optical lattices encompass both bosonic and fermionic statistics and are easy to scale up such that thousands of atoms can be manipulated at the same time, but it is harder to achieve long-range interactions and perform measurements. On the other hand, trapped ions can be controlled and their state can be read out superbly well. It is also possible to realize any kind of interactions with them. 
At the same time, however, scalability for trapped ions represents a great challenge and, because of their strong localization when cooled down to form a Coulomb crystal, quantum statistical effects are not visible unless they are emulated by implementing an involved sequence of quantum gates~\cite{CasanovaPRL12}. Hence, the combination of trapped ions and cold atoms can allow to use the best features of the two worlds. 

One particularly appealing idea for an ion-atom quantum simulator is to mimic electrons in a solid with ultracold fermionic atoms in interaction with an ion chain, which represents the crystal structure as proposed by \cite{BissbortPRL13}. Here the ion-atom interaction replaces the Coulomb one between an electron and an ion in the natural material. In Table~\ref{tab1} we illustrate a comparison of different energy and length scales between a typical solid-state and the combined ion-atom system. The parameters such as the ion separation $d$, the energy and length scale of the fermion-ion interaction as well as the corresponding Fermi and phonon energies are vastly different. On the other hand, the ratio $d/R^\star$ is comparable to solid-state systems, and therefore a band structure in an ion-atom system can be expected. For instance, for $^{171}$Yb$^+$ ions separated by 1.1~$\mu$m interacting with a $^6$Li atom one can have band gaps in the 10~kHz range. 
Furthermore, while in solid-state physics the Born-Oppenheimer approximation is a natural way for studying electronic dynamics (e.g., transport), in ion-atom systems with comparable masses (e.g., $^{40}$Ca$^+$+$^{40}$K) this becomes inapplicable. Hence, this scenario might enable to investigate a new regime that solid-state systems do not allow. We note that such interplay between long-range impurity-crystal interactions and variable mass ratio can be also obtained in systems of moving atoms in crystals of polar molecules~\cite{Pupillo2008}.

\begin{table}[b]
\caption{Comparison between a typical natural solid-state system and an ion-atom quantum simulator. The fermionic mass (electron/atom) and ionic core are denoted by $m_{\rm f}$ and $m_{\rm i}$, respectively. From~\cite{BissbortPRL13}.}
\begin{ruledtabular}
\begin{tabular}{llll}
 & Solid state & $^6$Li-$^{174}$Yb$^+$ & $^{40}$K-$^{40}$Ca$^+$ \\
\noalign{\smallskip}\hline
\noalign{\smallskip}
Lattice spacing $d$ (nm) & 0.3-0.6  & 10$^3$-10$^4$ & 10$^3$-10$^4$\\
Length scale $R^\star$ (nm) & 0.026 & 71 & 245\\
Energy scale $E^\star$ (kHz) & $10^{13}$ & 166 & 2.1 \\
$d/R^\star$ & 10-20  & 14-140 & 4-40\\
$m_{\rm i}/m_{\rm f}$ & 10$^{4}$-10$^5$ & $29$ & $1.0$\\
Fermi energy (MHz) & 10$^{8}$ & 0.02 & 0.02\\
Phonon energy (MHz) & 10$^{6}$ & 0.01-10 & 0.01-10 \\
\noalign{\smallskip}
\end{tabular}
\end{ruledtabular}
\label{tab1}
\end{table}
 
In the low-energy limit one can show that a Fr\"ohlich-type Hamiltonian can be derived, where the interaction describing the coupling between the moving atoms and the phonons excited in a homogenous ion chain has the form~\cite{BissbortPRL13} 
\begin{align}
\label{EQ:HAI1}
\hat H_\mathrm{ai}&=\sum_{k,k',s,D} \frac{\lambda_{k,k'}^{(s,D)}}{\sqrt{N_\mathrm{i}}} \, \hat \alpha_s \,  \hat c_k^\dag \, \hat c_{k'}^{\phantom{\dag}} \; + \; \mbox{H.c.}
\end{align}
Here $N_\mathrm{i}$ is the number of ions, $\hat c_k$ denotes the annihilation operator for an atom in a Bloch-state $\phi_k(x)$ of quasi-momentum $k$, and $\hat \alpha_s$ is the $s$-th phonon annihilation operator. Further, the atom-phonon coupling is given by
\begin{equation}
\label{EQ:atom_phonon_coupling}
\lambda_{k,k'}^{(s,D)}=4 C_4  \sum_n \frac{v_n^{(s)}-u_n^{(s)}}{\sqrt{2 m_\mathrm{i} \Omega_n /N_\mathrm{i}} } \int \hspace{-1mm} d^3 \br \, V_n(\br) \, \phi_k^*(\br) \, \phi_{k'}(\br),
\end{equation}
with $V_n(\br) ={(\br - \overline \bR_{j}) \cdot \mathbf{e}_{D}}/{|\br - \overline \bR_{j}|^6}$ and $D=x,y,z$, where $u_n^{(s)}$, $v_n^{(s)}$ are the phonon mode coefficients, and $\Omega_n$ is the bare phononic angular frequency. Here $\br$ denotes the atom position, whereby $\overline\bR_j$ the equilibrium position of the $j$-th ion. 

As an exemplary application of this system as a quantum simulator,~\cite{BissbortPRL13} considered the Peierls-type transition~\cite{Peierls:1991}. The basic idea  here is that it can be energetically favorable for a 1D crystal to undergo a transition 
from a linear to period-doubled zigzag arrangement when cooled below a critical temperature $T_c$. This is also manifested by opening a gap in the band structure, which means that the chain becomes an insulator. The energy that the linear ion string gains, as it is distorted into the zigzag configuration, is offset by decreasing the energy of the fermions. A sample band structure calculation for non-interacting ${}^{40}$K atoms near a $^{40}$Ca$^+$ ion string is shown in Fig.~\ref{FIG:zigzag}c, where a zigzag displacement of about $0.019$~$R^\star$ minimizes the total energy of the ion-atom system with phonon frequency 
$\omega_{k=\pi/d,x} \approx 2\pi\times60$~kHz, ion spacing $d=2$~$\mu$m~$\approx8$~$R^\star$ and atomic temperature $T_\mathrm{a}=15$~nK~$\approx 0.15$~$E^\star / k_B$. The phase diagram is shown in Fig.~\ref{FIG:zigzag}b for varying temperature and ion-atom trap separation. 
Interestingly, unlike the Peierls effect in solid-state materials, the optimal filling that generates the largest $T_c$ can be larger than $n=1/2$ (see Fig.~\ref{FIG:zigzag}d). 
We note that such phase transition can also be quantum simulated with atomic fermions interacting with bosons trapped in deep optical lattices by employing Feshbach resonances for generating large scattering lengths~\cite{Lan2014}. 
An interesting future perspective of the electron-phonon emulation with ion-atom systems is not only the investigation of the impact of the ionic micromotion on the atom-phonon coupling, but also the micromotion-induced coupling amplification, similarly to the parametrically enhanced superconductivity observed in compound solid materials~\cite{Mitrano2016} and theoretically described by phononic Floquet side bands~\cite{Babadi2017,Murakami2017}.

\begin{figure}[t!]
\includegraphics*[width=0.9\linewidth]{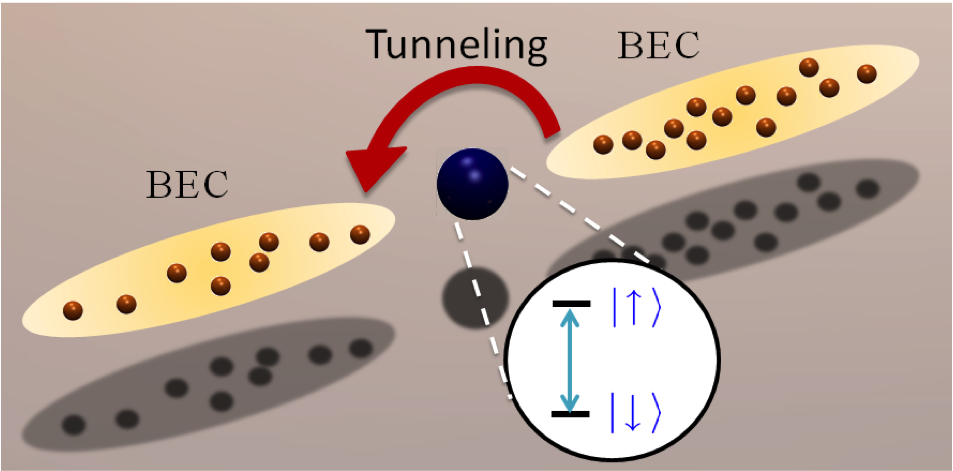}
\caption{Sketch of the atomic bosonic Josephson junction controlled by the internal spin state of a tightly trapped ion (large blue sphere).}
\label{fig:secIIIc03}
\end{figure}

A further example in which a single ion can be used to control the dynamics of a degenerate quantum gas is the bosonic Josephson junction. In this setup a weakly interacting ultracold Bose gas is confined in a double well potential. The most basic model of this system can be described in terms of just two parameters: the onsite energy $U$ and the hopping rate $J/\hbar$. The mean-field theory predicts that, if the inequality
$\Lambda = U N/(2J) >\Lambda_c = 2[(\sqrt{1 - z(0)^2}\cos[\phi(0)] + 1)]/z(0)^2$ is satisfied, the so-called macroscopic quantum self-trapping occurs~\cite{SmerziPRL97}. Here $z(0)$ is the initial population imbalance and $\phi(0)$ is the initial relative phase between the condensates 
in the left and right well. This effect is due to the interaction-induced detuning between the left and right well, which results in a strong suppression 
of the tunnelling between the two wells.
This process can be controlled by putting a tightly trapped ion in the centre of the double well (see Fig.~\ref{fig:secIIIc03}), which changes the tunnelling dynamics in a way which depends on its internal state. 
As shown in Sec.~\ref{sec:theor}, the short-range interaction contains a state-dependent part, which sets the ion-atom scattering length and influences the two-body wavefunction. In such a way the dynamics of the many-body system can be controlled by manipulating the state of the ion, e.g., with a laser field. This setting has been investigated by~\cite{GerritsmaPRL12} both at the single-particle and the many-body level. The latter has been analyzed within the two-mode approximation in which the bosonic quantum field operator has been expanded into the time-independent single-particle wavefunctions obtained as linear combinations of the ground and first excited state of the double well potential. Furthermore, a protocol for entangling the motional state of the condensate with the internal state of the ion has been proposed. A more detailed numerical analysis performed by using the multiconfigurational time-dependent Hartree method for bosons has confirmed the validity of the scheme for short operation times~\cite{SchurerPRA16} with finite temerature effects studied by~\cite{Ebgha2019}. At longer times, however, an additional mode (i.e., single particle orbital) was found to be significantly populated, rendering the two-mode description inadequate.
We note that a similar level of control over the dynamics can be also attained by controlling the motional state of the ion (i.e., phonon number in the ion trap), even though so far this analysis has been carried out at the single atom level only~\cite{JogerPRA14}.

The bosonic Josephson junction can be seen as a building block of a larger system in which an array of tightly trapped ions is used to control the dynamics of atoms in an optical lattice, e.g. by manipulating the ion internal state and therefore the atom hopping, as discussed previously. The ions can be addressed individually, which creates a unique opportunity of engineering the Hamiltonian of the atoms in a nontrivial way, both in space and time.~\cite{NegrettiPRB14} has shown that for an infinitely long and static ion chain, the ion-atom interaction forms a periodic potential for the atoms and 
their dynamics is described by an extended Hubbard Hamiltonian of the form
\begin{align}
\label{eq:BHsigma}
\hat H &=-\sum_k \hat J_ k\hat b^{\dag}_{k+1}\hat b_k+\frac{1}{2}\sum_k\hat U_k\hat n_k(\hat n_k -1) + \sum_k\epsilon_k\hat n_k
\end{align}
where $\hat{J}_k=J_{\downarrow,k}|\downarrow_k\rangle \langle \downarrow_k|+J_{\uparrow,k}|\uparrow_k\rangle \langle \uparrow_k|$, 
$\hat{U}_k =\sum_{\alpha,\beta=\uparrow,\downarrow}U_{\alpha,\beta;k}\vert\alpha_k\beta_{k+1}\rangle \langle \alpha_k\beta_{k+1}\vert$ 
with $\vert\uparrow_k\rangle,\,\vert\downarrow_k\rangle$ being the internal states of the ion at site $k$. Here both the hopping $\hat{J}_k$ and the onsite interaction $\hat{U}_k$ are replaced by operators, because of the aforementioned state-dependence. 
Although similar Hamiltonians have been derived for other systems, especially for ultracold atoms in optical lattices, the one given by Eq.~(\ref{eq:BHsigma}) results from the admixture of both $s$-wave and $p$-wave interactions among the scattering centres (the ions) and the moving atoms. With non-dipolar neutral atoms strong $p$-wave interactions are difficult to engineer, as they require $p$-wave Feshbach resonances. Hence, this offers additional controllability to the ion-atom system with respect to neutral atomic ones. In particular, the possibility to control independently the hopping and the onsite terms via optical control of the ion internal state is an interesting alternative to neutral-atom systems, where more elaborated schemes have been devised to engineer such state-dependent couplings. This enables the exploration of the physics of lattice models and entanglement generation between the moving particles and the scatterers in such hybrid systems similarly to~\cite{Ortner2009}. Furthermore, we note that the Hamiltonian~(\ref{eq:BHsigma}) resembles the one of an atomic ensemble in interaction with a field cavity mode. Such an interaction enables cavity-mediated long-range atom-atom interactions~\cite{MaschlerPRL05}. In the ion-atom system considered here, the quantum potential is provided by the ion-atom interaction where the atomic back-action on the quantum lattice potential may generate ion-atom entanglement via phonons~\cite{JogerPRA14,BissbortPRL13}. In addition to this, we note that with this setup one can also study the analogue of a single atom transistor~\cite{MicheliPRA06}, where one ion of the chain can eventually suppress the atomic tunnelling via the state-dependent ion-atom interaction, as 
well as quantum simulation of lattice gauge models. Indeed, as shown in~\cite{DehkharghaniArXiv17}, by engineering an energy offset in the double well of the $k$-th atom resonant to the Rabi frequency coupling the two ion internal states, one can shown that, within the rotating-wave approximation, the hopping term in Eq.~(\ref{eq:BHsigma}) can be recast into $J_{k}^{z} \hat b^\dag_{k}  \,\tilde\sigma^+_{k,k+1} \hat b_{k+1} + \text{H.c.}$ with $J_{k}^{z}=(J_{\uparrow,k} - J_{\downarrow,k})/2$, which has the desired local gauge invariance. Here, $\sigma^{+} = |+\rangle \langle -|$ with $\vert\pm\rangle$ being the eigenstates of the Pauli matrix $\hat\sigma_x$, whereas $\vert \uparrow\rangle$, $\vert\downarrow\rangle$ are the eigenstates of $\hat\sigma_z$. Such an approach to quantum simulation of lattice gauge models can offer interesting advantages compared to purely neutral settings~\cite{BanerjeePRL12}, where the boson-boson and fermion-boson interactions are chosen to be of the same order to fulfill gauge invariance. 

We conclude by mentioning that an exciting research direction with the ion-atom system is the quantum simulation of the Fermi and the Bose polaron, namely an impurity in either a fermionic or bosonic quantum bath, respectively (see Sec.~\ref{sec:impurity}). Polarons are attracting increasingly large interest in the ultracold quantum matter community~\cite{Casteels2013,Levinsen2015,SchmidtPRL16,Shchadilova2016,Ardila2016,JorgensenPRL16,HuPRL16,Parisi2017,CamargoPRL18,SchmidtPRA18,Ashida2018,Guenther2018} as the simplest example of nontrivial quantum field theory. In this arena, ion-atom systems could provide new insight into the problem, especially in the strong coupling regime~\cite{CasteelsJLTP11,TemperePRB09,Grusdt2017,SchurerArxiv16}, as well as allow to study the effects of long-range interactions.

\subsection{Quantum computation}
\label{subsec:gate}

\begin{figure}
\includegraphics[width=0.9\linewidth]{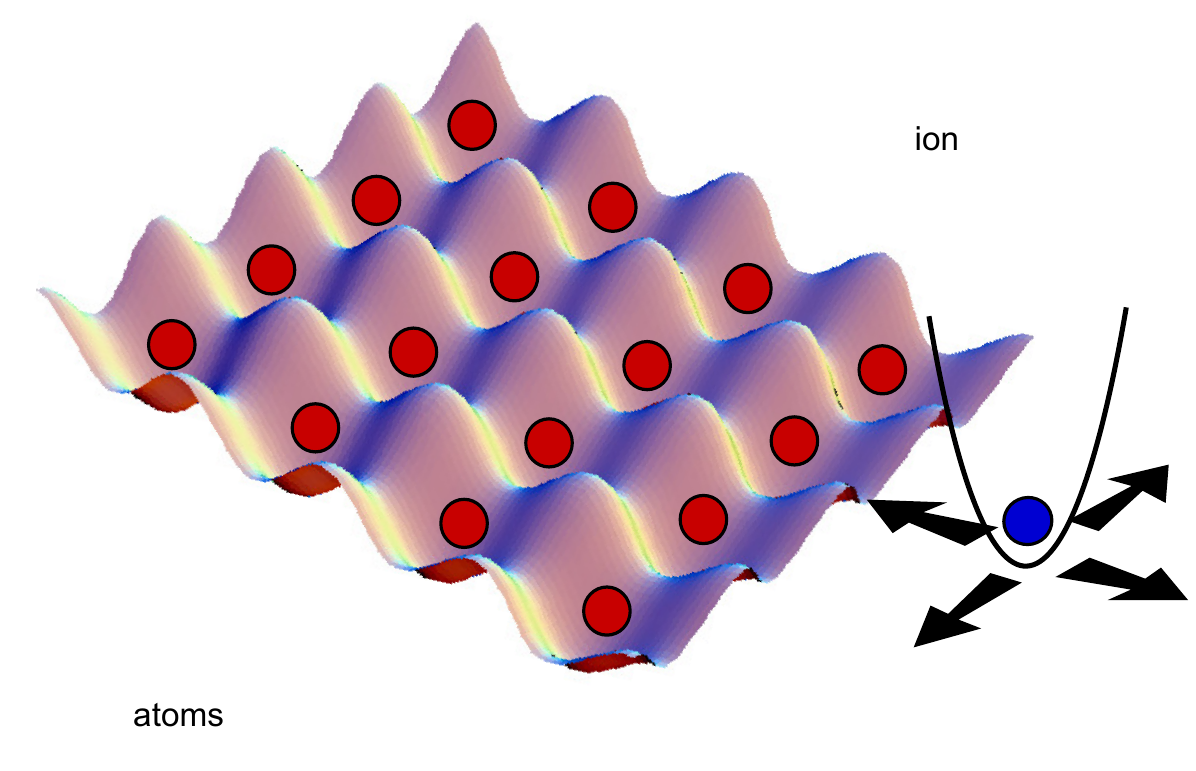}
\caption[Concept for Quantum Information with Atoms and Ions]{Concept for quantum computation with
atoms and ions: Atoms are prepared in an optical lattice in a Mott insulator phase. A movable ion is trapped in radio-frequency trap. From \cite{DoerkPRA10}.}
\label{fig1_DoerkPRA10}
\end{figure}

An important motivation for the interest in ultracold systems of trapped atoms and ions resides in their possible applications to quantum information processing. Among many possible physical implementations of the quantum computational schemes~\cite{CiracPRL95,LossPRA98,BarendsNP14,Zu2014,Nigg2014,Barends2016,Debnath2016}, hybrid systems of ultracold atoms and ions offer an attractive platform, potentially allowing to combine the long coherence times of neutral atoms with the short gate-operation times for charged particles due to the relatively strong interactions.

One of the possible setups for quantum computation with ultracold atoms and ions is presented schematically in Fig.~\ref{fig1_DoerkPRA10}~\cite{DoerkPRA10}. Here, the atoms are stored in an optical lattice in a Mott insulator phase such that each lattice site is occupied by exactly one atom. One movable ion confined in rf trap is used as a read/write head to create
long-distance entanglement between pairs of atoms and to perform quantum gates. The realization of the quantum gate is based on the internal state sensitive interaction between a single atom and a single ion.

By approximating the trapping potentials as harmonic, the spatial part of the system is effectively described by the Hamiltonian of Eq.~\eqref{H3D}.  
The qubit states are encoded in different hyperfine states of the atom and the ion, which in general provides state-dependent dynamics as long as the short-range phases (scattering lengths) are different for different scattering channels representing internal states.

\begin{figure}[tb!]
\center
\includegraphics[width=0.23\textwidth]{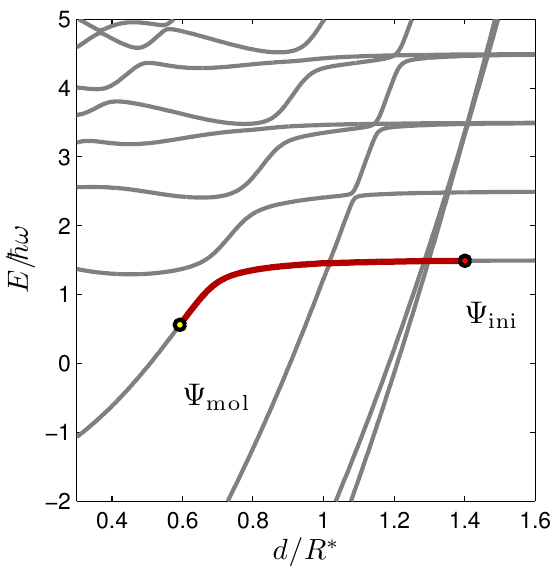}
\includegraphics[width=0.23\textwidth]{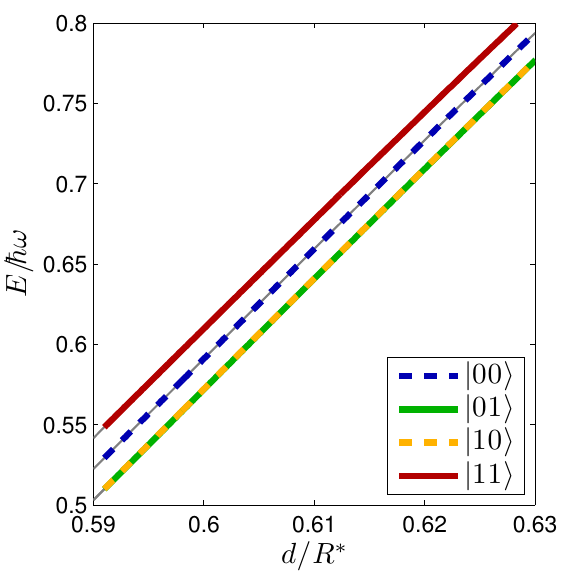}
\caption[Energy Curve in the Adiabatic Gate Process]{Energy spectrum for the $^{135}$Ba$^+$ ion and the $^{87}$Rb atom with the singlet and triplet scattering lengths: $a_s=0.90\, R^\star$ and $a_t=0.95\, R^\star$, respectively. Left panel shows the complete diagram for the $|11\rangle$-channel, while the right panel presents the close-up around $d=0.6\, R^\star$ for all the channels of the computational basis. Small differences between channels that can be observed are the basis for realizing the ion-atom phase gate. From \cite{DoerkPRA10}.}
\label{fig5_DoerkPRA10}
\end{figure}

The idea behind the quantum gate can be understood by considering the adiabatic dynamics and the correlation diagrams like Fig.~\ref{Fig:spectrum3D} and Fig.~\ref{fig5_DoerkPRA10}, connecting the eigenstates at small and large trap separations. The scheme is based on the adiabatic transfer from an initial oscillator state $\Psi_{\rm{ini}}$ to a molecular state $\Psi_{\rm mol}$, and back to the initial state
(see Fig.~\ref{fig5_DoerkPRA10}), which is achieved by variation of the relative trap distance $\mbf{d}(t)$.
During the transfer process each logical basis state acquires a different phase, since the energies of molecular states depend on the channel (see Fig.~\ref{fig5_DoerkPRA10}).
The particular shape function $d(t)$ allowing for perfect gate operation without losses can be found using optimal control algorithms \cite{DoerkPRA10}.
For the specific system of the $^{135}$Ba$^+$ ion and the $^{87}$Rb atom, the gate operation  of $346$ $\mu$s at fidelity $F=0.999$ fidelity has been predicted~\cite{DoerkPRA10}. In this scheme, radiative charge transfer losses as well as spin changing collisions are detrimental since the lead to occupation of states which are outside the computational subspace.

The above model was based on neglecting the micromotion effects and center of mass-relative motion coupling. The validity of these approximations has been studied in detail by~\cite{NguyenPRA12} within the Floquet framework. They found multiple avoided crossings with small energy gaps in the quasienergy spectrum below the critical separation, indicating that the proposed scheme would need to be done on longer time scales in order to avoid unwanted excitations.

The micromotion problem can be circumvented if the atom and ion are kept far enough apart to avoid the couplings. If their interaction is sufficiently strong, the gate implementation should still be possible. For this reason,~\cite{SeckerPRA16} investigated the case where the atom is dressed with the Rydberg state, which enhances its polarisability and makes the interaction stronger by many orders of magnitude. This enables the realization of the quantum gate of the M\o{}lmer-S\o{}rensen type~\cite{MolmerPRL99} with ms timescales and high fidelity, even when micromotion and motional state imperfection are taken into account.

\subsection{Probing quantum gases}
\label{subsec:probing}

\begin{figure}[tbh!]
\begin{center}
\includegraphics*[width=0.75\linewidth]{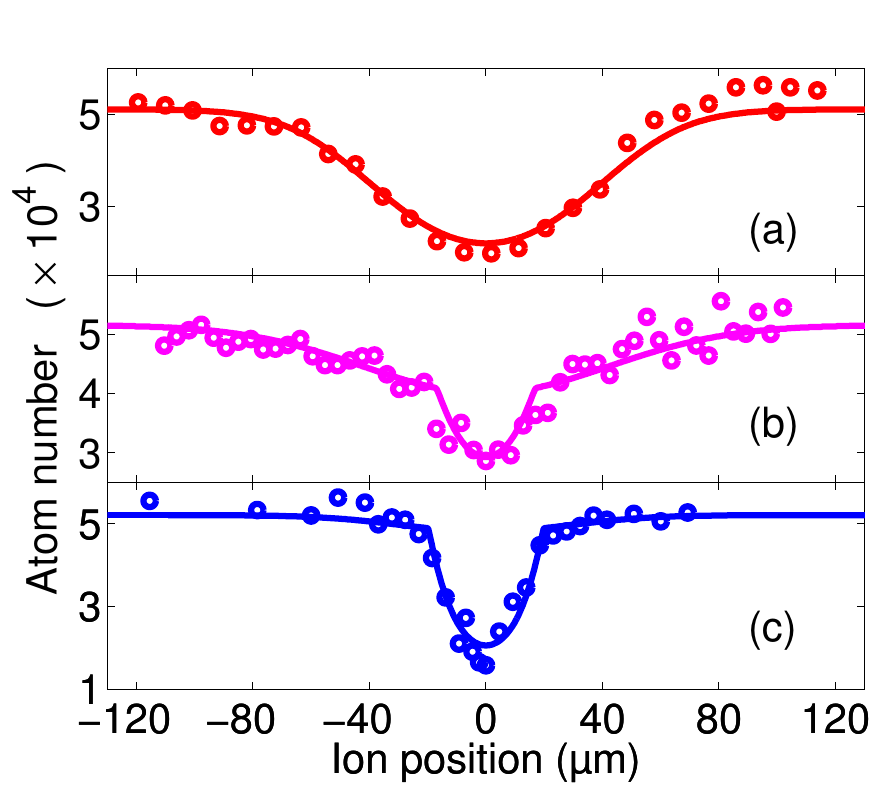}
\end{center}
\caption{Number of remaining $^{87}$Rb atoms as a function of the $^{87}$Rb$^+$ ion position relatively to the trap center: (a) thermal cloud; (b) partially condensed cloud; (c) an almost pure condensate. Solid lines correspond to fits, where the ion energy and atom temperature have been used as free parameters. From~\cite{SchmidPRL10}. 
}
\label{fig:secV03-2}
\end{figure}
\begin{figure}[tb!]
\begin{center}
\includegraphics*[angle=-90,width=0.85\linewidth]{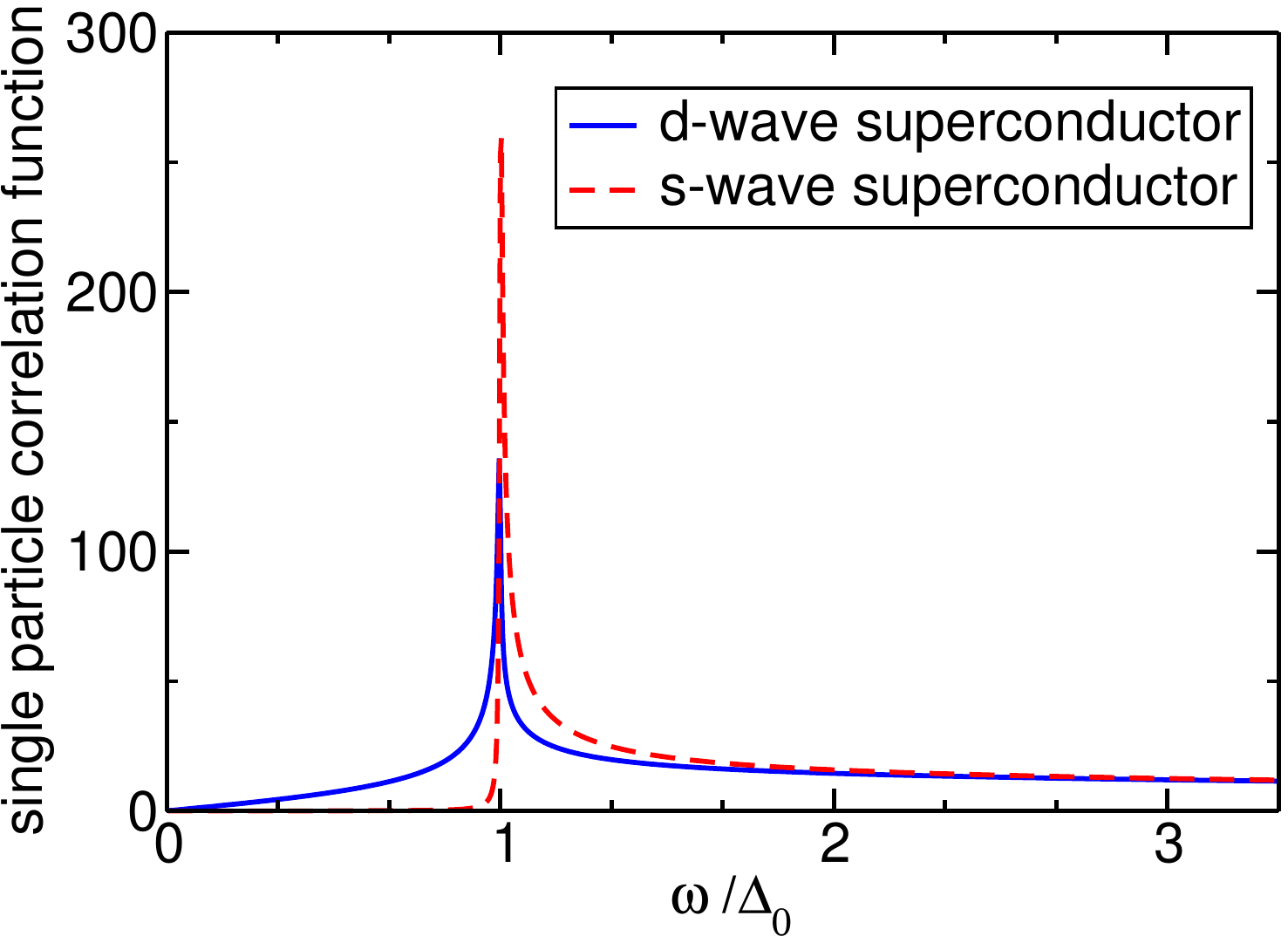}
\end{center}
\caption{Fourier transform of the temporal single-particle Green function $\langle \hat c_{\sigma,j}^\dag(t)\hat c_{\sigma_j}(0)\rangle$ for an s-wave and a d-wave superconductor. From~\cite{KollathPRA07}.
}
\label{fig:secV03-1}
\end{figure}

Another interesting application of the ion-atom systems is related to precision measurements. In particular, the ion can be used as a tool for detecting the spatial density profiles as well as density-density correlations and single particle Green functions of quantum degenerate atomic gases with the aim of inferring the nature of its many-body quantum state. Such measurements can be useful for the development of schemes for the preparation of exotic quantum phases. 

First steps in this direction have already been taken by performing measurements of the ion loss rate in different regions of the atomic cloud~\cite{ZipkesPRL10}, providing an estimate of the atomic density. Similar measurement has been demonstrated by~\cite{SchmidPRL10}, where the density profile for a thermal, partially condensed, and an almost pure condensate has been directly measured by looking at the number of remaining atoms as a function of the ion position relative to the atom trap centre. As Fig.~\ref{fig:secV03-2} clearly shows, all regimes are well distinguishable and in good agreement with theoretical models.  

\begin{figure*}[bt!]
\includegraphics*[width=0.3\linewidth]{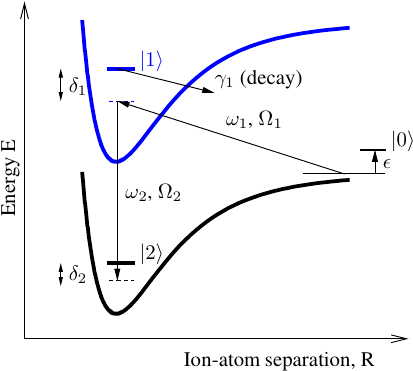}
\includegraphics*[width=0.4\linewidth]{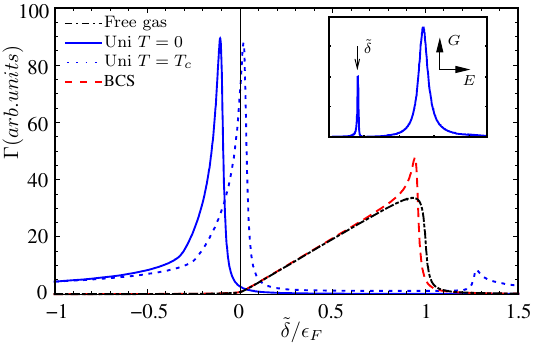}
\caption[]{Left panel: The Raman level scheme used for coupling an atom and an ion with initial energy $E$ such that a stable ion-atom molecule can be prepared with the aim of inferring the single-particle energy distribution of a Fermi gas. For further details see the text. Right panel: Photoassociation rate for an ion in an atomic Fermi gas as a function of the detuning $\tilde{\delta} = \delta_2 - \eta$ for different regimes of the gas: Free fermions at $T=0$ (dashed-dotted line), BCS regime at $T=0$ for $k_F^0 a_s^{\text{aa}} = -0.5$ (energy gap $\Delta_0 = 0.047 \epsilon_F$) with $k_F^0$ being the Fermi wavenumber and $a_s^{\text{aa}}$ the scattering length (dashed line), the unitarity limit at $T=0$ (solid line) and at $T=T_c$ (dotted line). The major peak is due to the breaking of fermionic pairs of atoms, while the minor one is attributed to thermally excited atoms. The inset shows the photoassociation rate density $G(\epsilon,E)$ (at a fixed initial ion energy) for the transition $\vert 0\rangle\rightarrow \vert 2\rangle$ 
for $\delta_2 = 2.5\delta_1$. The low peak is determined by the width $\gamma_2$ and can be controlled by the laser parameters (see text), whereas the maximal peak occurs when $E=\tilde{\delta}$. From~\cite{SherkunovPRA09}.
}
\label{fig:secV03-3}
\end{figure*}

Given the high controllability of the ion motion in a Paul trap, the coupling of a single ion to the atomic quantum system can even allow the spatial resolution to reach the nanometer scale. For instance, by assuming that atoms with different spin states $\vert\sigma\rangle$ trapped in an optical lattice are prepared in some many-body state $\vert \Psi_0\rangle$, the following protocol proposed by~\cite{KollathPRA07} can be implemented. Firstly, the ion in the state $\vert i\rangle$ is brought close to the $j$-th lattice site that we want to investigate.  A Raman pulse is then applied, coupling the ion-atom state $\vert a_j^\sigma,i\rangle$ to the weakly bound molecular ion $\vert a_j^\sigma+i\rangle$. The proper Hamiltonian describing this process is $\hat H_{\text{int}} =\sum_{\sigma}\Omega_{\sigma}\hat M_{\sigma}^\dag \hat I\hat c_{M_{\sigma},j} +$ h.c., where $\hat M_{\sigma}$ and $\hat I$ denote the annihilation operators for the molecular ion and the atomic ion, respectively, whereas $\Omega_{\sigma}$ is the effective two-photon Rabi frequency and $\hat c_{\sigma,j}$ is the annihilation operator of an atom in the spin state $\vert\sigma\rangle$ at the $j$-th lattice site.
The population of the molecular state is simply given by
$\langle\sum_{\sigma}\hat M_{\sigma}^{\dag}\hat M_{\sigma}\rangle = \sum_{\sigma}\sin^2(\Omega_{\sigma}\delta t)\langle\hat n_{\sigma,j}\rangle$ where $\delta t$ is the pulse duration. The molecular ion can be detected e.g. by measuring the oscillation frequency in the Paul trap. Such measurement process (photoassociation and detection) would take at least several hundred $\mu$s. Typical coherence times of a quantum gas experiment are on the hundred millisecond scale so tens of lattice sites could be measured sequentially.

A similar procedure with multiple ions could be utilized for density-density correlation measurements. Besides this, by using more elaborated sequences of Raman pulses with appropriated timings $\delta t$, one could in principle measure the difference $\Delta\langle \hat M_{\sigma}^{\dag}\hat M_{\sigma}\rangle$ corresponding to two separate molecule formations. As it can be analytically shown~\cite{KollathPRA07}, this quantity is proportional to the temporal correlation function $\langle \hat c_{\sigma,j}^\dag(t)\hat c_{\sigma_j}(0)\rangle$ (i.e, the single particle Green function), which reflects the nature of the system excitations. By computing the Fourier transform of $\langle \hat c_{\sigma,j}^\dag(t)\hat c_{\sigma_j}(0)\rangle$, one could in principle distinguish between different many-body quantum phases. In Fig.~\ref{fig:secV03-1}, we display as an example the Fourier transform of an s-wave and a d-wave superconducting state, which exhibit a different behavior at $\omega = \Delta_0$ which is the gap frequency. Thereby, such a measurement enables the understanding of the structure of the corresponding order parameter as well as the determination of the energy gap $\hbar\Delta_0$.

The scheme can also be adapted to measurements of the local single-particle energy distribution of a degenerate Fermi atomic gas.~\cite{SherkunovPRA09} have shown that when using the level scheme displayed in Fig.~\ref{fig:secV03-3} (left panel), the photoassociation rate of an atom and an ion is given by
\begin{equation}
\Gamma(\mathbf{R},E) = \frac{(2 m_\mathrm{a})^{2/3}}{3\pi} R^3_{\text{TF}} \int d\epsilon\sqrt{\epsilon}\,G(E,\epsilon) n_{\sigma}(\epsilon,\mathbf{R})\,,
\end{equation}
where
\begin{equation}
G(E,\epsilon) = \frac{\gamma_2\Omega^2(\epsilon)}{(\delta_2 - \eta - E)^2+\gamma_2^2/4}\,.
\end{equation}
Here $R_{\text{TF}}$ denotes the Thomas-Fermi radius of the gas in a harmonic trap, $E$ is the atom kinetic and ion trap energy of the combined (not molecular) atom and ion state $\vert 0\rangle$, $n_{\sigma}(\epsilon,\mathbf{R})$ is the momentum distribution of the atoms at the position $\mathbf{R}$ with energy $\epsilon = p^2/(2 m_\mathrm{a})$ and with spin index $\sigma$. The effective two-photon Rabi frequency $\Omega^2(\epsilon) = \Omega_1(\epsilon)\Omega_2/\delta_1$ is proportional to the corresponding Frank-Condon factors. The latter rely on the atom and ion wavefunctions, and therefore on the initial incoming atom energy $\epsilon$. The Rabi frequencies $\Omega_{1,2}$ are also proportional to the laser powers of the two light beams used to perform the transitions from the initial ion-atom state $\vert 0\rangle$ to the intermediated excited molecular state $\vert 1\rangle$ and from this to the final target molecular ground state $\vert 2\rangle$. By choosing properly the detunings $\delta_{1,2}$ of the two laser beams one can perform the transition $\vert 0\rangle\rightarrow \vert 2\rangle$ with negligible population of the state $\vert 1\rangle$ (i.e., only virtual transitions), whose lifetime is $\gamma_1^{-1}$. Besides this, the interaction of the laser fields with the ion-atom system results in an optical Stark shift $\eta = \Omega_2^2/\delta_1$ as well as a broadening of the molecular ground state $\gamma_2 = \Omega_2^2\gamma_1/\delta_1^2$. Here it has been assumed that $\Omega_1\ll \Omega_2$ so that the level shift of the state $\vert 0\rangle$ can be neglected. In Fig.~\ref{fig:secV03-3} (right panel) the photoassociation rate in different regimes of the Fermi gas is illustrated (see caption of Fig.~\ref{fig:secV03-3} for details). By fitting the measured photoassociation rate to the various theoretical models, it should be possible not only to extract the single-particle energy distribution $n_{\sigma}$, but also to determine the gas energy gap as a function of the position $\mathbf{R}$ in order to monitor the system state as it is tuned through the BEC-BCS crossover. 

While the schemes described above could also be implemented with a neutral impurity, the excellent resolution is essentially owed to the controllability of the ion motion in Paul traps. We note that in the case of a Bose condensate, the spontaneous capture process with the rate of $\sim 600$ s$^{-1}$~\cite{CotePRL02} would compete with the photoassociation rate discussed above. In the case of Fermi statistics this process can be safely neglected. 

Recent numerical study of a highly localized dissipative impurity, resembling an ion, interacting with a quasi-1D ensemble of bosonic atoms described by the Bose-Hubbard model and treated as an open Markovian quantum system has demonstrated how the atom loss due to the local impurity can be used to probe and manipulate the properties of the many-body quantum state~\cite{BarmettlerPRA11}. For example, it has been shown how to measure the local density of the atomic system and study impurity-induced excitations 
thereby suggesting a decoupling between the two parts of the bosonic system at the impurity location. 

\section{Formation and applications of cold molecular ions}
\label{sec:formation}

In this section, we present theoretical and first experimental studies of the formation of cold molecular ions in cold hybrid ion-atom systems, both spontaneous as well as controlled with external fields. We also report on the progress in sympathetic cooling of molecular ions and their applications.

\subsection{Spontaneous radiative association}
\label{sec:radiative_association}

In Section~\ref{sec:theor} we showed that the radiative association and charge transfer are the main channels of radiative losses for the ground state alkali-metal and alkaline-earth-metal ion-atom systems if the charge transfer process is energetically allowed. These losses can jeopardize applications of ion-atom systems in quantum simulation and computation. However, the radiative association can be used to produce cold molecular ions and related chemical reactions can be an interesting subject of studies on their own.

The formation of molecular ions in the spontaneous radiative association is governed by the Einstein coefficients  given by Eq.~\eqref{eq:Einstein}. They are proportional to the third power of the transition energy and the second power of the vibrationally averaged transition electric dipole moment. Since the transition electric dipole moment decays exponentially with the ion-atom distance, as discussed in Sec.~\ref{sec:theor}, the vibrationally averaged transition electric dipole moment is a result of two competing factors. When the binding energy of produced molecular ions is increasing, the Franck-Condon overlap between the entrance scattering state and the getting smaller molecular state is decreasing. At the same time, the value of relevant transition electric dipole moment function is the largest at small internuclear distance. As a result the association rates are the largest to relatively deeply bound vibrational states in the middle of the potential well. 

\begin{figure}[tb!]
\begin{center}
\includegraphics[width=0.95\columnwidth]{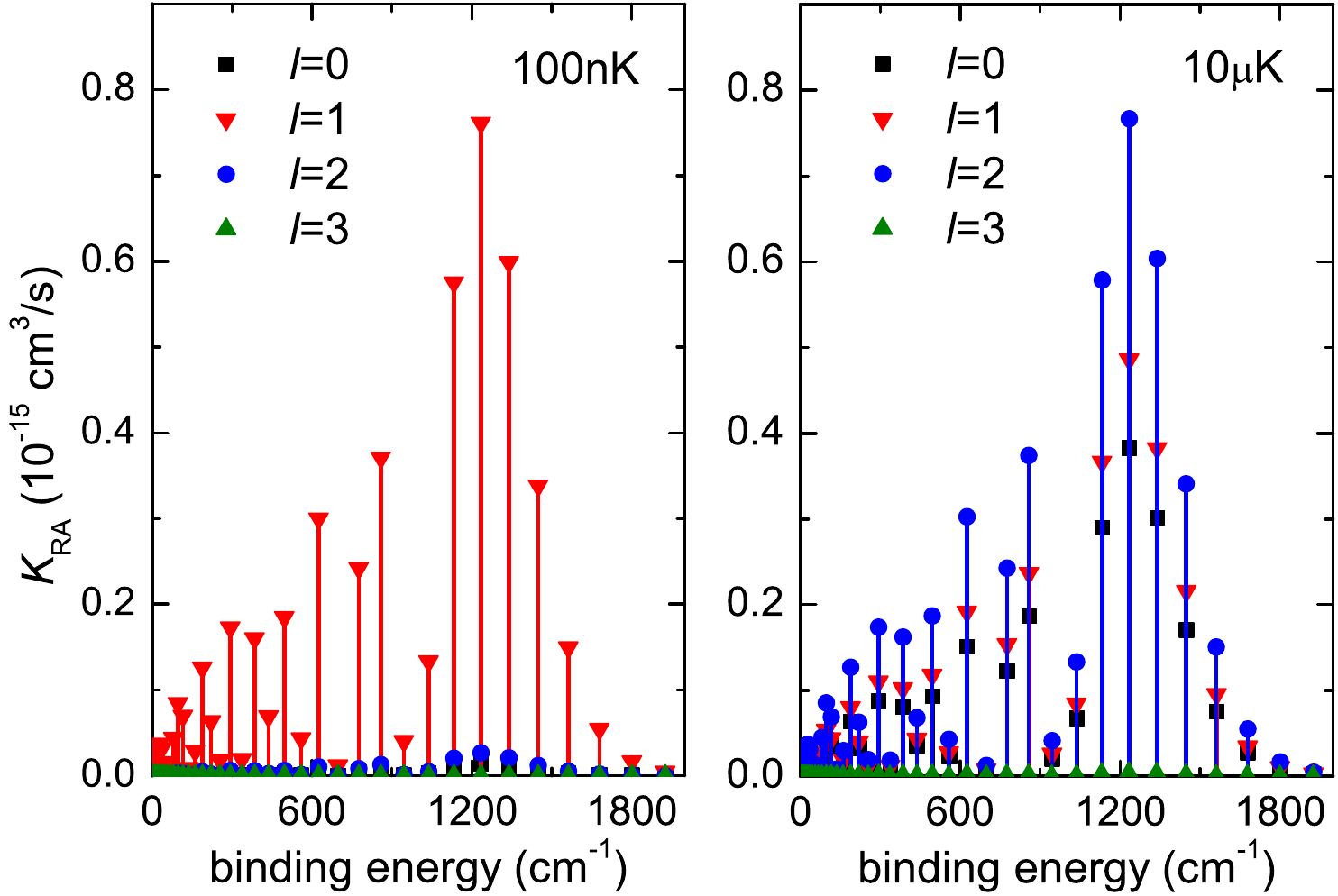}
\end{center}
\caption{Radiative association rate constant vs binding energy of the final ro-vibrational level in the  $X^1\Sigma^+$ electronic ground state for an Yb$^+$ ion colliding with a Li atom in the $A^1\Sigma^+$ electronic state plotted at a temperature corresponding to 100$\,$nK  (left panel) and 10$\,\mu$K (right panel). Rates to states with different rotational angular momentum $l$ are presented using different symbols and colors. From~\cite{TomzaPRA15a}.}
\label{fig7_TomzaPRA15}
\end{figure}

Figure~\ref{fig7_TomzaPRA15} presents the exemplary theoretical state-resolved radiative association rate constants for an Yb$^+$ ion colliding with a Li atom in the $A^1\Sigma^+$ electronic state~\cite{TomzaPRA15a}. The formation of a LiYb$^+$ molecular ion in the $X^1\Sigma^+$ electronic state with vibrational binding energy of around 1200$\,$cm$^{-1}$ is the most probable. The formation of weakly bound molecular ions, as mentioned in the previous paragraph, is suppressed in a characteristic way.
The formation of deeply bound molecular ions with binding energy larger than 2000$\,$cm$^{-1}$ (the depth of the $X^1\Sigma^+$ state is over 9000$\,$cm$^{-1}$) is suppressed by the presence of the repulsive wall of the $A^1\Sigma^+$ potential at internuclear distance significantly larger than the minimum position of the the $X^1\Sigma^+$ potential (cf.~Fig.~\ref{fig:LiYb+_curves}). The comparison of the calculated rotationally-resolved rate constants for two different collision energies reveals that already at temperature of 10$\,\mu$K p-wave collisions are important and result in the formation of molecular ions in a few rotational states.  

\begin{figure}[tb!]
\begin{center}
\includegraphics[width=0.8\columnwidth]{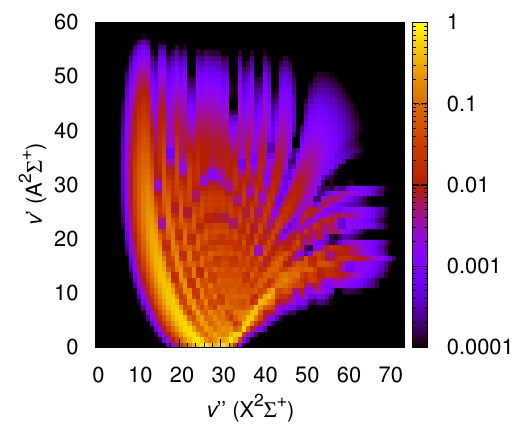}
\end{center}
\caption{Transition electric dipole moments between vibrational levels of the $X^2\Sigma^+$ ground and $A^2\Sigma^+$ excited electronic states of the LiRb$^+$ molecular ion. Adapted from~\cite{Tomza2017}.}
\label{fig_Tomza_LiRb+}
\end{figure}

Figure~\ref{fig_Tomza_LiRb+} shows the exemplary transition electric dipole moments between vibrational levels of the $X^2\Sigma^+$ ground and $A^2\Sigma^+$ excited electronic states of the LiRb$^+$ molecular ion $\langle \Psi_v |d(R) | \Psi_{v'} \rangle$~\cite{Tomza2017}. Such transition dipole moments govern both spontaneous radiative association and photoassociation followed by spontaneous or stimulated stabilization, which will be discussed in the next subsection. The picture we present is characteristic for all ion-atom systems when electronic states related to two well separated thresholds with different charge arrangements are considered. The transition probability between weakly bound vibrational levels of molecular ions is strongly suppressed. 

In Sec.~\ref{sec:theor} and Sec.~\ref{sec:app} we have shown that the spontaneous radiative association leading to the formation of molecular ions is expected to be the main channel of radiative losses in the case of cold collisions of ground-state alkali-metal and alkaline-earth-metal ions and atoms. If the ions or atoms are electronically excited or the entrance electronic state is not well separated from other states, the relative probability of association and charge transfer is system-specific and non-radiative charge transfer can dominate over radiative processes. The experimental investigations of the molecular ions formation were already discussed in Sec.~\ref{sec:app} in the context of ion-atom reactive collisions with the following molecular ions observed as products of cold collisions between respective ions and atoms:  RbCa$^+$~\cite{HallPRL11,HallMP13a}, RbBa$^+$~\cite{HallMP13b}, CaYb$^+$~\cite{RellergertPRL11}, CaBa$^+$~\cite{SullivanPRL12}.

\subsection{Photoassociation}

Cold molecular ions can also potentially be produced using photoassociation methods which have been successfully employed to produce ultracold neutral molecules both in the excited and ground states~\cite{JonesRMP06}. In neutral systems the atom pairs are associated by exciting them with a laser field from a scattering state to a bound molecular level below an excited-state dissociation threshold. Both strong dipole allowed transitions related to the strong atomic transition at the dissociation threshold in alkali-metal atoms and weak dipole-forbidden transitions related to the weak intercombination line transition in alkaline-earth-metal atoms have been employed in neutral systems. Molecules produced in excited states can spontaneously decay to the ground state or can be state-selectively transferred to the ground state by laser stimulated  transitions. 

In general, similar formation schemes can be employed in ion-atom systems with two distinct differences. First of all, the colliding ion-atom pairs can be in an excited electronic state with respect to the charge transfer process allowing for the photoassociation (stimulated emission) directly stabilizing the molecular ions down to the electronic ground state. Second of all, the transition dipole moments to states related to the charge-transfered thresholds are strictly vanishing with increasing internuclear distance, suppressing the probability of exciting weakly bound molecular ions to a much larger extent than in the case of the intercombination line transition (cf.~Fig.~\ref{fig_Tomza_LiRb+}).

\begin{figure}[tb!]
\begin{center}
\includegraphics[width=0.9\columnwidth]{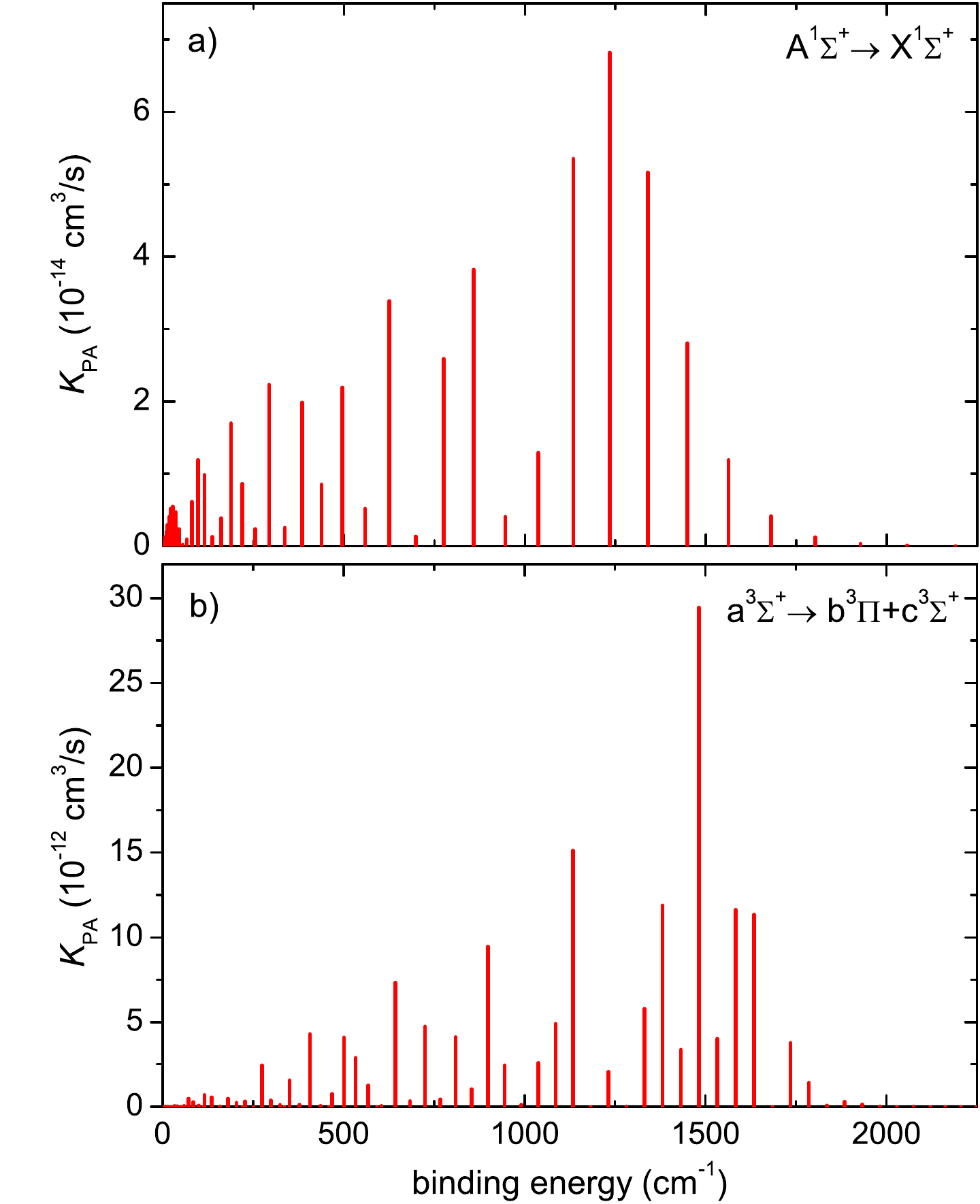}
\end{center}
\caption{Photoassociation rate constants into rovibrational levels of the $X^1\Sigma^+$ electronic ground state for collisions of $^{174}$Yb$^+$ ions with $^{6}$Li atoms in the $A^1\Sigma^+$ state (upper panel) and into  rovibrational levels of the $c^3\Sigma^+$ and $b^3\Pi$ excited electronic states for collisions in the $a^3\Sigma^+$ state (lower panel). From~\cite{TomzaPRA15a}.}
\label{fig9_TomzaPRA15a}
\end{figure}
\begin{figure}[tb!]
\begin{center}
\includegraphics[width=0.9\columnwidth]{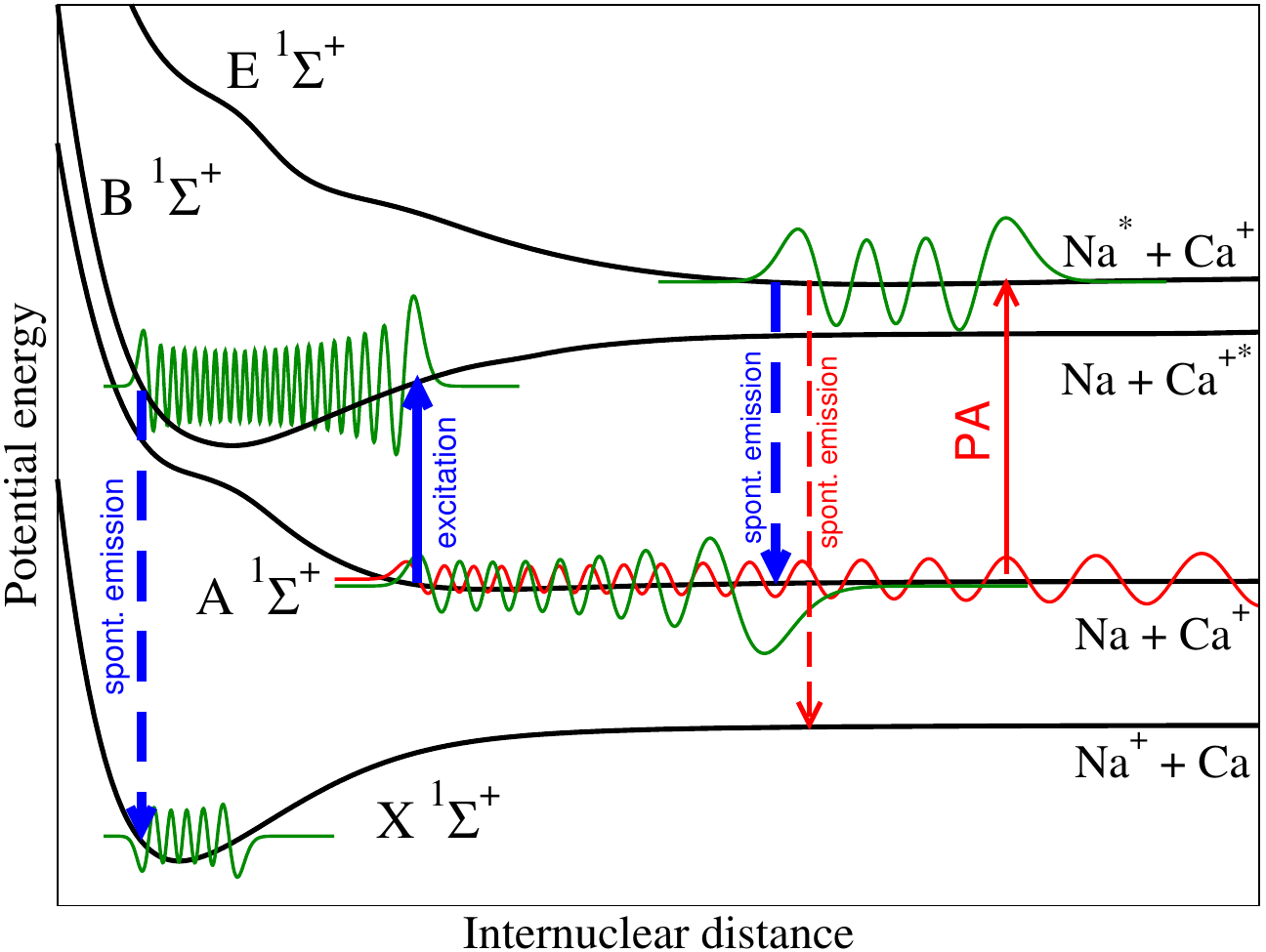}
\end{center}
\caption{Schematic representation of the optical pathways to produce NaCa$^+$ molecular ions in the electronic ground state.  From~\cite{GacesaPRA16}.}
\label{fig1_GacesaPRA16}
\end{figure}

\cite{TomzaPRA15a} investigated theoretically the photoassociation rates for both the above mentioned scenarios in ultracold collisions of Yb$^+$ ions and Li atoms. Figure~\ref{fig9_TomzaPRA15a} presents the results for the photoassociation via stimulated stabilization directly to molecular ions in the electronic ground state, $A^1\Sigma^+\to X^1\Sigma^+$, (upper panel) and the photoassociation to the lowest excited electronic states, $a^3\Sigma^+\to b^3\Pi+c^3\Sigma^+$, (lower panel). The shape of the spectra is typical for photoassociation in ion-atom systems, when the transitions to states related to charge-transfered thresholds are used, i.e.~the rates for the formation of weakly bound molecular ions are suppressed and the formation of relatively deeply bound molecular ions in the middle of the potential well is the most probable. Additionally, the overall shape of the photoassociation spectrum for the singlet symmetry is, as expected, very similar to the spontaneous radiative association spectrum presented in Fig.~\ref{fig7_TomzaPRA15} because they are governed by the same transition dipole moments. 

\cite{GacesaPRA16} studied theoretically the optical pathways for the formation of cold ground-state NaCa$^+$ molecular ions as schematically presented in Fig.~\ref{fig1_GacesaPRA16}. They considered photoassociation of Ca$^+$ ions and Na atoms colliding in the $A^1\Sigma^+$ electronic  state into weakly bound molecular levels in the $E^1\Sigma^+$ excited electronic state. Because the charge configuration for atomic thresholds related to the $A^1\Sigma^+$ and $E^1\Sigma^+$ electronic states is the same, the transition electric dipole moments do not vanish at large distances and this photoassociation is of the same nature as for neutral mixtures with significant formation rates into weakly bound molecular levels. Formed molecular ions spontaneously decay either to the ground electronic state or an intermediate state from which the population can be transferred to the ground state via an additional optical excitation. The efficiency of a two-photon scheme, via either the $B^1\Sigma^+$ or $C^1\Sigma^+$ potential, is sufficient to produce significant quantities of ground-state NaCa$^+$ molecular ions. 

\cite{RakshitPRA11} considered theoretically the photoassociative formation of LiBe$^+$ molecular ions, whereas \cite{SardarJPB16} explored the formation of ground-state LiCs$^+$ molecular ions by photoassociation and stimulated Raman adiabatic passage.

Controlled and selective photoassociation of molecular ions from a cold mixture of ions and atoms has not yet been realized experimentally. However, as we mentioned in the previous subsection, the control of spontaneous radiative association by exciting ions with a laser field was investigated by~\cite{HallPRL11}. \cite{SullivanPCCP11} observed the formation of $^{40}$Ca$_2^+$ molecular ions via photoassociative ionization of ultracold atoms in a hybrid $^{40}$Ca magneto-optical and ion trap system. Additionally, \cite{JyothiPRL16} studied the possibility of simultaneous trapping of $^{85}$Rb$^+_2$ molecular ions (formed by photoionization of photoassociated Rb$_2$ molecules) with ultracold $^{85}$Rb atoms in a magneto-optical trap and found that the photodissociation of Rb$^+_2$ molecular ions  by the cooling light of the MOT is the dominant mechanism for the loss of trapped molecular ions. This mechanism is predicted to be present in all alkali-metal molecular ions and thus implies that the use of a far-detuned optical dipole trap for ultracold atoms instead of a MOT is needed to produce and trap such molecular ions with atoms for a long time.

\subsection{Magnetoassociation}

In principle, magnetically tunable Feshbach resonances described in Sec.~\ref{secIII:Feshbach} can be used not only to tune the scattering properties of ion-atom systems but also can be employed to form weakly bound molecular ions using magnetoassociation in a similar manner as it was realized in neutral ultracold alkali-metal-atom gases~\cite{KohlerRMP06,ChinRMP10}. Potentially both an adiabatic time-dependent sweep of the magnetic field across the resonance and rf field association may work. However, to enable magnetoassociation the ultracold $s$-wave regime of ion-atom collisions has to be achieved, which implies that an optical trap rather than a Paul trap should be used. Additionally, the impact of the Landau quantization effects (the Lorenz force) in ultracold ion-atom collisions~\cite{SimoniJPB11} on the magnetoassociation has to be investigated, especially at stronger magnetic fields. Thus, significant technical developments and more detailed theoretical studies are still needed to anticipate the formation of ultracold weakly bound molecular ions using the magnetoassociation and their subsequent transfer to deeply bound vibrational states using the stimulated Raman adiabatic passage as it is realized for neutral molecules~\cite{QuemenerCR12}.


\subsection{Sympathetic cooling}

Sympathetic cooling, in which cold particles of one type cool particles of another type via elastic collisions, is a very efficient method for cooling trapped mixtures of different ions~\cite{LarsonPRL86} and is successfully used to cool mixtures of neutral atoms~\cite{MyattPRL97}. It is also believed to work for some mixtures of neutral molecules and atoms~\cite{WallisPRL09}. 
The sympathetic cooling of translational motion of ions by cold atoms in hybrid ion-atom systems using a Paul trap is limited by micromotion as we described in Sec.~\ref{sec:micromotion}.

Translationally cold molecular ions can be produced by sympathetic cooling in mixtures with laser-cooled atomic ions~\cite{WillitschPCCP08}. Low-energy collisions between ions are dominated by the Coulomb interaction which does not couple to the internal degrees of freedom. Therefore sympathetically cooled molecular ions can be rotationally and vibrationally warm. In such a scenario, internally warm samples can be further cooled using optical pumping schemes as demonstrated by~\cite{SchneiderNP10,StaanumNP10}. Molecular ions can also be prepared in selected rotational and vibrational states by (2+1) resonance-enhanced multiphoton ionization or threshold photoionization and cooled down translationally without heating internal motion~\cite{TongPRL10,TongPRA11}. However, even for state-selectively produced molecular ions, if they have an electric dipole moment, the blackbody-radiation can induce population redistribution and need for returning population to the rovibrational ground state by optical pumping~\cite{DebJCP14,StaanumNP10}. Another possibility is the use of cold or ultracold atoms to sympathetically cool molecular ions internal degrees of freedom to the rovibrational ground state~\cite{HudsonPRA09,HudsonEPJ16}. 

\cite{RellergertNature13} experimentally demonstrated sympathetic vibrational cooling of translationally cold BaCl$^+$ molecular ions immersed into an ultracold gas of Ca atoms. The quenching of the molecular ion vibrational motion by ultracold atoms was observed at a rate comparable to the classical Langevin rate, being over four orders of magnitude more efficient than traditional sympathetic cooling schemes. \cite{StoecklinNP16} theoretically explained this high efficiency by performing \textit{ab initio} calculations and they suggested that there exists a large class of systems exhibiting such an efficient vibrational cooling provided large polarizability of atoms and strong binding of molecular ions. 

Efficient rotational cooling of Coulomb-crystallized MgH$^+$ molecular ions by a helium buffer gas at cryogenic temperatures was realized by~\cite{HansenNature14}. It was achieved with very low buffer-gas collision rate (four to five orders of magnitude lower than in typical buffer-gas cooling settings), opening the way for translational sympathetic sideband cooling and investigations of laser-induced coherent processes with molecular ions.

In the context of buffer gas cooling, \cite{HauserNatPhys15} measured absolute scattering rates for rotational state-changing cold collisions of OH$^-$ ions with helium at cryogenic temperatures. The full quantum control of the rotationally inelastic collision was achieved by developing a method to manipulate molecular quantum states by non-resonant photodetachment. Reasonable agreement without adjustable parameters was found between experiment and quantum scattering calculations.

Rotational or vibrational quenching in cold and ultracold collisions of molecular ions with neutral atoms have been studied theoretically for several ion-atom systems in the context of cold or ultracold studies: He$_2^+$+He~\cite{BodoPRL02}, N$_2^+$+He~\cite{StoecklinPRA05,GuillonPRA07,GuillonEPJD08,GuillonPRA08}, OH$^-$+He~\cite{GonzalezJPB06}, OH$^+$+He~\cite{GonzalezEPJD07}, OH$^-$+Rb~\cite{GonzalezEPJD08,TacconiJCP09,GonzalezCP15,GonzalezNJP15}, CH$^+$+He~\cite{StoecklinEPJD08,HammamiJMS08}, LiH$^-$+He~\cite{LopezEPJD09,GonzalezJPB16}, NO$^+$+He~\cite{StoecklinJCP11}, MgH$^+$/BaH$^+$/SiO$^+$+He~\cite{PerezPRA16}, H$_2^+$/D$_2^+$+He~\cite{SchillerPRA17}, BaRb$^+$+Rb~\cite{Perez2018}.
Similarly, the prospects for sympathetic cooling of neutral molecules with atomic ions were theoretically investigated for systems of Li$^+$+D$_2$~\cite{BovinoPRA08}, Be$^+$/Mg$^+$/Ca$^+$+OH~\cite{RobicheauxPRA14}.

In addition,~\cite{YzombardPRL15} theoretically showed that laser cooling of molecular anions can be feasible and identified a number of potential candidates such as C$_2^-$ or BN$_2^-$, ZnO$^-$ or LiF$_2^-$. Realization of such a laser cooling of molecular anions can open the way to sympathetic cooling of other negatively charged atomic or molecular ions.

\subsection{Cold chemistry}

Using molecular ions instead of atomic ones in cold ion-neutral systems provides much larger number of possible species combinations to be explored and opens the way for investigating proper cold chemical reactions with rearrangement of atoms, which can be of interest in many areas of chemistry ranging from organic and inorganic chemistry to astrochemistry~\cite{McDaniel70,MikoschIRPC10}.

The ion-neutral interactions have a longer-range character as compared to neutral ones and ionic species can be manipulated and detected on the single particle level. This can allow one to investigate chemical reactions with particle densities of molecular ions much smaller and better controlled than with neutral molecules. The ionic products of chemical reactions can also be trapped thus opening the way for measuring product-state distributions and state-to-state rates~\cite{TongCPL12,HeazlewoodARPC15}.

The first attempts to cold chemistry with molecular ions are related to using multipole ion traps and cold He and H$_2$ buffer-gas cooling~\cite{ItanoPS95,GerlichPS95,GerlichFD09}. For example, \cite{OttoPRL08} studied the $\mathrm{NH}_2^-+\mathrm{H}_2\to \mathrm{NH}_3+\mathrm{H}^-$ reaction at temperature as low as 8$\,$K and observed significant enhancement of the reaction rate with decreasing temperature. \cite{MulinPCCP15} observed isotope exchange in cold reactions of OH$^-$ with D$_2$ and of OD$^-$ with H$_2$.
\cite{HauserNJP15} investigated complex formation and internal proton-transfer of OH$^-$+H$_2$ complexes at low temperature.

Another approach is based on monitoring the decay of laser-cooled ions in a Coulomb crystal colliding with velocity- or state-selected molecules in a beam. \cite{WillitschPRL08} studied chemical reactions between Ca$^+$ ions and quadrupole-guided and velocity-selected CH$_3$F molecules at collision energies as low as 1$\,$K. \cite{ChangScience13} observed specific chemical reactivities of 3-aminophenol conformers, spatially separated by electrostatic deflection, with cold Ca$^+$ ions. They measured a twofold larger rate constant for a more polar conformer as a consequence of conformer-specific differences in the long-range ion-molecule interaction potentials (different leading long-range electrostatic coefficients $C_{2,1}^\mathrm{elst}$ for conformers with different dipole moments). In a similar experiment, \cite{KilajNC18} observed different reactivities of H$_2$O molecules in para and ortho states in collisions with trapped diazenylium ions. A state-controlled reaction between laser cooled Ca$^+$ ions and warm NO~\cite{GreenbergPRA18} or O$_2$~\cite{Schmid2019} radical molecules were also investigated and the control over the reaction rates by tuning of the excited-state population of the laser cooled Ca$^+$ ions was demonstrated. 

\begin{figure}[tb!]
\begin{center}
\includegraphics[width=0.75\columnwidth]{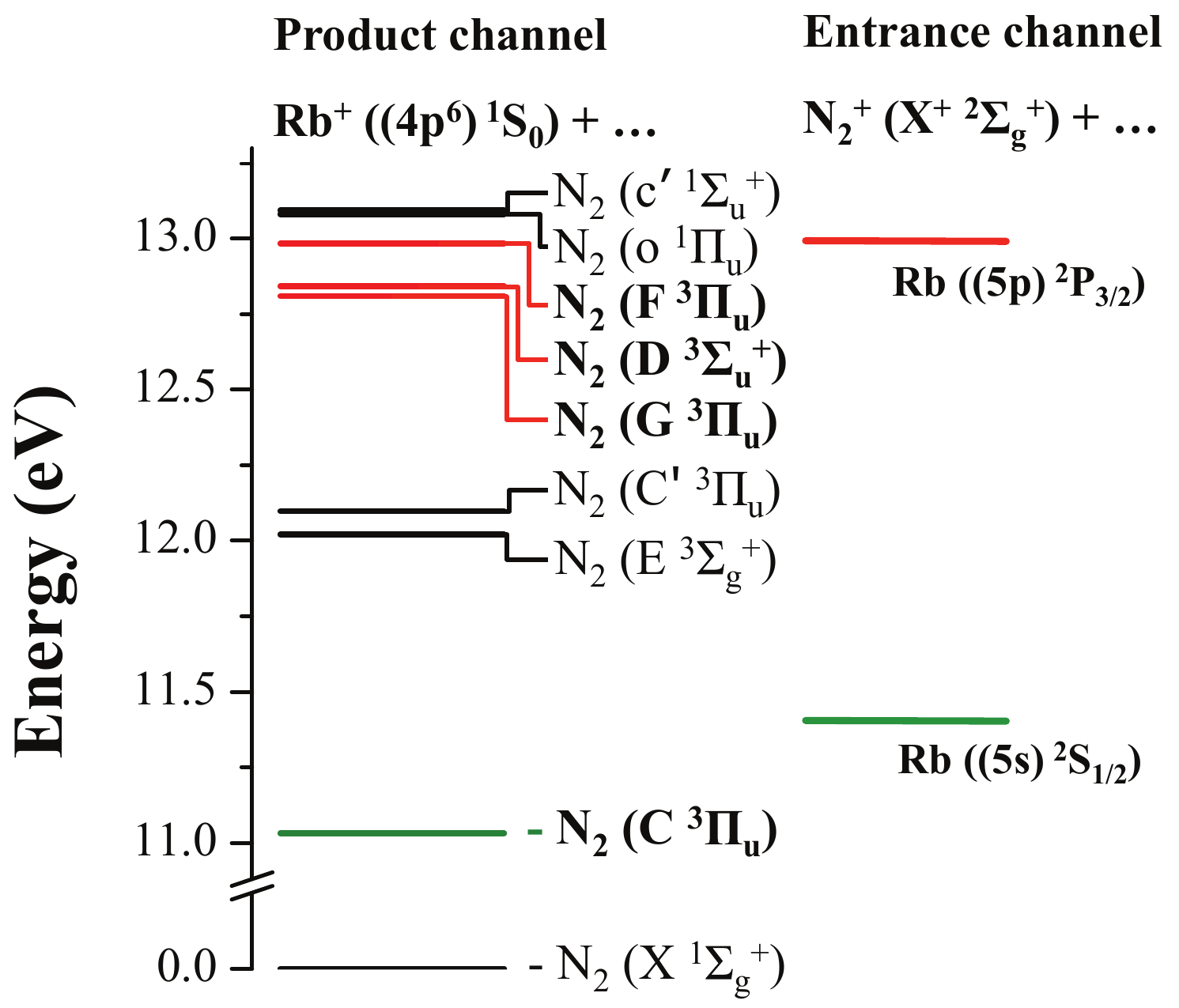}
\end{center}
\caption{Total electronic energies of N$_2^+$+Rb reactant (entrance) and Rb$^+$+N$_2$ product channels. The channels are characterized by the electronic states of the species as indicated. Charge transfer is most efficient in case of a near resonance between the entrance and product channels. The product channels closest in energy to the N$_2^+$+Rb (5s) and (5p) entrance channels are indicated in green and red, respectively. From~\cite{HallPRL12}.}
\label{fig4_FelixPRL12}
\end{figure}

The first experiments combining molecular ions with ultracold atoms in hybrid traps were realized for the N$_2^+$+Rb and OH$^-$+Rb systems by~\cite{HallPRL12} and ~\cite{DeiglmayrPRA12}, respectively. In the former, the reactive collisions between sympathetically cooled molecular ions and laser-cooled atoms in an ion-atom hybrid trap at collision energies as low as 20$\,$mK were investigated and a strong dependence of the reaction rate on the internal state of Rb atoms was found. Specifically, highly efficient charge exchange, four times faster than the Langevin rate, was observed with Rb in the excited $^2$P$_{3/2}$ state. This observation was rationalized by a capture process dominated by the charge-quadrupole interaction and a near resonance between the entrance and exit channels of the system (see.~Fig.~\ref{fig4_FelixPRL12}).

In the latter experiment, OH$^-$ anions were trapped in an octupole rf trap and Rb atoms in a magneto-optical trap. The associative electron detachment process,
\begin{equation}
\mathrm{OH}^-+\mathrm{Rb} \to \mathrm{RbOH}+e^-\,,
\end{equation}
was identified as the main loss mechanism and absolute rate coefficients for this process were measured. The experimental efforts for the OH$^-$+Rb system are accompanied by several related theoretical works~\cite{ByrdPRA13,KasJCP16,MidyaPRA16,KasJCP17,TomzaPCCP17}.

Recently, synthesis of mixed hypermetallic oxide BaOCa$^+$ from laser-cooled reagents in an cold ion-atom hybrid trap was  demonstrated~\cite{PuriScience17}. Cold barium methoxide ions (BaOCH$_3^+$) were first prepared and then exposed to ultracold Ca atoms. Mass spectral and theoretical analyses revealed a barrierless reaction pathway in which triplet-state Ca($^3P_J$) atoms displaces the methyl group.  The reaction kinetics as a function of collision energy over the range 0.005\,K to 30 K and of individual Ca fine-structure levels were investigated and compared favorably with calculations based on long-range capture theory.


Cold and ultracold chemical reactions of molecular ions with neutral atoms have also been theoretically studied for few other molecular ion-neutral systems: MgH$^+$+Rb~\cite{TacconiEPJD09}, $^3$He$^4$He$^+$+$^4$He~\cite{BodoPS09}, HeH$^+$+He~\cite{BovinoJPCA11}, D$^+$+H$_2$~\cite{LaraJCP15,LaraPRA15}.

\subsection{Precision spectroscopy}

Cooling translational motion of atoms, ions or molecules reduces Doppler broadening of spectral lines, and cooling their internal motion can bring them into a single quantum state. All these significantly facilitate high accuracy spectroscopic measurements. Cold and ultracold molecules are promising candidates for high precision measurements to probe both electronic structure theory as well as fundamental aspects of quantum physics such as the electron's electric dipole moment or the time-variation of fundamental constants, e.g.~the electron-proton mass ratio and the fine-structure constant~\cite{CarrNJP09}. Molecular ions formed at ultralow temperatures or cooled down via buffer gas or sympathetic cooling have an additional advantage of possibly long storage time in a Paul trap. 

The first precision spectroscopy and measurement of the electron's electric dipole moment through electron spin resonance spectroscopy on the $^{180}$Hf$^{19}$F$^+$ molecular ions in their metastable $^3\Delta_1$ electronic state were demonstrated by~\cite{LohScience13,Cairncross17}. 

Electric-dipole-forbidden infrared transitions in cold molecular ions were observed by \cite{GermannNatPhys14}. This measurement was enabled by the very long interrogation times afforded by the sympathetic cooling of individual quantum-state-selected molecular ions trapped in a Coulomb crystal.

The high-sensitivity transitions for the precision measurement of time-variation of the proton-to-electron mass ratio were theoretically identified e.g.~in N$_2^+$~\cite{KajitaPRA14} and O$_2^+$~\cite{HannekePRA16} molecular ions. 
The precision measurement of the rotational Lamb shift of molecular ion's rotational structure, while immersed in a Bose-Einstein condensate, was theoretically proposed by~\cite{MidyaPRA16}, as described in Sec.~\ref{sec:impurity}.

Larger number of precision measurement proposals and experimental realizations  can be expected when molecular ions are routinely formed at or sympathetically cooled to low or ultralow temperatures.

\subsection{Laboratory astrochemistry}

One of the early motivations to study cold ion-atom and ion-molecule collisions, especially with hydrogen and helium at cryogenic temperatures, was the astrochemical context. Typical temperatures of the interstellar clouds are between few and several Kelvins. For the collisions with very light partners, like helium and hydrogen, these temperatures can results in a similar number of only a few partial waves (similar collision energy in units of $E^\star$) as for cold hybrid ion-atom systems employing heavy alkali- and alkaline-earth-metal ions and atoms operating at miliKelvin temperatures. Thus, the knowledge gained about cold collisions and especially chemical reactions using cold hybrid ion-atom experiments can be instructive for better understanding chemical transformations of ions in space~\cite{PetrieMSR07,Snow2008,LarssonRPR12}. 

Surprisingly, a number of cations, but just a few molecular anions, have been conclusively detected in the interstellar space~\cite{MillarCR17}. Precision spectroscopy and collisional studies of cold molecular anions can shed new light on their stability, properties of valence and dipole-bound excited states, which in turn can help to understand better their abundance in the universe.

The knowledge of rates for chemical reactions between simple atoms, molecules, and ions present in different astronomical environments is very important for simulating and understanding evolution of astronomical objects~(see.~e.g,~\cite{SmithARAA11,SnowARAA06,WakelamAJ12,McElroyAA13}). The state-selective measurements with selected cold ion-neutral systems may provide necessary reactive collisions rates with adequate accuracy.

Finally, many astronomical observations are based on the identification of molecular lines in a broad range of wavelengths (see~e.g~\cite{MullerAA01,SchoierAA05}). Precision spectroscopy of molecular ions sympathetically cooled in hybrid ion-atom traps may provide necessary spectra with adequate accuracy.

\section{Conclusions and outlook}
\label{sec:conclusions}

Over the past ten years, cold hybrid ion-atom systems have come a long way from a handful of pioneering experiments to an established field of research. The capabilities as well as technical limitations of currently used setups are now well understood. Useful information about charged impurity dynamics, inelastic collisions, and chemical reactions have been obtained. At the same time, multiple theoretical proposals for implementations of quantum technologies have yet to be realized. Here we describe some possible future directions for the next generation of experiments.

{\it Control of inelastic collisions}. The understanding of two-body and three-body inelastic processes in ion-atom mixtures is the first step towards controlling them. While it is possible to choose systems in which radiative charge exchange is not present~\cite{TomzaPRA15b}, it looks more appealing to be able to control the collision rates with external fields, which requires lowering the collision energy.

{\it Reaching the s-wave limit}. In state-of-the-art experimental setups the ion-atom collisions take place at energies for which many partial waves are involved. This is one of the reasons for the lack of observation of scattering resonances (e.g.~Feshbach resonances), as contributions from different partial wave channels cannot be tuned independently. To achieve full control of the collisional properties it will be necessary to reach the single partial wave limit. To this end, it is most convenient to work with combinations of heavy ions and light atoms such as Yb$^+$+Li. It is possible that novel experiments which do not use the Paul trap will play an important role in completing this task.

{\it Realization of cold hybrid ion-atom systems without the Paul trap}. As we described in detail, Paul traps impose strong limitations on the degree of control over the ion-atom system due to the inherent micromotion. Moving to a purely optical setup as described in Sec.~\ref{subsec:alt} would thus open new opportunities, especially in the context of quantum engineering. This wound require mastering the great experimental challenge of keeping the ion in a shallow optical trap in the presence of atoms for a long time.

{\it Combining ions with Rydberg atoms}. Another strategy to overcome the micromotion issues is to enhance the interaction strength so that the ion and the atom can be kept away from each other. This can be achieved by dressing the atoms with Rydberg states~\cite{SeckerPRA16,SeckerarXiv16}. Rydberg excitation of ions and exploring possibilities given by Rydberg molecules can also open new research directions. 

{\it State-to-state chemistry}. After reaching the $s$-wave regime and full control over quantum states of colliding particles, the realization of controlled state-to-state chemical reactions with detection of product-state distributions would be the next step. This requires the development of new methods of selective trapping and monitoring reaction products. On the other hand, universality of ion trapping and sympathetic cooling can open the way for experiments with $p$-block, transition-metal, or lanthanide ions. 

{\it More complex systems}. Cold ion-molecule systems and molecular ions are of great interest from the chemical point of view (in particular for astrochemical studies and chemistry of atmospheres). Simple molecular ions are abundant in interstellar clouds along with hydrogen molecules, and precise knowledge of their reaction rates is crucial for our understanding of the evolution of the universe. Especially, the knowledge about molecular and atomic anions and their chemical transformations is limited. Multiply charged atomic and molecular ions are also completely unexplored in the context of cold hybrid ion-atom systems.  

{\it Larger systems}. Periodic arrangements of both atoms and ions in cold ion-atom systems are not well studied yet. Combining cold ions with atoms in an optical lattice in the context of quantum information storing and processing is still awaiting experimental realization. Similarly, combinations of atoms in an optical lattice and crystallized ions in one- or two-dimensional arrays can bring many opportunities for interesting quantum simulations~\cite{SchneiderRPP12}. 

{\it Quantum technologies}. Finally, cold hybrid ion-atom systems are also a candidate to be a platform for engineering novel applications within  emerging quantum technologies especially in the context of quantum computation, simulation, and sensing.
Even through they are less advanced than other practical technological solutions based on trapped ions or on solid-state systems, the unique properties of cold hybrid ion-atom systems can provide important proof-of-principle experimental demonstrations and in some cases (see Sec.~\ref{seq:quant_sim}) enable the exploration of specific regimes that are not easily accessible with other platforms.

\begin{acknowledgments}

We are grateful to U.~Bissbort, R.~C\^ot\'e, M.~Deiss, O.~Dulieu, M.~Drewsen, A.~Gl\"atzle, F.~Grusdt, B.~Heazlewood, J.~Hecker~Denschlag, E.~Hudson, C.~P.~Koch, S.~Kotochigova, F.~Meinert, V.~Melezhik, R.~Moszynski, T.~Mukaiyama, R.~Ozeri, T.~Pfau, S.~Rangwala, T.~Secker, T.~Sch\"atz, P.~Schmelcher, F.~Schmidt-Kahler, J.~Schurer, A.~Simoni, W.~Smith, T.~Wasak, R.~Wester, S.~Willitsch for valuable discussions about cold hybrid ion-atom systems and comments on the manuscript.

We acknowledge financial support: M.T.~from the National Science Centre Poland (Grants No.~2015/19/D/ST4/02173 and 2016/23/B/ST4/03231) and the Foundation for Polish Science within the Homing programme co-financed by the EU Regional Development Fund, K.J.~from the Alexander von Humboldt Foundation, R.G.~from the European Union via the European Research Council (Starting Grant 337638) and the Netherlands Organization for Scientific Research (Vidi Grant 680-47-538), A.N.~from the cluster of excellence the Hamburg Centre for Ultrafast Imaging of the Deutsche Forschungsgemeinschaft, T.C. from the Deutsche Forschungsgemeinschaft via SFB/TRR~21 Co.Co.Mat., Z.I.~from the National Science Centre Poland (Grant No.~2014/14/M/ST2/00015).

\end{acknowledgments}

\bibliography{ion-atom}

\end{document}